\documentclass[twocolumn,showpacs,preprintnumbers,amsmath,amssymb]{revtex4}
\usepackage{graphics}% Include figure files
\usepackage{graphicx}% Include figure files
\usepackage{dcolumn}% Align table columns on decimal point
\usepackage{bm}% bold math
\usepackage{epsfig}

\begin{document}

%*****************************************************

%LA-UR-00-5654
 
\title{Study of the anomalous  acceleration   of Pioneer 10 and 11} 
 
\author{{John D. Anderson},\footnote{Electronic address:
        john.d.anderson@jpl.nasa.gov}$^a$  
{Philip A. Laing},\footnote{Electronic address: Philip.A.Laing@aero.org}$^b$ 
{Eunice L. Lau},\footnote{Electronic address: 
        Eunice.L.Lau@jpl.nasa.gov}$^a$ \\
{Anthony S. Liu},\footnote{Deceased (13 November 2000).}$^c$
{Michael Martin Nieto},\footnote{Electronic address: mmn@lanl.gov}$^d$ 
        and 
{Slava G. Turyshev}\footnote{turyshev@jpl.nasa.gov}$^a$ \\
~~~~ }

\affiliation{$^a$Jet Propulsion Laboratory, California Institute
of  Technology, Pasadena, CA 91109 \\
$^b$The Aerospace Corporation, 2350 E. El Segundo Blvd., 
El Segundo,  CA 90245-4691\\
$^c$Astrodynamic Sciences, 2393 Silver Ridge Ave., Los Angeles, CA 90039 \\
$^d$Theoretical Division (MS-B285), 
Los Alamos National Laboratory,\\
University of California, Los Alamos, NM 87545}

\date{11 April 2002}

%%************************ABSTRACT

\begin{abstract} 
Our previous analyses of radio Doppler and ranging data from  distant
spacecraft in  the solar system indicated that an apparent anomalous
acceleration is acting on  Pioneer 10 and 11,  with a magnitude  
$a_P\sim  8\times 10^{-8}$ cm/s$^2$,  directed towards the Sun.
% \cite{anderson,moriond}.  
 Much effort has been expended looking for
possible systematic origins of  the residuals, but none has been found.
A detailed
investigation of effects both external to and internal to the
spacecraft, as well as those due to modeling and computational
techniques, is provided.  We also discuss  the methods, theoretical
models, and experimental techniques used to detect and study small
forces acting on   interplanetary spacecraft.  These include the methods
of radio Doppler data collection, data editing, and data reduction.   
  
There is now further data for the Pioneer 10 orbit determination.   The
extended Pioneer 10 data set spans 3 January 1987 to 22 July 1998.   [For
Pioneer 11 the shorter span goes from 5 January 1987 to the time of  loss
of coherent data on 1 October 1990.]  With these data sets  and more
detailed studies of  all the systematics, we now give a result, of  
$a_P = (8.74 \pm 1.33) \times 10^{-8} ~~{\rm cm/s}^2$.   (Annual/diurnal
variations on top of $a_P$, that leave $a_P$ unchanged, are also
reported and discussed.)
\end{abstract}    


%\keywords{Suggested keywords}%Use showkeys class option if keyword
                              %display desired
\maketitle

%********************* CONTENTS 
 
%\tableofcontents
%\listoffigures
%\listoftables
 
%*********************1) INTRODUCTION

\section{\label{intro}INTRODUCTION}

Some thirty years ago, on 2 March  1972, Pioneer 10 was launched on
an Atlas/Centaur rocket from  Cape Canaveral.  Pioneer 10 was Earth's
first space probe to an outer planet.  Surviving intense radiation, it
successfully  encountered Jupiter on 4 December 1973 
\cite{science}-\cite{pioweb}.   In trail-blazing the exploration of the
outer solar system, Pioneer 10  paved the way for, among others,  Pioneer
11 (launched on  5 April 1973), the Voyagers, Galileo, Ulysses, and the
upcoming  Cassini encounter with Saturn. After Jupiter and  (for Pioneer
11) Saturn encounters, the two spacecraft followed hyperbolic orbits
near  the plane of the ecliptic to  opposite sides of the solar
system.   Pioneer 10 was also the first mission  to enter the edge of
interstellar space. That major event occurred in June 1983, when Pioneer
10 became the first spacecraft to  ``leave the solar system'' as it 
passed beyond the orbit of the farthest known planet. 

The scientific data collected by Pioneer 10/11 has yielded unique
information about the outer region of the solar system.   This is due  in
part to the spin-stabilization  of the Pioneer spacecraft.   At launch
they  were spinning at approximately  4.28 and 7.8  revolutions per
minute (rpm),  respectively, with the spin axes running through the
centers of the dish  antennae.   Their spin-stabilizations  and   great
distances from the Earth  imply     a minimum number of Earth-attitude
reorientation maneuvers are required.  This permits precise acceleration
estimations, to the level of
$10^{-8}$ cm/s$^2$ (single measurement accuracy averaged  over 5 days).
Contrariwise, a Voyager-type three-axis stabilized spacecraft is not well
suited for a precise celestial mechanics experiment as its numerous 
attitude-control maneuvers can  overwhelm the signal of a small external
acceleration. 

In summary, Pioneer spacecraft represent an ideal  system to perform
precision celestial mechanics experiments.  It is relatively  easy to
model the spacecraft's behavior and, therefore, to study  small forces
affecting its  motion in the dynamical environment of the solar system.
Indeed,  one of the main objectives of the Pioneer extended missions   
(post Jupiter/Saturn encounters) \cite{extended}  was   to perform
accurate celestial mechanics experiments.  For instance, an attempt  was
made to  detect the presence of  small bodies in the solar system,
primarily in the Kuiper belt. It was hoped that a small perturbation of
the spacecraft's trajectory would reveal the  presence of these objects
\cite{jdakuiper}-\cite{pulsar}.    Furthermore, due to extremely precise
navigation and a high quality tracking data, the Pioneer 10 scientific 
program also included a search for low frequency gravitational waves 
\cite{anderson85,anderson93}. 

Beginning in 1980, when at a distance of 20 astronomical units (AU) from
the Sun the solar-radiation-pressure  acceleration on Pioneer 10 {\it
away} from the Sun  had decreased to  $< 5 \times 10^{-8}$ cm/s$^2$, we 
found that the largest systematic error in the acceleration residuals
was a constant bias, $a_P$,   directed {\it toward} the Sun. Such
anomalous data have been  continuously received ever since.  Jet
Propulsion Laboratory (JPL) and The Aerospace Corporation produced
independent orbit determination   analyses of the Pioneer data extending
up to July 1998. We ultimately concluded \cite{anderson,moriond},  that
there is an unmodeled  acceleration, $a_P$, towards the Sun  of $\sim 8
\times 10^{-8}$  cm/s$^2$ for both Pioneer 10 and Pioneer 11.

The purpose of this paper is to present a detailed explanation of the
analysis of the  apparent anomalous, weak,  long-range   acceleration  of
the Pioneer spacecraft  that we detected in the outer regions of the
solar system. We attempt to survey all sensible forces  and to  estimate
their contributions to the anomalous acceleration.  We will  discuss the
effects of these small non-gravitational forces (both generated on-board
and external to the vehicle) on the motion of the distant spacecraft
together with the methods used to collect and process the radio Doppler
navigational data.  

We begin with descriptions of the spacecraft and other systems and 
the strategies for obtaining and analyzing information from them. 
%  1)Intro is Sec. 1.
%  2)
In Section \ref{pioneer} we  describe the Pioneer 
(and other) spacecraft. We  
provide the reader with  important technical information on  the
spacecraft, much of which  is not easily accessible.   
%  3)
In Section \ref{Exp_tech} we describe how raw data is obtained  
and analyzed 
%  4)
and in Section \ref{navigate} we discuss the basic elements of a
theoretical foundation for spacecraft navigation in the solar system.

The next major part of this manuscript is a description and 
analysis of the results of this investigation.  
% 5) 
We first describe how the anomalous acceleration was originally identified   
from the data of all the spacecraft in Section \ref{results}
\cite{anderson,moriond}.  
%  6) 
We then give  our recent results in Section \ref{recent_results}.
%%%  7), 8), AND 9)
In the following three sections  we discuss possible  experimental
systematic origins for the  signal. These include systematics generated
by physical phenomena from sources external to  (Section
\ref{ext-systema}) and internal to  (Section
\ref{int-systema}) the spacecraft.  This is followed by Section
\ref{Int_accuracy}, where the  accuracy of the solution for $a_P$ is
discussed.  In the process  we go over possible numerical/calculational
errors/systematics. 
%  10) 
Sections \ref{ext-systema}-\ref{Int_accuracy} are then summarized in 
the total error budget of Section \ref{budget}.

We end our presentation by first considering 
%  11)
possible unexpected physical origins for the anomaly (Section \ref{newphys}).
%  12)
In our conclusion, Section \ref{disc},  we  summarize our results and
suggest  venues for further study of the discovered anomaly.

%************************2) PIONEER SPACECRAFT AND MISSION**********
%\newpage

\section{\label{pioneer}THE PIONEER AND OTHER SPACECRAFT}

In this section we describe in some detail the Pioneer 10 and 11 spacecraft 
and their missions.  We concentrate on those spacecraft 
systems that play important roles in maintaining the continued 
function of the vehicles and in determining their dynamical behavior in
the solar system. Specifically we  present an overview of propulsion and
attitude control systems, as well as thermal and communication systems.  

Since our analysis addresses certain results from  the
Galileo and Ulysses missions, we also give short  
descriptions of these missions in the final subsection.    

%************************** Pioneer Spacecraft  

\subsection{General description of the Pioneer spacecraft}
\label{sec:pio_description}

\begin{figure}[h]
\begin{center}
\noindent \vskip  4pt   
\psfig{figure=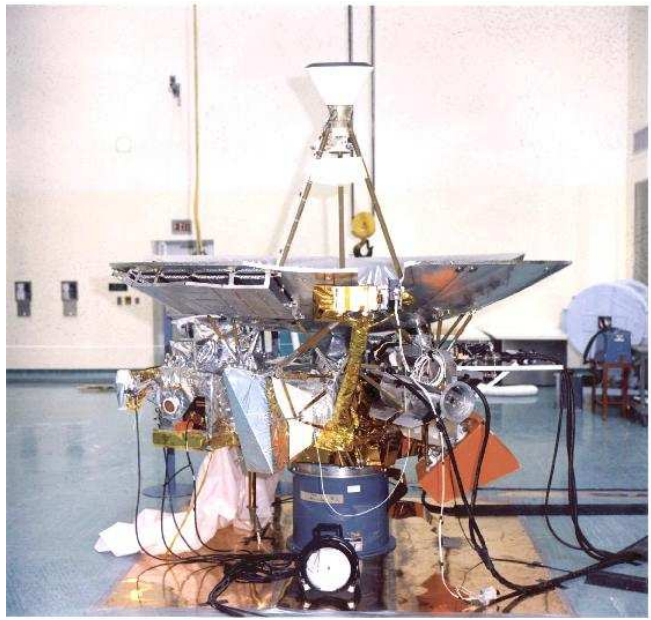,width=86mm}%,height=90mm}
\end{center}
\caption[NASA photo \#72HC94, with caption ``The Pioneer F spacecraft 
during a checkout with the launch vehicle third stage at Cape Kennedy.'']
{NASA photo \#72HC94, with caption ``The Pioneer F spacecraft 
during a checkout with the launch vehicle third stage at Cape Kennedy.''
Pioneer F  became Pioneer 10. 
\label{fig:pio_design}}
\end{figure}

\begin{figure*}[ht]
\begin{center}\noindent    
\psfig{figure=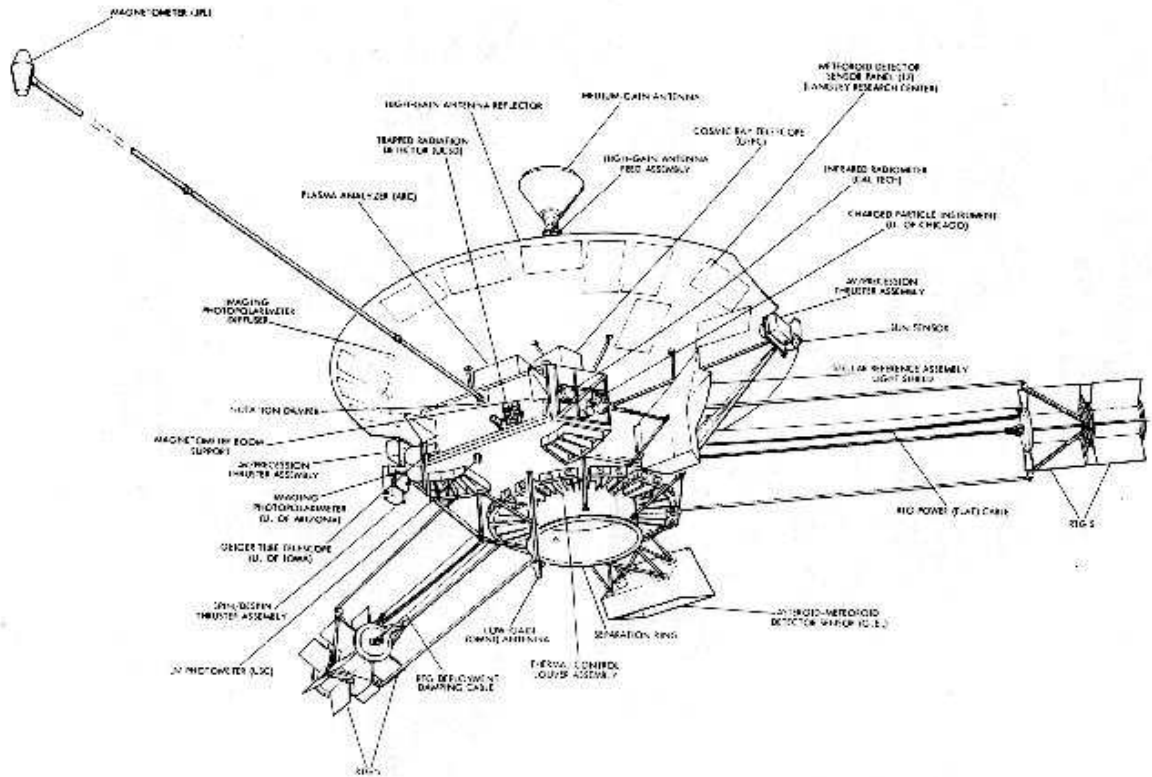,width=155mm}%,height=90mm}
\end{center}\vskip -10pt
\caption{A drawing of the Pioneer spacecraft.  
 \label{fig:trusters}}
\end{figure*}

Although some of the more precise  details are often difficult to
uncover,  the general parameters of the Pioneer spacecraft are known
and  well documented \cite{science}-\cite{pioweb}. The two spacecraft
are identical in design \cite{design}. At   launch  each had a ``weight''
(mass) of  259 kg.  The   ``dry weight'' of the total module was 223 kg 
as there were  36 kg of hydrazine propellant \cite{mass,gasuse}.  The
spacecraft were designed to fit within the three meter diameter shroud of an
added third stage to the Atlas/Centaur launch vehicle. Each spacecraft is 
2.9 m long from its base to its cone-shaped medium-gain antenna.  
The high gain antenna (HGA) is  made of aluminum honeycomb
sandwich material.  It is 2.74 m in diameter and 46 cm deep in the shape
of a parabolic dish. (See Figures 
\ref{fig:pio_design} and \ref{fig:trusters}.)

%************
%************

%************

%************

The main equipment compartment is 36 cm deep.  The hexagonal flat top and
bottom have 71 cm long sides.  The equipment compartment  provides a
thermally controlled environment for scientific instruments. Two
three-rod trusses, 120 degrees apart, project from two sides of the
equipment compartment. At their ends,  each holds two SNAP-19  (Space
Nuclear Auxiliary Power, model 19) RTGs  (Radioisotope Thermoelectric
Generators) built by Teledyne Isotopes for the Atomic Energy Commission.
These RTGs are situated about 3 m from the center of the spacecraft and
generate its electric power.  [We will go into more detail on
the RTGs in  Section \ref{int-systema}.] A third single-rod boom, 120
degrees from the other two, positions a magnetometer about 6.6 m from the
spacecraft's center. All three booms were extended after launch. With the
mass of the magnetometer  being 5 kg and the mass of each of the four
RTGs being  13.6 kg, this configuration defines the  main  moment of
inertia along the $z$-spin-axis.  It is  about 
${\cal I}_{\tt z} \approx 588.3$ kg m$^2$ \cite{vanallen}. 
[Observe that this all left only about 164 kg for the main
bus and superstructure, including the antenna.]

Figures \ref{fig:pio_design} and \ref{fig:trusters}  show the arrangement
within the spacecraft equipment compartment.  The majority of the
spacecraft electrical assemblies are located in the central hexagonal
portion of the compartment, surrounding a 16.5-inch-diameter
spherical hydrazine tank.  Most of the scientific instruments' electronic
units and internally-mounted sensors are in an instrument bay (``squashed''
hexagon) mounted on one side of the central hexagon. The equipment
compartment is in an aluminum honeycomb structure.  This 
provides support and meteoroid protection. It is  covered with insulation
which, together with louvers under the platform, provides 
passive thermal control.  [An  exception is from off-on control by
thermal power dissipation of some subsystems. (See  
Sec. \ref{int-systema}).]

%*****************

\subsection{Propulsion and attitude control systems}
\label{sec:prop}

Three pairs of these rocket thrusters near the rim
of the HGA provide a threefold function of spin-axis precession,
mid-course trajectory correction,  and spin control.  
Each of the three thruster pairs develops its repulsive jet force from 
a catalytic decomposition of liquid hydrazine in a small rocket 
thrust chamber attached to the oppositely-directed nozzle.
The resulted hot gas  is then expended through six individually controlled
thruster nozzles to effect spacecraft maneuvers. 

The spacecraft is attitude-stabilized by spinning about an axis which is
parallel to the axis of the HGA.  The nominal spin rate for Pioneer 10 is
4.8 rpm. Pioneer 11 spins at approximately 7.8 rpm because a
spin-controlling  thruster malfunctioned during the spin-down
shortly after launch.  [Because of the danger that the thruster's valve
would not be able to close again, this particular thruster has not been
used since.] During the mission an Earth-pointing attitude is
required to illuminate the  Earth with  the narrow-beam HGA. Periodic
attitude adjustments are required throughout the mission to compensate
for the variation in the heliocentric longitude of the Earth-spacecraft
line.  [In addition, correction of launch vehicle injection errors were
required to provide the desired Jupiter encounter trajectory and Saturn
(for Pioneer 11) encounter trajectory.]  These velocity vector
adjustments involved reorienting the spacecraft to direct the thrust in
the desired direction.

There were no anomalies in the engineering telemetry from the propulsion
system, for either spacecraft, during any mission phase from launch to 
termination of the Pioneer mission in March 1997. From the viewpoint of
mission operations at the NASA/Ames control center, the propulsion system
performed as expected, with no catastrophic or long-term pressure drops in
the propulsion tank.  Except for the above-mentioned Pioneer 11
spin-thruster incident, there was no  malfunction of the propulsion nozzles,
which were only opened every few months by ground command. The fact that
pressure was maintained in the tank has been used to infer that no impacts
by Kuiper belt objects occurred, and a limit has been placed on the size
and density distribution of such objects
\cite{jdakuiper}, another useful scientific result.

For attitude control, a star sensor (referenced to Canopus) and two
sunlight sensors provided reference for orientation and roll maneuvers.
The star sensor on Pioneer 10 became inoperative at Jupiter
encounter, so the sun sensors were used after that.  
For Pioneer 10, spin calibration was done by the DSN until
17 July 1990.    From 1990 to 
1993 determinations  were made by analysts 
using data from the  Imaging Photo Polarimeter (IPP).  
After the 6 July 1993 maneuver, 
there was not enough power left to support the IPP.  
But approximately every six months analysts still could get a rough 
determination using information obtained from  conscan maneuvers 
\cite{conscan} on an uplink signal.
When using conscan, the high gain feed is off-set. 
Thruster firings are used  to spiral in to the correct pointing of the
spacecraft antenna to  give the maximum signal strength. To run this
procedure (conscan and attitude) it is  now necessary to  turn off the
traveling-wave-tube (TWT) amplifier. So far, the power and tube life-cycle
have worked and the Jet Propulsion Laboratory's (JPL) Deep Space Network
(DSN) has been  able to reacquire the signal. It takes about 15 minutes
or so to do a maneuver. 
[The magnetometer boom incorporates a hinged, viscous, damping
mechanism at its attachment point, for passive nutation control.]

In the extended mission phase, after Jupiter and Saturn encounters, the
thrusters  have been used for precession maneuvers only. Two pairs of 
thrusters at opposite sides of the spacecraft have nozzles directed along
the spin axis, fore and aft  (See Figure \ref{fig:trusters}.)   In 
precession mode, the thrusters are fired by opening  one nozzle in each
pair.  One fires to the front and the other fires to the rear of the
spacecraft  \cite{rearfront}, in brief thrust pulses.   Each thrust pulse
precesses the spin axis a few tenths of a degree until the desired
attitude is reached. 

The two nozzles of the third thruster pair, no longer in use,  are aligned
tangentially to the antenna rim.  One points in the direction  opposite to
its (rotating) velocity vector and the other with it.  These were used
for spin control. 

%*************

\subsection{Thermal system and on-board power}
\label{sec:onboard}

Early on the spacecraft instrument compartment is thermally controlled 
between $\approx$ $0$ F and 90 F. This is done with the aid  of
thermo-responsive louvers located at the bottom of the  equipment
compartment. These louvers are adjusted by bi-metallic springs.    They
are completely closed below $\sim40$ F and completely  open above 
$\sim 85$ F.  This  allows controlled heat to escape in the
equipment compartment.  Equipment is kept within an operational range of
temperatures by multi-layered blankets of insulating aluminum plastic. 
Heat is provided by electric heaters, the heat from the instruments 
themselves,  and by twelve one-watt radioisotope heaters powered directly
by non-fissionable plutonium 
($^{238}_{~94}$Pu$ \rightarrow ^{234}_{~92}$U$+{}^4_2$He).

$^{238}$Pu, with a half life time of 87.74 years, also provides the
thermal source for the thermoelectric devices in the RTGs. Before launch,
each spacecraft's four RTGs delivered a total of  approximately 160 W of
electrical power \cite{tele,Rconf}.   Each of the four space-proven
SNAP-19 RTGs converts 5 to 6 percent of the heat released from  plutonium
dioxide fuel to electric power. RTG power is greatest at 4.2 Volts; an
inverter boosts this to 28 Volts for distribution. RTG life is degraded
at low currents; therefore, voltage is regulated by shunt dissipation of
excess power. 

The power subsystem controls and regulates the RTG power output with
shunts, supports the spacecraft load, and performs battery load-sharing.
The silver cadmium battery consists of eight cells of 5 ampere-hours
capacity each.  It supplies pulse loads in excess of RTG capability 
and may be used for sharing peak loads. 
The battery voltage is often discharged and charged.  This can be seen
by telemetry of the battery discharge current and charge current

At launch each RTG supplied about 40 W to the input of the $\sim 4.2$ V
Inverter Assemblies.  (The output for other uses includes the DC bus 
at 28 V and the AC bus at 61 V)  Even 
though electrical power degrades with time
(see Section \ref{subsec:mainbus}),  at $-41$ F the essential 
platform temperature as of the year 2000 is still between the acceptable 
limits of $-63$ F to 180 F. The RF power output from the 
% TWT-A 
traveling-wave-tube amplifier is still operating normally. 

The equipment compartment is insulated from extreme heat influx with
aluminized mylar and kapton blankets. Adequate warmth is provided by
dissipation of 70 to 120 watts of electrical power by electronic units
within the compartment; louvers regulating the release of this heat below
the mounting platform maintain  temperatures in the vicinity of the
spacecraft equipment and scientific instruments within operating limits.
External component temperatures are controlled, where necessary, by
appropriate coating and, in some cases, by radioisotope or electrical
heaters.

The energy production from the radioactive decay obeys an exponential
law. Hence, 29 years after launch, the radiation from Pioneer 10's RTGs
was about 80 percent of its original intensity. However the electrical
power delivered to the equipment compartment has decayed at a faster rate
than the $^{238}$Pu decays radioactively.   Specifically, 
the electrical power first decayed very quickly and then slowed  to
a still fast linear decay  \cite{lasher}. By 1987 the 
degradation  rate was about $-2.6$  W/yr for Pioneer 10 and even greater 
for the sister spacecraft.

This fast depletion rate  of electrical power from the RTGs
is caused by normal deterioration of the thermocouple
junctions in the thermoelectric devices.

The spacecraft needs 100 W to power all systems, including 26 W for the
science instruments.  Previously, when the available electrical power was
greater than 100 W,  the excess power was either thermally radiated into
space by a  shunt-resistor radiator or it was used to charge a battery in
the equipment compartment. 

At present only about 65 W of power is available to Pioneer 10
\cite{theorypower}.  Therefore,  all the instruments are no longer able to
operate simultaneously.   But the power subsystem  continues to provide
sufficient power to support the current spacecraft load: transmitter,
receiver, command and data handling, and the Geiger Tube Telescope  (GTT)
science instrument. As pointed out in Sec. \ref{subs:pioneer},  
the science package and transmitter are turned off in extended cruise mode 
to provide enough power to fire the attitude control thrusters.

%*****************

\subsection{Communication system}

The Pioneer 10/11 communication systems use S-band 
($\lambda\simeq 13$ cm) Doppler  frequencies
\cite{sband}. The communication uplink from Earth is at 
approximately 2.11 GHz.  The
two spacecraft transmit continuously at a power of eight watts.   They
beam their signals, of approximate frequency 2.29 GHz,
to Earth by means of the parabolic 2.74 m
high-gain antenna. Phase coherency with the ground transmitters,
referenced to H-maser frequency standards, is maintained by means of an
S-band transponder with  the 240/221 frequency turnaround ratio (as
indicated by the values of the above mentioned frequencies).  

The communications subsystem provides for: i) up-link and down-link
communications; ii) Doppler coherence of the down-link carrier signal; 
and iii) generation of the conscan \cite{conscan} signal for closed loop
precession of the spacecraft spin axis towards Earth. S-band carrier
frequencies, compatible  with  DSN,  are used in
conjunction with a telemetry modulation of the down-link signal. The
high-gain antenna is used to maximize the telemetry data rate at extreme
ranges. The coupled medium-gain/omni-directional antenna with fore and aft
elements respectively, provided broad-angle communications at
intermediate and short ranges. For DSN acquisition, these three antennae
radiate a non-coherent RF signal, and for Doppler tracking, there is a
phase coherent mode with a frequency translation ratio of 240/221.

Two frequency-addressable phase-lock receivers are connected to the two 
antenna systems through a ground-commanded transfer switch and two
diplexers, providing access to the spacecraft via either signal path. The
receivers and antennae are interchangeable through the transfer switch by
ground command or automatically, if needed.

There is a redundancy in the communication systems, with two receivers
and two transmitters coupled to two traveling-wave-tube  amplifiers. 
Only one of the two redundant systems has been used for the extended
missions, however.

At launch,  communication  with the   spacecraft was at  a data rate  256
bps  for Pioneer 10  (1024  bps for Pioneer 11).   Data rate degradation
has been  $-1.27$ mbps/day for Pioneer 10  ($-8.78$ mbps/day for Pioneer
11). The DSN still continues to provide good data with the received signal
strength of about $-178$ dBm (only a few dB from the receiver threshold). 
The data signal to noise ratio  is still mainly under 0.5 dB.  The data
deletion rate is often between 0 and 50 percent, at times more.   However,
during the test of  11 March 2000, the average deletion rate was about 8
percent.  So, quality data are still available.  

\subsection{Status of the extended mission}
\label{subs:pioneer}

\begin{figure}[h]
\begin{center}\noindent    
\epsfig{figure=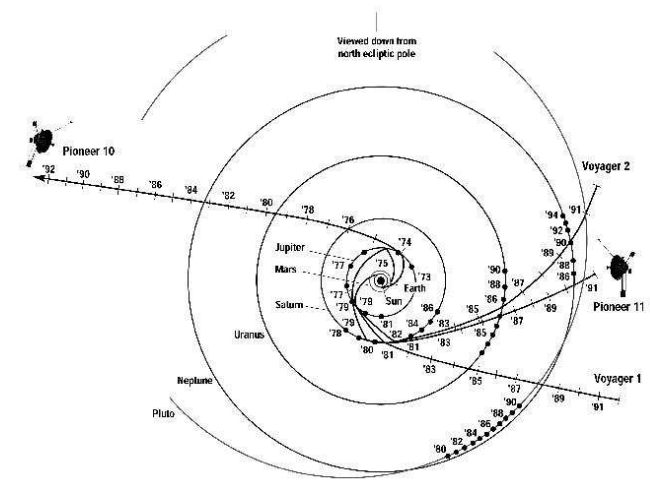,width=86mm}%,height=90mm}
\end{center}
\caption
{Ecliptic pole view of Pioneer 10, Pioneer 11, and Voyager trajectories.  
Pioneer 11 is traveling approximately in the direction of the 
Sun's orbital motion about the galactic center.  The galactic center 
is approximately in the direction of the top of the figure.   
[Digital artwork by T. Esposito. NASA ARC Image \# AC97-0036-3.]
 \label{fig:pioneer_path}}
\end{figure}

The  Pioneer 10 mission officially ended on 31 March 1997 when it was 
at a distance of 67  AU from the Sun.  
(See Figure \ref{fig:pioneer_path}.)  
At a now nearly constant velocity relative 
to the Sun of  
$\sim$12.2 km/s, Pioneer 10 will continue its motion 
into interstellar space, heading generally for the red star Aldebaran,
which forms the eye of  Taurus (The Bull) Constellation. 
Aldebaran is about 68 light years away  and it would be expected to
take Pioneer 10 over 2 million years to reach its neighborhood.

%************

%************

A switch failure in the Pioneer 11 radio system   on 1 October 1990
disabled the generation of coherent Doppler signals.  So,  after that
date, when the spacecraft was $\sim 30$ AU away from the Sun, no useful
data have been generated  for our scientific investigation.  Furthermore, 
by September 1995, its power source was nearly exhausted. Pioneer 11 could 
no longer make any scientific observations, and routine mission operations
were terminated.  The last communication from Pioneer 11 was received in
November 1995,  when the spacecraft was at distance of $\sim 40$ AU from
the Sun.  (The relative Earth motion carried it out of view of the
spacecraft antenna.)  The spacecraft is headed toward the constellation
of  Aquila  (The Eagle), northwest of the constellation of Sagittarius, 
with a velocity relative to the Sun of  $\sim$11.6 km/s
Pioneer 11 should pass  close to the nearest star in the constellation 
Aquila in about 4 million years \cite{pioweb}. 
(Pioneer 10 and 11 orbital parameters are given in the Appendix.)

However, after mission termination  the Pioneer 10 radio system was still
operating  in the coherent mode when commanded to do so from the Pioneer
Mission Operations center at the NASA Ames Research Center (ARC).  As a
result,  after 31 March 1997,  JPL's DSN was still able to  deliver
high-quality coherent data to us on a regular schedule from distances
beyond 67 AU. 

Recently, support of the Pioneer spacecraft has been on a non-interference
basis to other NASA projects.  It was used for the purpose of training 
Lunar Prospector controllers  in DSN  coordination of tracking
activities.  Under this training program,  ARC has been
able  to maintain contact with Pioneer 10.  This has required careful
attention to the DSN's ground system, including the installation of advanced
instrumentation, such as low-noise digital receivers.  This extended  the
lifetime of Pioneer 10 to the present. [Note  that the DSN's early
estimates, based on instrumentation in place in 1976, predicted  that
radio contact would be lost about 1980.]

At the present time  it is mainly the drift of the spacecraft relative to
the solar velocity that necessitates maneuvers to continue keeping 
Pioneer 10   pointed towards the Earth.  
The latest successful precession maneuver to point  the spacecraft to
Earth was accomplished on 11 February 2000,  when Pioneer
10 was at a distance from the Sun of 75 AU. 
[The distance from the Earth was $\sim 76$ AU 
with a corresponding  round-trip light time of about 21 hour.] 
The  signal level increased
0.5-0.75 dBm \cite{dBm}  as a result of the maneuver. 

This  was the seventh
successful maneuver that has been done in the blind since  26 January
1997.  At that time it had been determined that the electrical  power to
the spacecraft had degraded to the point where the spacecraft transmitter
had to be turned off to have   enough power to perform the maneuver.
After 90 minutes in the blind  the transmitter was turned back on again.
So, despite the continued weakening  of Pioneer 10's signal,
radio Doppler measurements were still available. 
The next attempt at a maneuver, on 8 July 2000, 
turned out in the end to be 
successful.  Signal was tracked on 
9 July 2001.  Contact was reestablished on the
30th anniversary of launch, 2 March 2002.
% 9 July 2001.  Another maneuver attempt
% will not be needed before early 2002.  

%************************

\subsection{The Galileo and Ulysses missions and spacecraft}
\label{othercraft}

\subsubsection{The Galileo mission}
\label{galileocraft}

The Galileo mission to  explore the Jovian system \cite{johnson}  was
launched  18 October 1989 aboard  the Space Shuttle Discovery.   Due to
insufficient launch power to reach its final destination  at 5.2 AU, a
trajectory was chosen with planetary flybys to gain   gravity assists.
The spacecraft flew by Venus  on 10 February 1990 and twice by the Earth,
on  8 December 1990 and on   8 December 1992.  The current Galileo
Millennium Mission continues  to study  Jupiter and its moons, and  
coordinated  observations with the Cassini flyby in December 2000.

The dynamical properties of the Galileo spacecraft are very well known. 
At launch the orbiter had a mass of 2,223 kg. This included  925 kg  
of usable propellant, meaning over 40\% of the orbiter's mass at  launch
was for propellant!  The science payload was 118 kg and the  probe's
total mass was 339 kg.  Of this latter,  the probe descent module was 121
kg, including a 30 kg science payload.  The tensor of inertia  of the
spacecraft had the following components at launch: $J_{\tt xx}= 4454.7, 
J_{\tt yy}= 4061.2, J_{\tt zz}= 5967.6, J_{\tt xy}= -52.9, J_{\tt xz}=
3.21,  J_{\tt yz}= -15.94$ in units of kg m$^2$. Based on  the area of the
sun-shade plus the booms and the RTGs we obtained a  maximal
cross-sectional area of  19.5 m$^2$. Each of the two of the Galileo's 
RTGs at launch delivered  of 285 W of electric power to the subsystems.

Unlike previous planetary spacecraft, Galileo featured an innovative  
``dual spin'' design: part of the orbiter would rotate constantly at
about  three rpm  and part of the spacecraft would remain  fixed in
(solar system) inertial space.  This means that the orbiter could easily
accommodate magnetospheric experiments (which need to made   while the
spacecraft is sweeping) while also providing stability and a fixed
orientation for cameras and other sensors.  The spin rate could  be
increased to 10 revolutions per minute for additional stability during
major propulsive maneuvers.

Apparently there was a mechanical problem between the spinning  and
non-spinning sections. Because of this, the project decided to often use 
an  all-spinning mode,  of about 3.15 rpm.  This was   especially true
close to the Jupiter Orbit Insertion (JOI), when  the entire spacecraft
was  spinning (with a slower rate, of course). 

Galileo's original design called for a deployable  high-gain antenna
(HGA) to unfurl. It would   provide approximately 34 dB of gain at
X-band (10 GHz) for a 134 kbps  downlink of science and priority
engineering data.  However,  the  X-band HGA failed to unfurl on 11 April
1991. When it again did not deploy following the Earth fly-by in 1992, 
the spacecraft was  reconfigured to utilize the S-band, 
8 dB, omni-directional  low-gain antenna (LGA) for 
downlink. 

The S-band frequencies are 2.113 GHz - up and 2.295 GHz 
- down, a conversion factor of 240/221  at the Doppler  frequency
transponder.  This configuration yielded much lower data rates 
than originally scheduled,  8-16 bps through JOI \cite{LGA}.
Enhancements at the DSN and reprogramming the flight computers on Galileo
increased telemetry bit rate to 8-160 bps,   starting in the spring of
1996. 
   
Currently, two types of Galileo navigation data are
available, namely  Doppler and range measurements.   As mentioned before,
an instantaneous comparison between the ranging signal that  goes up with
the ranging signal that  comes down would yield an ``instantaneous''
two-way  range delay. Unfortunately, an instantaneous comparison was not
possible in this case. The reason is that the signal-to-noise ratio on the
incoming ranging signal is small and a long integration time (typically 
minutes) must be used (for correlation purposes).  During such  long
integration times, the range to the spacecraft is constantly changing. It
is therefore necessary to ``electronically freeze'' the range delay long
enough to permit an integration to be performed.  The result represents
the range at the moment of freezing
\cite{anderson75,Kinman92}.

%****************************************************

\subsubsection{The Ulysses mission}
\label{ulyssescraft}

Ulysses was launched  on 6 October 1990,  also from  the  Space Shuttle
Discovery, as a cooperative project of NASA and the European Space Agency
(ESA).   JPL manages the US portion of the mission for NASA's Office of
Space Science.   Ulysses' objective was to characterize the heliosphere as
a function of  solar latitude \cite{genU}.  To reach high solar latitudes,
its voyage took  it to Jupiter on 8 February 1992. As a result, its orbit
plane was rotated about 80 degrees out of the ecliptic plane.

Ulysses explored the heliosphere over the Sun's south pole between June and
November,  1994, reaching  maximum Southern latitude of 80.2 degrees   on
13 September   1994.  It continued in its orbit out of the plane of the 
ecliptic, passing perihelion in March 1995 and over the north solar  pole 
between June and September 1995. It returned again to  the Sun's south
polar region in late 2000. 

The total mass at launch was the sum of two parts: a dry mass of   333.5 kg
plus a propellant mass of 33.5 kg. The tensor of inertia is given by its
principal components  $J_{\tt xx} =371.62, J_{\tt yy} = 205.51,   
J_{\tt zz} = 534.98$ in units kg m$^2$. The maximal cross section is
estimated to be 10.056 m$^2$.  This estimation is based on the radius of the
antenna 1.65 m (8.556 m$^2$)  plus the areas of the RTGs   and part of
the science compartment   (yielding an additional $\approx$ 1.5 m$^2$).  
The spacecraft was spin-stabilized at 4.996 rpm. The electrical power is
generated by modern RTGs, which  are located much closer to the main bus
than are those of the Pioneers.  The power generated  at launch was  285
W.

Communications with the spacecraft  are performed at X-band (for downlink
at 20 W with a conversion factor of 880/221) and  S-band  (both for uplink
2111.607 MHz and downlink 2293.148 MHz, at 5 W with a conversion factor of
240/221).  Currently both Doppler and range data are available for both
frequency bands. While the main communication link is S-up/X-down, the
S-down link was   used only for radio-science purposes. 

Because of Ulysses' closeness to the Sun and also because of  its
construction, any hope to model Ulysses for small forces might  appear to
be doomed by  solar radiation pressure and internal heat radiation from the
RTGs.    However, because the Doppler signal direction is towards the Earth 
while the radiation pressure varies with distance and has a  direction
parallel  the Sun-Ulysses line,   in principle these effects could be
separated.  And again, there was range data.  This all would  make it
easier to model  non-gravitational acceleration components normal to the
line of sight, which usually  are poorly and not significantly determined.

The Ulysses spacecraft spins at $\sim 5$ rpm around its antenna axis (4.996
rpm initially).  The angle of the spin axis 
with respect to the spacecraft-Sun line varies from near zero at Jupiter
to near 50 degrees at perihelion. 
Any on-board forces that could perturb the spacecraft trajectory are
restricted to a direction along the spin axis. [The other two components
are canceled out by the spin.] 

As the spacecraft and the Earth travel around the Sun, the direction from
the spacecraft to the Earth changes continuously. Regular changes of the
attitude of the spacecraft are performed throughout the mission to keep
the Earth within the narrow beam of about one degree full width of the
spacecraft--fixed parabolic antenna.

%*************************3) DATA ACQUISITION AND PREPARATION

\section{\label{Exp_tech}DATA ACQUISITION AND PREPARATION}

Discussions of radio-science experiments with spacecraft in the
solar system requires at least a general knowledge of the sophisticated
experimental techniques used at the DSN complex.  Since its 
beginning in  1958 the DSN complex has undergone  a number of  major
upgrades and additions. This was
necessitated by the needs of  particular space missions. [The last such
upgrade was conducted for the Cassini mission when the  DSN capabilities
were extended to cover the Ka radio frequency bandwidth. For more
information on  DSN methods, techniques, and present capabilities, see
\cite{dsn}.]  For the purposes of the present analysis one will need a
general knowledge of the methods and techniques implemented in the
radio-science subsystem of the DSN complex.   

This section reviews the techniques that are used to obtain the radio
tracking data from  which, after analysis,  results are generated. 
Here we will briefly discuss the DSN hardware that plays a pivotal 
role for our study of the anomalous acceleration.

%********************************

\subsection{Data acquisition}
\label{data-acquisition}

The Deep Space Network (DSN) is the network of ground stations that are
employed to track  interplanetary spacecraft \cite{dsn,dsn82}. There are
three  ground DSN complexes, at  Goldstone, California,  at Robledo de
Chavela, outside  Madrid, Spain,  and at Tidbinbilla, outside Canberra, 
Australia.

There are many antennae, both existing and decommissioned, that have 
been used by the DSN for spacecraft navigation.    For our four
spacecraft (Pioneer 10, 11, Galileo,  and Ulysses), depending on the time
period involved, the following  Deep Space Station (DSS) antennae  were
among those used:  (DSS 12, 14, 24) at the California  antenna
complex;   (DSS 42, 43, 45, 46) at the Australia complex;   and (DSS 54,
61, 62, 63) at the Spain complex.    Specifically, the Pioneers used (DSS
12, 14, 42, 43, 62, 63),  Galileo used (DSS 12, 14, 42, 43, 63), and  
Ulysses used (DSS 12, 14, 24, 42, 43, 46, 54, 61, 63).  

The DSN tracking system is a phase coherent system. 
By this we mean that
an ``exact'' ratio  exists between the transmission and reception
frequencies;  i.e., 240/221 for S-band or 880/221 for X-band
\cite{sband}.  (This is in distinction to the usual concept of 
coherent radiation used in atomic and astrophysics.)

Frequency is an average frequency, defined as the number of cycles per
unit time. Thus, accumulated phase is the integral of frequency. High
measurement precision is attained by maintaining the frequency accuracy
to 1 part per $10^{12}$ or better (This is in agreement with the 
expected Allan deviation for the S-band signals.)

\begin{figure}[h]
\begin{center}\noindent 
\epsfig{figure=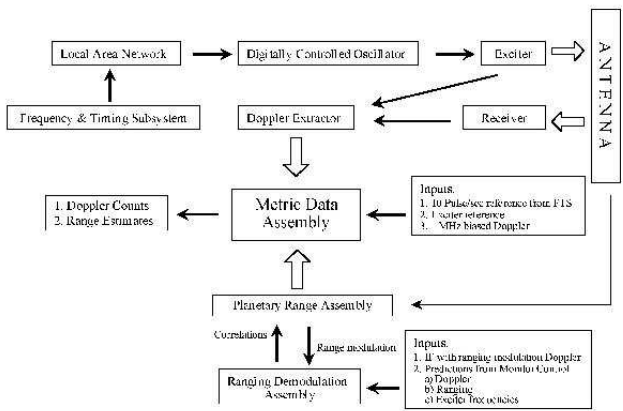,width=85mm}%,height=90mm}
\end{center}\vskip -10pt
\caption
{Block-diagram of the DSN complex as used for 
radio Doppler tracking of an interplanetary spacecraft.
For more detailed drawings and technical specifications see 
Ref. \cite{dsn}. 
 \label{fig:dsn_block}}
\end{figure}

{ \bf The DSN Frequency and Timing System (FTS): ~}
The DSN's FTS is the source for the high
accuracy just mentioned (see Figure \ref{fig:dsn_block}).  At its center is
an hydrogen maser that produces a precise and stable reference frequency  
\cite{barnes,vessot74}.  These devices have Allan deviations
\cite{SFJ98}  of approximately $3\times 10^{-15}$ to
$1\times 10^{-15}$ for integration times of $10^2$ to $10^3$ seconds,
respectively.

%************

%************

These masers are good enough so that the quality of Doppler-measurement
data is limited by thermal or plasma noise, and not by the inherent
instability of the frequency references.  Due to the extreme accuracy of
the hydrogen masers, one can very precisely characterize the spacecraft's
dynamical variables using  Doppler and range techniques. 
The FTS generates a  5 MHz and 10 MHz reference frequency which is sent
through the local area  network to the Digitally Controlled Oscillator
(DCO).

{\bf The Digitally Controlled Oscillator (DCO) and Exciter:~} Using
the highly  stable output from the FTS, the DCO, through digitally controlled
frequency  multipliers, generates the Track Synthesizer Frequency (TSF) of
$\sim 22$ MHz.   This  is then sent to the Exciter Assembly.  The Exciter
Assembly multiplies  the TSF by 96 to produce the S-band carrier signal at
$\sim 2.2$ GHz.  The signal power is amplified by Traveling Wave Tubes
(TWT) for  transmission. If ranging data are required, the Exciter Assembly
adds the  ranging modulation to the carrier. 
[The DSN tracking system has undergone many upgrades during the 29 years
of tracking Pioneer 10.  During this period internal frequencies have
changed.]

This S-band frequency is sent to the antenna where it is amplified and
transmitted to the spacecraft. The onboard receiver tracks the up-link
carrier using a phase lock loop. To ensure that the reception signal does
not interfere with the transmission, the spacecraft (e.g., Pioneer) has a
turnaround transponder with a ratio of 240/221. 
The spacecraft
transmitter's local oscillator is phase locked to the up-link carrier.
It  multiplies the received frequency by the above ratio and then
re-transmits the signal to Earth.

%************
{\bf Receiver and Doppler Extractor:~}
When the two-way \cite{way} signal reaches the ground, the receiver locks
on to the signal and tunes the Voltage Control Oscillator (VCO) to null out 
the phase error. The signal is sent to the Doppler Extractor.  
At the Doppler Extractor 
the current transmitter signal from the Exciter is multiplied by 240/221 
(or 880/241 for X-band)) and a bias, of 1 MHz for S-band 
or 5 MHz for X-band \cite{sband}, is added to the Doppler. 
The Doppler data is no longer modulated at
S-band but has been reduced as a consequence of the bias to an
intermediate frequency of 1 or 5 MHz

Since the light travel time to and from Pioneer 10 is long
(more than 20 hours), 
the transmitted frequency and the current transmitted frequency
can be different. The difference in frequencies are recorded  separately
and are accounted for in the orbit determination programs we discuss in
Section \ref{results}.
 
%************
 {\bf Metric Data Assembly (MDA):~} The MDA consists of computers
and Doppler  counters  where continuous count Doppler data are generated.
The  intermediate frequency (IF) of 1 or 5 MHz with a Doppler modulation is 
sent to the Metric Data Assembly (MDA). From the FTS a 10 pulse per second 
signal is also  sent to the MDA for timing. 
At the MDA, the IF and the resulting 
Doppler pulses are counted at a rate of 10 pulses per second. At each tenth 
of a second, the number of Doppler pulses are counted.  A second counter 
begins at the instant the first counter stops.  The result is 
continuously-counted Doppler data. (The Doppler data is a biased Doppler of
1 MHz, the bias later being removed by the analyst to obtain the true
Doppler counts.)   The Range data (if present) together with the Doppler
data is sent  separately to the Ranging Demodulation Assembly. The
accompanying Doppler  data is  used to rate aid (i.e., to ``freeze'' the
range signal) for demodulation and cross correlation. 

{\bf Data Communication:~}
The total set of tracking data is sent by local area network to the 
communication center.  From there it is transmitted to 
the Goddard Communication Facility 
via commercial phone lines or by government leased lines.  
It then goes  to JPL's Ground Communication Facility where it is 
received and recorded by the Data Records Subsystem.

%*****************************
%******************************

\subsection{Radio Doppler and range techniques}
\label{Dopp_tech}

Various radio tracking strategies are available for determining the
trajectory parameters of interplanetary spacecraft.   However,  radio
tracking Doppler and range techniques  are the most commonly used methods
for navigational purposes.  The position and velocities  of the DSN
tracking stations must be known to high accuracy. The transformation from
a Earth fixed coordinate system to the International Earth Rotation
Service (IERS) Celestial System is a complex series of rotations that
includes precession, nutation, variations in the Earth's rotation
({\tt UT1-UTC}) and polar motion. 

Calculations of the motion of a spacecraft are made on the basis of the
range time-delay and/or the Doppler shift in the signals. This type of
data was used to determine the positions, the velocities, and the
magnitudes of the orientation maneuvers for the Pioneer, Galileo, and
Ulysses spacecraft considered in this study. 

Theoretical modeling of the group delays and phase delay rates are done
with the orbit determination software we describe in the next
section. 

%**********

{\bf Data types:}  Our data describes the observations that are  the basis
of the results of this paper.   We receive our data from DSN in
closed-loop mode, i.e., data that has been  tracked with phase lock loop
hardware. (Open loop data is tape recorded but  not tracked by phase lock
loop hardware.)  The closed-loop data constitutes our Archival Tracking
Data File (ATDF),  which we copy \cite{datatapes}
to the National Space Science Data
Center (NSSDC)  on magnetic tape.  The ATDF files are stored on
hard disk in the RMDC (Radio Metric Data Conditioning group)  of JPL's
Navigation and Mission Design Section.   We access these files and run
standard software to produce an Orbit Data File for input into the orbit
determination programs which we  use.  (See Section \ref{results}.)  

The data types are two-way  and three-way \cite{way} Doppler  and two-way
range. (Doppler and range are defined in the following two subsections.)
Due to unknown clock offsets between the stations, three-way range is
generally not taken or used. 

The Pioneer spacecraft only have two- and three-way S-band \cite{sband} 
Doppler. Galileo also has S-band range data near the Earth.  Ulysses has
two- and three-way S-band up-link and X-band \cite{sband}  down-link
Doppler and range as well as S-band up-link and S-band down-link, 
although we have only processed the Ulysses S-band up-link and X-band 
down-link Doppler and range.  

%**********************************************************
\subsubsection{Doppler experimental techniques and strategy}
\label{sec:doppler}

In Doppler experiments a radio signal transmitted from the Earth to the
spacecraft is coherently transponded and sent back to the Earth.  Its
frequency change is measured with great precision, using the
hydrogen masers at the DSN stations.   The observable is the DSN 
frequency shift \cite{drift}
\begin{equation}
 \Delta \nu(t)={\nu_0}\,\frac{1}{c}\frac{d \ell}{dt}, 
\label{eq:doppler} 
\end{equation}
where $\ell$ is the overall optical distance (including
diffraction effects) traversed by a photon in both directions. [In the
Pioneer Doppler experiments, the stability of the fractional drift at the 
S-band is on the order of $\Delta \nu/\nu_0\simeq10^{-12}$,  for integration 
times on the order of $10^3$ s.]   Doppler measurements provide the
``range rate'' of the spacecraft and therefore are affected by all the
dynamical phenomena in the volume between the Earth and the spacecraft.   

Expanding upon what was discussed in Section \ref{data-acquisition}, 
the received signal and the transmitter frequency (both are at S-band) as 
well as a 10 pulse per second timing reference from the FTS are fed to
the  Metric Data Assembly (MDA). There the Doppler phase (difference
between  transmitted  and received phases plus an added bias) is
counted.  That is,   digital counters at the MDA record the zero
crossings of the difference  (i.e., Doppler, or alternatively the beat
frequency of the received frequency  and the exciter frequency). After
counting, the bias is removed so that the true  phase is produced.   

The system produces ``continuous count Doppler''  and it uses two
counters. Every tenth of a second, a Doppler phase count is recorded from
one of the counters. The other counter continues the counts. The
recording alternates between the two counters to maintain a continuous
unbroken count. The Doppler counts are at 1 MHz for S-band or 5  MHz for
X-band. The wavelength of each S-band cycle is about 13 cm. Dividers  or
``time resolvers'' further subdivide the cycle into 256 parts, so that
fractional cycles are measured with a resolution of 0.5 mm. This accuracy
can only be maintained if the Doppler is continuously counted (no breaks
in the count) and coherent frequency standards are kept throughout the
pass. It should be noted that no error is accumulated in the phase count
as long as lock is not lost. The only errors are the stability of the
hydrogen maser and the resolution of the ``resolver.''  

Consequently, the JPL Doppler records are not frequency measurements.  
Rather, they are digitally counted measurements of the Doppler phase
difference between the transmitted and received S-band frequencies, 
divided by the count time. 

Therefore, the Doppler observables, we will refer to, have units of cycles
per second or Hz. Since total count phase observables are Doppler
observables multiplied by the count interval T$_c$, they have units of
cycles. The Doppler integration time refers to the total counting  of the elapsed
periods of the wave with the reference frequency of the hydrogen maser. The usual
Doppler integrating times for the  Pioneer Doppler signals refers to the data
sampled over intervals of 10 s, 60 s, 600 s,  or  1980 s.

%****************

\subsubsection{Range measurements}

A range measurement is made by phase modulating a signal onto the up-link
carrier and having it echoed by the transponder. The transponder
demodulates this ranging signal, filters it, and then re-modulates it
back onto the down-link carrier. At the ground station, this returned
ranging signal is demodulated and filtered.  An instantaneous comparison
between the outbound ranging signal and the returning ranging signal
that comes down would yield the two-way delay.  Cross correlating the
returned phase modulated signal with a ground duplicate yields the time
delay.  (See \cite{anderson75} and references therein.)  As the range code
is repeated over and over, an ambiguity can exist. The orbit determination
programs  are then used to infer  (some times with great difficulty)
the number of range codes that exist between a particular transmitted
code  and its own  corresponding received code.  

Thus, the ranging data are independent of the Doppler data, which
represents a frequency shift of the radio carrier wave without
modulation. For example, solar plasma introduces a group delay in the
ranging data but a phase advance in the Doppler data. 

Ranging data can also be used to distinguish an actual range change from
a  fictitious range change seen in Doppler data that is caused by a
frequency error \cite{falsedop}. The Doppler frequency integrated over
time  (the accumulated phase) should equal the range change except for
the  difference introduced by charged particles

%*********************************
%*******************

\subsubsection{Inferring position information from Doppler} 

It is also possible to infer the position in the sky of a spacecraft 
from the Doppler data. This is accomplished by examining the diurnal 
variation imparted to the Doppler shift by the Earth's rotation. As the 
ground station rotates underneath a spacecraft, the Doppler shift is 
modulated by a sinusoid. The sinusoid's amplitude depends on the
declination  angle of the spacecraft and its phase depends upon the right
ascension. These  angles can therefore be estimated from a record of the
Doppler shift that is  (at least) of several days duration.  This  
allows for a determination of the  distance to the spacecraft through the 
dynamics of spacecraft motion using standard orbit theory contained in
the  orbit determination programs.

%**************%********************************
%****************************

\subsection{Data preparation}
\label{Data_edit}

In an ideal system, all scheduled observations would be used in
determining parameters of physical interest. However, there are
inevitable  problems that occur in data collection and processing that  
corrupt the data.  So, at  various stages of the signal processing one
must remove or ``edit'' corrupted data.  Thus, the need arises for
objective editing criteria.    Procedures have been developed which
attempt to excise corrupted data on the basis of objective criteria. 
There is always a temptation to eliminate data that is not well explained
by existing models, to thereby ``improve'' the agreement between theory
and experiment.  Such an approach may, of course, eliminate the very data
that would indicate deficiencies in the  {\it a~priori} model.  This would
preclude the discovery of improved models.

In the processing stage that fits the Doppler samples, checks are made to
ensure  that there are no integer cycle slips in the data stream that
would corrupt the phase. This is done by considering the difference of
the phase observations taken  at a high rate (10 times a second) to
produce Doppler. Cycle slips often are dependent on tracking loop
bandwidths, the signal to noise ratios, and predictions of frequencies.  
Blunders due to out-of-lock can be determined by  looking at the original
tracking data.   In particular,  cycle slips due to loss-of-lock stand
out as a 1 Hz blunder point for each cycle slipped.

If a blunder point is observed, the count is stopped and  a Doppler point
is generated by summing the preceding points. Otherwise  the count is
continued until a specified maximum duration is reached.  Cases where
this procedure detected the need for cycle corrections were flagged in
the database and often  individually examined by  an analyst.  Sometimes 
the data was corrected,  but nominally the  blunder point was just
eliminated.  This  ensures that the data is consistent over a  pass.
However, it does not guarantee that the pass is good,  because other 
errors can affect the whole pass and remain undetected until the orbit 
determination is done.  

To produce an input data file for an orbit determination program,
JPL has a software package known as the Radio Metric Data Selection,
Translation, Revision, Intercalation, Processing and Performance
Evaluation Reporting (RMD-STRIPPER) Program.  As we discussed in  Section
\ref{sec:doppler}, this input file has data that can be integrated over  
intervals with different durations:  10 s, 60 s, 600 s and 1980 s. This 
input Orbit Determination File (ODFILE) obtained from the RMDC group is
the initial data set with which  both the  JPL and  The Aerospace
Corporation groups started their analyses. Therefore, the initial data file
already contained some common data editing that the RMDC group  
had implemented through  program flags, etc.  The data set  we
started with had already been compressed to 60 s.  So,  perhaps there were
some blunders that had already been removed using the initial STRIPPER
program.  

The orbit analyst manually edits the remaining corrupted data points.
Editing  is done either by plotting the data residuals and deleting them
from the fit  or plotting weighted data residuals. That is, the residuals
are divided by  the standard deviation assigned to each data point and
plotted. This gives  the analyst a realistic view of the data noise during
those times when the  data was  obtained while looking through the solar
plasma. Applying an ``$N$-$\sigma$'' ($\sigma$ is the standard deviation)
test, where $N$ is the choice of the analyst (usually 4-10) the analyst
can delete those points that lie outside the
$N$-$\sigma$  rejection criterion without being biased in his selection.
The $N$-$\sigma$ test, implemented in CHASMP,  is  very useful for data 
taken near solar conjunction since the solar  plasma adds considerable
noise to the data. This criterion later was changed to a similar criteria
that rejects all data with residuals in the fit extending for more than
$\pm 0.025$ Hz from the mean.  Contrariwise, the  JPL analysis edits only
very corrupted  data; e.g., a blunder due to a phase lock loss, data with
bad spin calibration, etc. Essentially the Aerospace  procedure eliminates
data in the tails of the Gaussian probability frequency distribution
whereas the JPL procedure accepts this data.

If needed or desired, the orbit analyst  can choose to perform an
additional data compression of the original navigation data.   The JPL
analysis does not apply any additional data compression  and uses all the
original data from the ODFILE as opposed to Aerospace's approach.
Aerospace makes an additional  compression of data within CHASMP.    It
uses the longest available data integration times which can be composed
from either summing up adjacent data intervals or by using   data spans
with duration  $\ge 600$ s.   (Effectively Aerospace prefers  600 and 1980
second data intervals and applies a low-pass filter.) 

The total count of  corrupted data points is about 10\% of
the total raw data points. The analysts' judgments play an important role
here and is one of the main reasons that JPL and Aerospace have slightly
different results.   (See Sections \ref{results}and
\ref{recent_results}.)   In Section \ref{results}we will show a typical
plot  (Figure \ref{fig:aerospace} below) with  outliers  present in the
data. Many more outliers are off the plot. One would expect that the two
different strategies of data compression used by the two teams 
would result in significantly different numbers of  total  data points 
used in the two independent analyses. The influence of this fact on the
solution estimation accuracy will be addressed in Section
\ref{recent_results} below.

%******************* DATA WEIGHT

\subsection{Data weighting}
\label{dataweight} 

Considerable effort has gone into accurately
estimating measurement errors in the observations. These errors provide
the data weights necessary to accurately estimate the parameter
adjustments and their associated uncertainties. To the extent that
measurement errors are accurately modeled, the parameters extracted from
the data will be unbiased and will have accurate sigmas assigned to
them.   Both JPL and Aerospace assign a standard uncertainty of 1 mm/s
over a 60 second count time for the S--band Pioneer  Doppler data. 
(Originally the JPL team was weighting the data by 2 mm/s
uncertainty.)

A change in the DSN antenna elevation angle also directly affects the
Doppler  observables due to tropospheric refraction. Therefore, 
to correct for the influence of the Earth's troposphere
the data can also be deweighted for low elevation angles.  
The phenomenological  range  correction  is given as 
\begin{equation}
\sigma= \sigma_{\tt nominal} \left(1+\frac{18}{(1+\theta_E)^2}\right),
\label{eq:sig_aer0}
\end{equation}
where $\sigma_{\tt nominal}$ is the basic standard deviation (in Hz) 
and $\theta_E$ is the  elevation angle in degrees \cite{cane}. Each leg is
computed separately and summed.    For Doppler the same
procedure is used.  First, 
Eq. (\ref{eq:sig_aer0}) is multiplied by $\sqrt{60 \,{\rm s}/T_c}$, 
where  $T_c$ is the count time. Then a numerical time 
differentiation of Eq. (\ref{eq:sig_aer0}) is performed.  That is, 
Eq. (\ref{eq:sig_aer0})  is differenced and divided by the count time,
$T_c$. (For more details on this standard technique see Refs.
\cite{Moyer71}-\cite{MuhlemanAnderson81}.) 

There is also the problem of data weighting for data influenced by the 
solar corona.  This will be discussed in Section \ref{corona+wt}.   

%***************

\subsection{Spin calibration of the data}
\label{spincalibrate} 

The radio signals used by DSN to communicate with spacecraft are
circularly polarized. When these signals are reflected from 
spinning spacecraft antennae  
a Doppler bias is introduced that is a function of the spacecraft spin rate.
Each revolution of the spacecraft adds one cycle of phase to the up-link
and the down-link.  The up-link cycle is multiplied by the turn around
ratio 240/221  so that the bias  equals (1+240/221) cycles per revolution
of the spacecraft.  

High-rate spin data is available for Pioneer 10 only  up to July 17,
1990,  when the DSN ceased doing spin  calibrations. (See Section
\ref{sec:prop}.)  After this date,  in order to reconstruct the spin
behavior for the entire data span  and thereby  account for the spin bias
in the Doppler signal,  both analyses modeled the spin by performing
interpolations  between the data points. The JPL interpolation was 
non-linear with  a high-order polynomial fit of the data.   (The
polynomial was from second up to sixth order, depending on the data
quality.)  The CHASMP   interpolation was linear between the spin data
points.   

%(Essentially it truncated the non-linear polynomial 
%fit implemented by JPL.)

After a maneuver in mid-1993,  there was not enough power left to support
the IPP.  But analysts  still could get a
rough determination approximately every six months  using information
obtained from the conscan maneuvers.  No spin determinations were made
after 1995.     However, the archived  conscan data could still yield
spin data at every maneuver time  if such work was approved.  Further,   
as the phase center of the main antenna is slightly offset from the spin
axis, a very small (but detectable) $\sin$e-wave signal appears in the 
high-rate Doppler data.  In principle, this could  be used to
determine the spin rate for passes taken after 1993,    but it has not
been attempted.  Also, the failure of one of the spin-down thrusters
prevented  precise  spin calibration of the Pioneer 11 data.  

Because the spin rate of the Pioneers was changing over the data span, the
calibrations also provide an indication of gas leaks that affect the 
acceleration of  the spacecraft.  A careful  look at the records shows how
this can be a problem.   This will be discussed in Sections 
\ref{spinhistory} and \ref{sec:gleaks}.

%************************ 4) SPACECRAFT NAVIGATION

\section{\label{navigate}BASIC THEORY OF SPACECRAFT NAVIGATION}

Accuracy of modern radio tracking techniques has  provided the means
necessary to explore the gravitational environment in the solar system up to
a limit never before possible \cite{massprog}. 
The major role in this quest belongs to
relativistic celestial mechanics experiments with planets (e.g., passive
radar ranging) and interplanetary spacecraft (both Doppler and range
experiments).  Celestial mechanics experiments with spacecraft have been
carried out by JPL since the early 1960's \cite{anderson74,dsn86}.    
The motivation was to improve both  the ephemerides of solar system
bodies and also the knowledge of the  solar system's dynamical
environment. This has become possible due to  major improvements in 
the accuracy of spacecraft navigation, which is still a critical
element for a number of space missions.   
The main objective of spacecraft navigation  is to
determine the  present position and velocity of a spacecraft and to
predict its future trajectory.  This is usually done by measuring
changes in the  spacecraft's position and then,  using those
measurements,   correcting (fitting and adjusting) the  predicted
spacecraft  trajectory.    

In this section we will discuss the theoretical foundation that is
used for the analysis of tracking data from interplanetary spacecraft.  We
describe the basic physical models   used  to determine a
trajectory, given the data.

%*****************

\subsection{Relativistic equations of motion}

The spacecraft ephemeris, generated by a numerical integration
program, is a file of spacecraft   positions and velocities as functions
of ephemeris (or coordinate) time ({\tt ET}). The integrator requires the
input of various parameters. These  include adopted constants ($c$, $G$,
planetary mass ratios, etc.) and parameters that are estimated from fits
to observational data (e.g., corrections to planetary orbital elements).

The ephemeris programs use equations for point-mass relativistic
gravitational  accelerations.  They are derived from the variation of a
time-dependent,  Lagrangian-action integral that is referenced to a
non-rotating, solar-system,  barycentric, coordinate frame. In addition
to modeling point-mass interactions, the  ephemeris programs contain 
equations of motion that model  terrestrial and lunar figure effects,
Earth tides, and lunar physical librations 
\cite{Newhall83}-\cite{Standish95a}.  The programs treat the Sun, the
Moon, and the nine planets  as point masses in the isotropic,
parameterized post-Newtonian, N-body metric  with Newtonian
gravitational perturbations from large, main-belt asteroids.   

Responding to the increasing demand of the navigational accuracy, the
gravitational field in the solar system is modeled to include a number of
relativistic effects that are predicted by the  different metric theories
of gravity.    Thus, within the accuracy of modern experimental
techniques, the   parameterized  post-Newtonian (PPN) approximation  of 
modern theories of gravity  provides a useful starting point not only for
testing these predictions, but also for describing the motion of
self-gravitating bodies and test particles.  As discussed in detail in  
\cite{Will93}, the accuracy of the PPN  limit (which is slow motion and
weak field) is adequate for all foreseeable solar system tests of general
relativity and a number of other metric theories of  gravity. (For the
most general formulation of the PPN formalism,  see the works of Will and
Nordtvedt \cite{Will93,WillNordtvedt72}.) 

For each body $i$ (a planet or spacecraft anywhere in the solar system),
the point-mass acceleration is  written  as 
\cite{Moyer71,Moyer00,Newhall83,estabrook69,Moyer81}
\begin{widetext}
\begin{eqnarray}\nonumber
\ddot{\bf r}_i&=&\sum_{j\not=i}\frac{\mu_j({\bf r}_j-{\bf r}_i)}
{r^3_{ij}}\Bigg(1-\frac{2(\beta+\gamma)}{c^2}\sum_{k\not=i}
\frac{\mu_k}{r_{ik}}-\frac{2\beta-1}{c^2}\sum_{k\not=j}
\frac{\mu_k}{r_{jk}}-\frac{3}{2c^2}\Big[\frac{({\bf r}_j-{\bf
r}_i)\dot{\bf r}_j}{r_{ij}}\Big]^2+\frac{1}{2c^2} ({\bf r}_j-{\bf
r}_i)\ddot{\bf r}_j-\frac{2(1+\gamma)}{c^2}
\dot{\bf r}_i\dot{\bf r}_j+\\ 
&+&\gamma\left(\frac{v_i}{c}\right)^2+
(1+\gamma)\left(\frac{v_j}{c}\right)^2\Bigg)+ 
\frac{1}{c^2}\sum_{j\not=i}\frac{\mu_j}{r^3_{ij}}
\Big([{\bf r}_i-{\bf r}_j)]\cdot[(2+2\gamma)\dot{\bf r}_i-(1+2\gamma)
\dot{\bf r}_j]\Big)(\dot{\bf r}_i-\dot{\bf r}_j)+ 
\frac{3+4\gamma}{2c^2}\sum_{j\not=i}
\frac{\mu_j\ddot{\bf r}_j}{r_{ij}}
\label{eq:rdotdot}
\end{eqnarray}
\end{widetext}
where  $\mu_i$ is the ``gravitational constant'' of body {\it i}.  It
actually is its mass times the Newtonian  constant: $\mu_i={\it G}m_i$. 
Also,  ${\bf r}_i(t)$ is the barycentric position of body $i$, 
$r_{ij}=|{\bf r}_j-{\bf r}_i|$ and $v_i=|\dot{\bf r}_i|$. For  
planetary motion, each of these equations depends on the others. So they
must be iterated in each step of  the integration of the equations of
motion. 

The barycentric acceleration of each body $j$ due to Newtonian effects
of the remaining bodies and the asteroids is denoted by
$\ddot{\bf r}_j$.  In Eq.~(\ref{eq:rdotdot}), $\beta$ and $\gamma$ are
the  PPN parameters \cite{Will93,WillNordtvedt72}.   General relativity
corresponds to $\beta = \gamma = 1$, which we choose for our study.   The
Brans-Dicke theory is the most famous among the alternative theories of
gravity.  It contains, besides the metric  tensor,  a scalar field
$\varphi$  and an arbitrary coupling constant $\omega$, related to the
two PPN parameters as $\gamma= \frac{1+\omega}{2+\omega},  ~\beta=1$.
Equation  (\ref{eq:rdotdot}) allows the consideration of any problem in
celestial mechanics within the PPN framework.   

%******************************

\subsection{Light time solution and time scales}
\label{sec:time_scales}

In addition to planetary equations of motion Eq.~(\ref{eq:rdotdot}), one
needs to solve the relativistic  light propagation  equation in order to
get the solution for the total light time travel. In the solar system,
barycentric, space-time frame of reference this equation is given by:
\begin{eqnarray}\nonumber
t_2-t_1&=&\frac{r_{21}}{c}+\frac{(1+\gamma)\mu_\odot}{c^3}
\ln\bigg[\frac{r_1^\odot+r_2^\odot+r_{12}^\odot}
{r_1^\odot+r_2^\odot-r_{12}^\odot}\bigg]+\\
&+& \sum_{i} \frac{(1+\gamma)\mu_i}{c^3}
\ln\bigg[\frac{r_1^i+r_2^i+r_{12}^i}
{r_1^i+r_2^i-r_{12}^i}\bigg], 
\label{eq:lt}
\end{eqnarray}
where $\mu_\odot$ is the gravitational constant of the Sun and
$\mu_i$ is the gravitational constant of a planet, an outer planetary
system, or the Moon.  
$r_1^\odot, r_2^\odot and r_{12}^\odot$ are the heliocentric distances
to the point of RF signal emission on Earth, to the
point of signal reflection at the spacecraft, and the relative distance 
between these two  points.  
Correspondingly, $r_1^i, r_2^i,$ and $r_{12}^i$ are similar
distances relative to a particular $i$-th body in the solar system.
In the spacecraft light time solution, $t_1$ refers 
to the transmission time at a tracking station on Earth, and
$t_2$ refers to the reflection time at the spacecraft or, for one-way
\cite{way}  data, the transmission time at the spacecraft. The reception
time at the tracking station on Earth or at an Earth satellite is
denoted by $t_3$. Hence, Eq.~(\ref{eq:lt}) 
is the up-leg light time equation.  The
corresponding down-leg light time equation is obtained by replacing
subscripts as follows: $1\rightarrow 2 $ and  $2\rightarrow 3 $. 
(See the details in \cite{Moyer00}.)

The spacecraft equations of motion relative to the solar system
barycenter are essentially the same as given by Eq. (\ref{eq:rdotdot}). 
The gravitational constants of the Sun, planets and the planetary  systems
are the values associated with the solar system barycentric frame of
reference, which are obtained from the planetary ephemeris \cite{Moyer81}.
We treat a distant spacecraft as a point-mass particle. The spacecraft
acceleration is integrated numerically to produce the spacecraft
ephemeris. The ephemeris is interpolated at the ephemeris time ({\tt ET})
value of the interpolation epoch.  This is the time coordinate $t$  in 
Eqs. (\ref{eq:rdotdot}) and (\ref{eq:lt}), i.e., $t\equiv\,{\tt ET}$. 
As such, ephemeris time  means coordinate time in the chosen celestial
reference frame. It is  an independent variable for the
motion of celestial bodies, spacecraft, and light rays. The scale of {\tt
ET} depends upon which  reference frame is selected and one may use a
number of time scales depending on the  practical applications. It is
convenient to express {\tt ET} in terms of  International Atomic Time
({\tt TAI}). {\tt TAI} is based upon the second in the International
System of Units ({\tt SI}).  This second 
is defined to be the duration of 9,192,631,770 periods
of the radiation corresponding to the transition between two hyperfine
levels of the ground state of the cesium-133 atom \cite{exp_cat}.
 
The differential equation relating ephemeris time ({\tt ET}) in the solar
system barycentric reference frame to {\tt TAI} at a tracking station on 
Earth or on Earth satellite can be obtained directly from the Newtonian
approximation to the N-body metric \cite{Moyer81}. This expression has
the form 
\begin{eqnarray}
\frac{d \,{\tt TAI}}{d\, \tt ET}= 
1-\frac{1}{c^2}\Big(U-\langle U\rangle +
\frac{1}{2}v^2-\frac{1}{2}\langle v^2 \rangle\Big) 
+{\cal O}(\frac{1}{c^{4}}),
\label{eq:tai_et}
\end{eqnarray}
where $U$ is the solar system gravitational potential evaluated at
the tracking station and $v$ is the solar system barycentric velocity of
the tracking station. The brackets $\langle ~\rangle$ on the right side
of Eq. (\ref{eq:tai_et}) denote long-time average  of the quantity
contained within them. This averaging amounts to integrating out periodic
variations in the gravitational potential, $U$, 
and the barycentric velocity, $v^2$, at the location of a tracking
station.  The desired time scale transformation is then obtained by using
the planetary ephemeris to calculate the terms in Eq. (\ref{eq:tai_et}). 

The vector  expression for the ephemeris/coordinate time ({\tt ET}) in
the solar system barycentric frame of reference minus the {\tt TAI} obtained
from an atomic clock at a tracking station on Earth has the
form \cite{Moyer81} 
%\begin{widetext}
\begin{eqnarray}
{\tt ET-TAI} &=& 32.184~{\rm s}+
\frac{2}{c^2}(\dot{\bf r}^\odot_{\tt B}\cdot {\bf r}^\odot_{\tt B})+
\frac{1}{c^2}(\dot{\bf r}^{\tt SSB}_{\tt B}
\cdot {\bf r}^{\tt B}_{\tt E})+ \nonumber\\
&+&\frac{1}{c^2}(\dot{\bf r}^{\tt SSB}_{\tt E}\cdot 
{\bf r}^{\tt E}_A)+\frac{\mu_J}{c^2(\mu_\odot+\mu_J)}
(\dot{\bf r}^\odot_J\cdot  {\bf r}^\odot_J)+\nonumber\\ 
&+&
\frac{\mu_{Sa}}{c^2(\mu_\odot+\mu_{Sa})}
(\dot{\bf r}^\odot_{Sa}\!\cdot{\bf r}^\odot_{Sa})+   
\frac{1}{c^2}(\dot{\bf r}^{\tt SSB}_\odot\!\!\cdot 
{\bf r}^\odot_{\tt B}),
\label{eq:time}\hskip 14pt
\end{eqnarray} 
%\end{widetext}
where ${\bf r}^j_i$ and $\dot{\bf r}^j_i$ position and velocity vectors
of point $i$ relative to point $j$ (they are functions of {\tt ET});
superscript or subscript ${\tt SSB}$ denotes solar system barycenter;
$\odot$ stands for the Sun; ${\tt B}$ for the Earth-Moon barycenter; $ 
E, J, Sa $ denote the Earth, Jupiter, and Saturn correspondingly, and $A$
is for the location of the atomic clock on Earth which reads {\tt TAI}.  
This approximated analytic result  contains the clock synchronization 
term which depends upon the location of the atomic clock and five
location-independent periodic terms. 
There are several alternate expressions 
that have up to several hundred additional periodic terms which
provide greater accuracies than  Eq. (\ref{eq:time}). The use of these
extended expressions provide transformations of {\tt ET} -- {\tt TAI} to 
accuracies of 1 ns \cite{Moyer00}.

For the purposes of our study the Station Time ({\tt ST}) is especially
significant.  This time is the atomic time {\tt TAI} at a DSN tracking
station on Earth,  {\tt ST}={\tt TAI}$_{\tt station}$. This atomic time
scale  departs by a small amount from the ``reference time scale.''  
The reference time scale for a DSN tracking station on Earth is the 
Coordinated Universal Time ({\tt UTC}).  This last  is standard time for
0$^\circ$ longitude.   (For more details see \cite{Moyer00,exp_cat}.)  

All the vectors in Eq.~(\ref{eq:time}) except the geocentric   position
vector of the tracking station on Earth can be interpolated from the
planetary ephemeris or computed from these quantities. Universal Time
({\tt UT}) is the measure of time which is the basis for all civil
time keeping. It is an observed time scale.  The specific version used
in JPL's Orbit Determination Program (ODP) is {\tt UT1}. This is used to
calculate mean sidereal time, which is the Greenwich hour angle of the mean
equinox of date measured in the true equator of date.  Observed
{\tt UT1} contains 41 short-term terms with periods between 5 and 35 days.
They are caused by long-period solid Earth tides. When the sum 
of these terms, {\tt $\Delta$UT1}, is subtracted from {\tt UT1}  
the  result is called {\tt UT1R}, where {\tt R} means regularized. 

Time in any scale is represented as seconds past
1 January 2000, 12$^{\tt h}$, in that time scale. This epoch is
J2000.0, which is the start of the Julian year 2000. The Julian Date
for this epoch is JD 245,1545.0.  Our analyses used the standard
space-fixed  J2000 coordinate system, which is provided by 
the International
Celestial Reference Frame (ICRF).  This is a quasi-inertial reference 
frame defined from the radio positions of 212 extragalactic sources
distributed over the entire sky \cite{Ma98}. 
 
The variability of the earth-rotation vector relative to the body of the
planet or in inertial space is caused by the gravitational torque exerted
by the Moon, Sun and planets, displacements of matter in different parts of
the planet and other excitation mechanisms. The observed oscillations can
be interpreted in terms of mantle elasticity, earth flattening, structure
and properties of the core-mantle boundary, rheology of the core,
underground water, oceanic variability, and atmospheric variability on time
scales of weather or climate.  

Several space geodesy techniques contribute to the continuous monitoring
of the Earth's rotation by the International Earth Rotation Service (IERS). 
Measurements of the Earth's rotation presented in the form of time
developments 
of the so-called Earth Orientation Parameters ({\tt EOP}). Universal time
({\tt UT1}), polar motion, and the celestial motion of the pole
(precession/nutation) are determined by Very Long-Baseline
Interferometry (VLBI). Satellite geodesy techniques, such as 
satellite laser ranging (SLR) and using the 
Global Positioning System (GPS), determine
polar motion and rapid variations of universal time. The satellite 
geodesy programs used in the IERS allow determination of the time
variation of the Earth's gravity field.  This 
variation reflects the evolutions of the
Earth's shape and of the distribution of mass in the planet. The programs 
have also detected changes in the location of the center of mass of the
Earth relative to the crust. It is possible to investigate other 
global phenomena such as the mass redistributions of the atmosphere, oceans,
and solid Earth.

Using the above experimental techniques,
Universal time and polar motion are available daily with an accuracy of
about 50 picoseconds (ps).  They are determined from VLBI astrometric 
observations with an accuracy of 0.5 milliarcseconds (mas). 
Celestial pole motion is available every five to
seven days at the same level of accuracy.  These estimations of accuracy
include both short term and long term noise. Sub-daily variations in
Universal time and polar motion are also measured on a campaign basis.

In summary, this dynamical model accounts for a number of post-Newtonian
perturbations in the motions of the planets, the Moon, and spacecraft. 
Light propagation is  correct to order
$c^{-2}$.  The equations of motion of extended celestial bodies are valid
to order  $c^{-4}$.  Indeed, this dynamical model has been  good enough
to perform tests of general relativity 
\cite{anderson75,Will93,WillNordtvedt72}. 

%**************

\subsection{Standard modeling of small, non-gravitational forces}
\label{sec:syst0}

In addition to the mutual gravitational interactions of the various
bodies in the solar system and the gravitational forces acting on a
spacecraft as a result of presence of those bodies, it is also important
to consider a number of non-gravitational forces which are important for
the motion of a spacecraft. (Books and lengthy reports have been written
about practically all of them. Consult Ref.~\cite{milani,longuski}
for a general introduction.)

The Jet Propulsion Laboratory's ODP accounts for many  sources of 
non-gravitational accelerations. Among them, the most relevant to this
study,  are: i) solar radiation pressure, ii)  solar wind pressure,  
iii) attitude-control maneuvers together with a  model for  unintentional
spacecraft mass expulsion due to  gas leakage of the propulsion system. We
can also account for possible influence of the interplanetary media and DSN
antennae contributions to the spacecraft radio tracking data  and consider
the torques produced by above mentioned forces.  The Aerospace CHASMP code
uses a model for  gas leaks that can be adjusted to include the
effects of the recoil force due to emitted radio power and anisotropic
thermal radiation of the spacecraft. 

In principle, one could set up complicated engineering  models to predict
at least some of the effects.   However,  their residual uncertainties
might be unacceptable for the experiment,  in spite of the significant
effort required. In fact, a constant acceleration produces a linear
frequency drift that can be  accounted for in the data analysis by a
single unknown parameter. 

The figure against which we compare the effects of  non-gravitational
accelerations on the Pioneers' trajectories is the expected error in the
acceleration error estimations.  This is on the order of 
\begin{equation}
\sigma_0 \sim 2\times 10^{-8} ~~{\rm cm/s^2},
\label{eq:req}
\end{equation}
\noindent where  $\sigma_0$ is a single determination accuracy
related to acceleration measurements averaged over number of days. 
This would contribute to our result as $\sigma_N\sim\sigma_0/\sqrt{N}$.
Thus, if  no systematics are involved then $\sigma_N$ will just tend 
to zero as time progresses. 

Therefore, the important thing is to know that these effects
(systematics)  are not too large, thereby overwhelming any possibly
important signal  (such as our anomalous acceleration).  This  will be 
demonstrated in Sections \ref{ext-systema} and  \ref{int-systema}. 

%*************** CORONA + WEIGHT *******

\subsection{Solar corona model and weighting}
\label{corona+wt}

The electron density and density gradient in the solar atmosphere  influence
the propagation of radio waves through the medium. So, both range and Doppler
observations at S-band are affected by  the electron density in the
interplanetary medium and outer solar corona. Since, throughout the experiment,
the closest approach  to the center of the Sun of a radio ray path  was greater
than 3.5 $R_\odot$, the medium may be regarded as collisionless. The {\it one
way} time delay associated with a plane wave passing through the solar corona
is obtained 
\cite{MuhlemanAnderson81,anderson74,MuhlemanEspositoAnderson77}  by integrating
the group velocity of propagation along the ray's path, $\ell$:
\begin{eqnarray}\nonumber
 \Delta t &=& \pm \frac{1}{2c\,n_{\tt crit}(\nu)}
 \int_\oplus^{SC}d\ell~n_e(t, {\bf r}),  \\
n_{\tt crit}(\nu) &=& 1.240\times 10^4~ 
\Big(\frac{\nu}{1~{\rm MHz}}\Big)^2~~{\rm cm}^{-3},
\label{eq:sol_plasma}
 \end{eqnarray}
where 
 $n_e(t, {\bf r})$ is the free electron density in the solar plasma, 
$c$ is the speed of light,   and 
$n_{\tt crit}(\nu)$ is the critical plasma density for the radio carrier
frequency $\nu$. 
The plus sign is applied for ranging data and the minus sign
for Doppler data \cite{slavanote}. 

Therefore, in order to
calibrate the plasma contribution, one should know the electron density
along the path. One usually decomposes the electron density, $n_e$,
into a static,
steady-state part, $\overline{n}_e(\bf{r})$, plus a fluctuation  
$\delta n_e(t, {\bf r})$, i.e.,
$n_e(t, {\bf r})= \overline{n}_e({\bf r})+ \delta n_e(t, {\bf r})$. 
The physical properties of the second term are hard to
quantify.  But luckily, its effect on Doppler observables and,
therefore, on our results is small.  (We will address this issue in Sec.
\ref{solarwind}.)  On the contrary, the steady-state corona  behavior is
reasonably well known and several plasma models can be found in the
literature \cite{MuhlemanEspositoAnderson77}-\cite{bird}. 

Consequently, while  studying the effect of a systematic error from
propagation of the S-band carrier wave through the solar plasma, both
analyses adopted the following model for the electron density profile 
\cite{MuhlemanAnderson81}:
\begin{eqnarray}
 n_e(t, {\bf r})= 
A\Big(\frac{R_\odot}{r}\Big)^2+
B\Big(\frac{R_\odot}{r}\Big)^{2.7}
       e^{-\left[\frac{\phi}{\phi_0}\right]^2}+
C\Big(\frac{R_\odot}{r}\Big)^6.
\label{corona_model_content}
\end{eqnarray}
$r$ is the heliocentric distance to the immediate ray trajectory 
and $\phi$ is the helio-latitude normalized by the reference latitude 
of $\phi_0=10^\circ$. The parameters $r$ and $\phi$ are determined 
from the trajectory coordinates on the tracking link being modeled. 
The parameters $A, B, C$ are  parameters chosen to 
describe the solar electron density.  (They are commonly given in two 
sets of units, meters or cm$^{-3}$ \cite{scunits}.)
They can be treated as stochastic parameters, 
to be determined from the fit.  But in both analyses we ultimately chose 
to use the values determined from the recent solar corona studies done
for the Cassini mission.  These newly obtained values are:  $A=
6.0\times 10^3, B= 2.0\times 10^4, C= 0.6\times 10^6$,  all in meters
\cite{Ekelund}. [This is what we will refer to as the ``Cassini corona
model.''] 

Substitution of Eq. (\ref{corona_model_content}) into Eq.
(\ref{eq:sol_plasma}) results in the following steady-state solar corona
contribution to the range model that we used in our analysis:
\begin{eqnarray}\nonumber
\Delta_{\tt SC}{\rm range}&=& \pm \Big(\frac{\nu_0}{\nu}\Big)^2\bigg[
A\Big(\frac{R_\odot}{\rho}\Big)F+ \\
&+& B\Big(\frac{R_\odot}{\rho}\Big)^{1.7}
       e^{-\left[\frac{\phi}{\phi_0}\right]^2}+
C\Big(\frac{R_\odot}{\rho}\Big)^{5}\bigg]. \hskip 10pt 
\label{corona_model}
\end{eqnarray}
$\nu_0$ and $\nu$ are a reference frequency and the actual frequency of
radio-wave [for Pioneer 10 analysis $\nu_0=2295$ MHz], $\rho$ is the
impact parameter with respect to the Sun and
$F$ is a light-time correction factor. For  distant spacecraft this
function is given as follows:
%\begin{widetext}
\begin{eqnarray}
F&=&F(\rho, r_T, r_E)=\\\nonumber
&=&\frac{1}{\pi}\Bigg\{ 
{\sf ArcTan}\Big[\frac{\sqrt{r_T^2-\rho^2}}{\rho}\Big]+
{\sf ArcTan}\Big[\frac{\sqrt{r_E^2-\rho^2}}{\rho}\Big]\Bigg\},
\label{eq:weight_doppler*}
\end{eqnarray}
%\end{widetext}
\noindent where $r_T$ and $r_E$ are the heliocentric radial distances to the
target and to the Earth, respectively. Note that the sign of the 
solar corona range correction is negative for Doppler and positive for
range. The Doppler correction is obtained from Eq. (\ref{corona_model})
by simple time differentiation. 
Both analyses  use the same physical model,  Eq. (\ref{corona_model}), 
for the steady-state solar corona effect on the
radio-wave propagation through the solar plasma.  
Although the actual implementation of the model in the two codes is
different, this turns out not to be significant. (See  Section
\ref{Ext_accuracy}.)

CHASMP can also consider the effect of  temporal variation in the 
solar corona by  using the recorded history of solar
activity.  The  change in solar activity is linked to the variation  of
the total number of sun spots per year as observed  at a particular 
wavelength of the solar radiation, $\lambda$=10.7 cm.  The actual data
corresponding to this variation  is given in Ref. \cite{F10-7}.  CHASMP
averages this data over 81 days and normalizes the value of the flux by
150. Then it is used as a time-varying scaling factor in Eq.
(\ref{corona_model}). The result is referred to as the   ``F10.7 model.''

Next we come to corona data weighting.  JPL's
ODP does not apply corona weighting.  On the other hand, 
Aerospace's CHASMP can apply corona weighting if desired. 
Aerospace uses a standard weight augmented by a weight function that
accounts  for noise introduced by solar plasma and low elevation. 
The weight values are 
adjusted so that  i) the post-fit weighted sum of the squares is close to
unity  and  ii) approximately
uniform noise in the residuals is observed throughout the fit span. 

Thus, the corresponding solar-corona weight function is:
\begin{equation}
\sigma_{\tt r}= \frac{k}{2} \Big(\frac{\nu_0}{\nu}\Big)^2
\Big(\frac{R_\odot}{\rho}\Big)^{\frac{3}{2}},  \label{weightrange}
\end{equation} 
where, for range data, $k$ is an input constant 
nominally equal to 0.005 light seconds, 
$\nu_0$ and $\nu$ are a reference frequency  and the actual  frequency,
$\rho$ is the trajectory's impact parameter with respect to the Sun in km,
and $R_\odot$ is the solar radius in km \cite{Muhleman}. 
The solar-corona weight function for Doppler is essentially the same, 
but obtained by numerical time differentiation of Eq.~(\ref{weightrange}).   

%***************  MODELING OF MANEUVERS

\subsection{Modeling of maneuvers}
\label{model-maneuvers}

There were 28 Pioneer 10 maneuvers during our data  interval from 3
January 1987 to 22 July 1998. Imperfect coupling of the hydrazine
thrusters used for the spin orientation maneuvers produced integrated
velocity changes of a few millimeters per second. The times and durations
of each maneuver were provided by NASA/Ames.  JPL  used this data as
input to ODP.  The Aerospace team used a slightly different approach. In
addition to the original data, CHASMP used the spin-rate data file to
help determine the times and duration of maneuvers.  The CHASMP
determination mainly agreed with the data used by JPL. [There were  minor
variations in some of the times, one maneuver was split in two, and one
extraneous maneuver was added to Interval II to account for data not
analyzed (see below).] 

Because the effect on the spacecraft acceleration could not be determined
well enough from the engineering telemetry, JPL included a single unknown
parameter in the fitting model for each maneuver.  In JPL's ODP analysis,
the maneuvers were modeled by instantaneous velocity increments at the
beginning  time of each maneuver (instantaneous burn model).  
[Analyses of individual maneuver fits show the residuals to be small.] In
the CHASMP analysis,  a constant acceleration acting over the duration of
the maneuver was included as a parameter in the fitting model (finite
burn model).    Analyses of individual maneuver fits show the residuals
are small. Because of the Pioneer spin, these accelerations are important
only along the Earth-spacecraft line, with the other two components
averaging out over about 50 revolutions of the spacecraft over a typical
maneuver duration of 10 minutes.  

By the time Pioneer 11 reached Saturn, the pattern of the
thruster firings was understood. Each maneuver caused a change in
spacecraft spin and a velocity increment in the spacecraft trajectory,
immediately followed by two to three days of gas leakage, large enough to
be observable in the Doppler data
\cite{null81}.  

Typically the Doppler data is time averaged over 10 to 33 minutes, which
significantly reduces the high-frequency Doppler noise. The residuals
represent our fit.   They are converted from units of Hz to Doppler
velocity by the formula \cite{drift}
\begin{equation}
[\Delta v]_{\tt DSN} 
= \frac{c}{2}   \frac{[\Delta \nu]_{\tt DSN}}{\nu_0},
\label{hztodoppler}
\end{equation}
where $\nu_0$ is the downlink carrier frequency, 
$\sim 2.29$  GHz, $\Delta \nu$ is the
Doppler residual in Hz from the fit, and $c$ is the speed of light.
 
\begin{figure}[!ht]
%\begin{center}
\noindent  
\vskip -10pt
\epsfig{figure=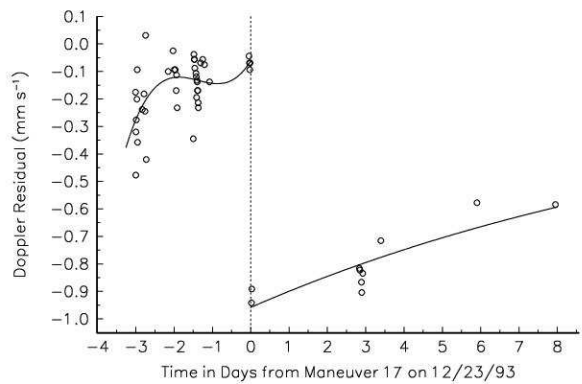,width=86mm}%,height=90mm}
%\end{center}
\vskip -10pt
\caption{The Doppler residuals after a fit for maneuver \# 17 on 23
December 1993.
 \label{fig:man17}}
\end{figure}

As an illustration, consider the fit to one of the Pioneer 10 maneuvers,
\# 17, on 22 December 1993, given in Figure \ref{fig:man17}.  This  was
particularly well covered by  low-noise 

%************

%************

\noindent Doppler data near solar
opposition.  Before the start of the maneuver, there is a systematic
trend in the residuals which is represented by a cubic polynomial in 
time. The standard error in the residuals is 0.095 mm/s. After the
maneuver, there is a relatively small velocity discontinuity of $-0.90
\pm 0.07$  mm/s.   The discontinuity arises because the model fits the
entire data interval. In fact,  the residuals increase after the
maneuver.  By 11 January 1994, 19 days after the maneuver, the residuals
are scattered about their pre-maneuver mean of $-0.15$ mm/s.

For purposes of characterizing the gas leak immediately after the maneuver,
we fit the post-maneuver residuals by a two-parameter exponential curve,
\begin{equation}
\Delta v = -v_0 \exp\Big[-\frac{t}{\tau}\,\Big] - 0.15 ~~~{\rm mm/s}.
\end{equation} 
The best fit yields $v_0 = 0.808$ mm/s and the time constant
$\tau$ is 13.3 days, a reasonable result. The time derivative of the
exponential curve yields a residual acceleration immediately after the
maneuver of 7.03 $\times$ 10$^{-8}$ cm/s$^{2}$. This is close to the
magnitude of the anomalous acceleration inferred from the Doppler data, but
in the \emph{opposite} direction. However the gas leak rapidly decays and
becomes negligible after 20 days or so.

%**************** ORBIT DETERMINATION

\subsection{Orbit determination procedure}
\label{sec:OD}

Our orbit determination procedure first determines the spacecraft's
initial position and velocity in a data interval. For each data interval,
we then estimate the magnitudes of the orientation maneuvers, if any. The
analyses are modeled to include the effects of planetary perturbations,
radiation pressure, the interplanetary media, general relativity, and
bias and drift in the Doppler and range (if available). Planetary
coordinates and solar system masses are obtained using JPL's Export
Planetary Ephemeris DE405, where DE stands for the Development
Ephemeris. [Earlier in the study, DE200 was used.  
See Section \ref{subsec:accel}.] 

We include models of  precession, nutation, sidereal rotation, 
polar motion, tidal  effects, and tectonic plates drift.   Model values of
the tidal deceleration, nonuniformity of rotation, polar motion, Love
numbers, and Chandler wobble are obtained observationally, by means of
Lunar and Satellite Laser Ranging (LLR and SLR) techniques and VLBI.  
Previously they were combined into a common publication by either the
International Earth Rotation Service (IERS)  or by the United States Naval
Observatory (USNO).   Currently this information is provided by the ICRF.  
JPL's Earth Orientation Parameters (EOP) is a major 
source contributor to the ICRF.

The implementation of the J2000.0 reference coordinate system in
CHASMP involves only rotation from the Earth-fixed to the J2000.0
reference frame and the use of JPL's ~ DE200 ~ planetary ephemeris  
\cite{Laing91}. The rotation from ~ J2000.0 ~ to Earth-fixed is computed
from a series of   rotations which include precession, nutation, the
Greenwich hour angle, and pole wander. Each of these general
categories is also a multiple rotation and  is treated separately by
most software. Each separate rotation matrix is  chain multiplied to
produce the final rotation matrix. 
 
CHASMP, however, does not separate precession and nutation.  Rather, 
it combines them into a single  matrix operation.  
This is achieved by using a different set of  angles  to describe precession
than is used in the ODP.  (See a description of the standard  set of  angles in
\cite{Lieske76}.)  These angles separate luni-solar precession from  planetary
precession.  Luni-solar precession, being the linear term of the nutation
series for the   nutation in longitude, is combined with the nutation in
longitude from the DE200 ephemeris tape  \cite{Standish82}.   

Both JPL's ODP and The Aerospace Corporation's CHASMP use the JPL/Earth
Orientation Parameters (EOP) values.    This could be a source of common
error.  However the comparisons between EOP and IERS show an 
insignificant difference. Also, only secular terms, such as precession, can
contribute errors to the anomalous acceleration. Errors in short period
terms  are not correlated  with the anomalous acceleration.

%***************************** PARAMETER ESTIMATE

\subsection{Parameter estimation strategies}
\label{sec:PE}

During the last few decades, the algorithms of orbital analysis
have been extended to incorporate  Kalman-filter estimation procedure 
that is based on the concept of ``process noise''  (i.e., random,
non-systematic 
forces, or random-walk  effects).  This was
motivated by the need to respond to the significant improvement  in
observational accuracy and, therefore, to the  increasing sensitivity to 
numerous small perturbing factors of a stochastic nature that
are responsible for  observational noise.  This approach is well
justified when one needs to make accurate predictions of the 
spacecraft's future behavior  using only the spacecraft's past hardware
and electronics state history as well as the dynamic environment
conditions in the  distant craft's vicinity. Modern navigational
software often uses Kalman filter estimation 
since it more easily allows determination of 
the temporal noise history than does the  weighted least-squares
estimation.  

To take advantage of this while obtaining JPL's original results
\cite{anderson,moriond} discussed in  Section \ref{results}, JPL used  
batch-sequential methods with variable batch sizes and process noise
characteristics.  That is, a  batch-sequential
filtering and smoothing algorithm with process noise was  used with ODP. 
In this approach any small anomalous forces may be treated as 
stochastic parameters affecting the spacecraft trajectory. As
such, these parameters are also responsible for the stochastic
noise in the observational data. To better characterize these noise
sources, one splits the data interval into  a number of 
constant or variable size batches and makes  assumptions 
on possible statistical properties of these noise factors.  One then
estimates the mean values of the unknown parameters within the
batch and also their second statistical moments.  

Using batches has the advantage of dealing with a smaller number of
experimental data segments.  We  experimented with a number of different
constant  batch sizes; namely,  0, 5, 30, and 200  day batch sizes. 
(Later we also used 1 and 10 day batch sizes.)  
In each batch one estimates the
same number of desired parameters.  So, one expects that the smaller the
batch size the larger the resulting statistical errors.  This is because a
smaller number of data points is used to estimate the same number of
parameters.  Using the entire data interval as a single batch while
changing the process noise {\it a priori} values is expected 
in principle (see below) to
yield  a result identical to the least-squares estimation.  In the single
batch case, it would produce only one solution for the anomalous
acceleration. 

There is another important parameter that was taken into account in the
statistical data analysis reported here.  This is the expected correlation
time for the underlying stochastic processes (as well as the process
noise) that may be  responsible for the anomalous  acceleration. For
example,  using a  zero correlation time is useful in  searches for an 
$a_P$ that is generated by a random process.   One therefore 
expects that an $a_P$ estimated from one batch is statistically
independent (uncorrelated) from those estimated from other batches.  Also,
the use of finite correlation times indicates one is considering an 
$a_P$ that may show a temporal variation within the data interval.  
We experimented with a number of possible correlation times and will
discuss the corresponding assumptions when needed. 
 
In each batch one estimates solutions for the set of desired
parameters at a specified epoch within the batch.  One usually
chooses to report solutions  corresponding to the
beginning, middle, or  end of the batch. General coordinate
and  time transformations (discussed in Section
\ref{sec:time_scales}) are then used  to report the solution 
in the epoch chosen for  the entire data interval. One may also 
adjust the solutions among adjacent batches by accounting for 
possible correlations. This process produces a smoothed solution for the
set of solved-for parameters.  More details on this so
called ``batch--sequential algorithm with smoothing filter''  are
available in Refs. \cite{Moyer71}-\cite{Gelb}.

Even without process noise, the inversion algorithms of the 
Kalman formulation and the weighted least-squares method seem radically
different.  But as shown in \cite{sherman}, if one uses a single batch 
for all the data and if one uses certain assumptions 
about, for instance,  the process noise and the smoothing algorithms,
then the two methods are mathematically identical.  
When  introducing process noise, an additional
process noise matrix is also added into the solution algorithm. The
elements of this matrix are chosen by the user as prescribed by  standard
statistical techniques used for navigational data processing. 

For the recent results reported in Section \ref{recent_results},  JPL
used  both the batch-sequential and  the weighted least-squares 
estimation approaches. JPL originally implemented only the 
batch-sequential method, which yielded the detection (at a level smaller
than could be detected with any other  spacecraft) of an annual
oscillatory term smaller in size than the anomalous acceleration
 \cite{moriond}. (This term is discussed in Section \ref{annualterm}.)  
The recent studies  included weighted least-squares estimation to see
if this annual term was a calculational anomaly. 
 
The Aerospace Corporation uses only the weighted least-squares approach
with its CHASMP software.    A $\chi^2$ test is used as an indicator of
the quality of the fit.  In this case, the anomalous acceleration is
treated  as a constant parameter over the entire data interval. To solve
for $a_P$ one estimates the statistical weights   for the data points and
then uses these in a general  weighted  least-squares fashion.   Note
that the weighted least-squares method can obtain a result similar  to
that from a batch-sequential approach (with smoothing filter, zero
correlation time and  without process noise)  by cutting the data
interval into smaller pieces  and then  looking at the temporal variation
among the individual solutions.  

As one will see in the following, in the end, both programs yielded very 
similar results.  The differences between them can be  mainly attributed
to  (other) systematics.  This gives  us confidence that both programs
and their implemented estimation algorithms are correct to the accuracy
of this investigation.

%***************** SECTION 5 --- FIRST RESULTS **********
%\newpage

\section{\label{results}ORIGINAL DETECTION OF THE ANOMALOUS ACCELERATION}

\subsection{Early JPL studies of the anomalous Pioneer Doppler
residuals}
\label{subsec:accel}

As mentioned in the introduction, by 1980 Pioneer 10 was at 20 AU,  so
the solar radiation pressure  acceleration had decreased to $< 5\times
10^{-8}$ cm/s$^2$. Therefore, a search for unmodeled accelerations 
(at first with the further out Pioneer 10) could begin at this level.  
With the acceptance of a proposal of two of us (JDA and ELL) to
participate in the Heliospheric Mission on Pioneer 10 and 11, such a
search began in  earnest \cite{jpl}.  

The JPL analysis of unmodeled  accelerations used the JPL's Orbit 
Determination  Program (ODP)  \cite{Moyer71}-\cite{Moyer00}.  Over the
years the data continually indicated that the largest systematic error in
the  acceleration residuals is a constant bias of 
$a_P \sim (8\pm 3) \times 10^{-8}$ cm/s$^2$,  directed {\it toward} the
Sun (to within the beam-width of the Pioneers' antennae \cite{sunearth}).  

As stated previously, the analyses were  modeled to include the effects
of planetary perturbations, radiation pressure, the interplanetary 
media, general relativity, together with  bias and drift in the Doppler
signal.  Planetary coordinates and the solar system masses  were taken
from JPL's Export Planetary  Ephemeris DE405,  referenced to ICRF.  The
analyses used the standard space-fixed J2000 coordinate system with its
associated JPL planetary ephemeris DE405 (or earlier, DE200). The
time-varying Earth orientation in J2000 coordinates is defined by a 1998
version of JPL's EOP file, which accounts for the inertial
precession and nutation of the Earth's spin axis, the geophysical
motion of the Earth's pole with respect to its spin axis, and the Earth's
time varying spin rate. The three-dimensional locations of the tracking
stations in the Earth's body-fixed coordinate system (geocentric radius,
latitude, longitude) were taken from a set recommended by ICRF for JPL's
DE405.  

Consider  ${\nu}_{\tt obs}$,  the frequency  of
the re-transmitted signal observed by a DSN antennae, and 
$\nu_{\tt model}$,  the predicted frequency  of that signal.  
The observed, two-way anomalous effect can be expressed 
to first order in $v/c$ as \cite{drift}
\begin{eqnarray}\nonumber
\left[\nu_{\tt obs}(t)- \nu_{\tt model}(t)\right]_{\tt DSN} 
= - \nu_{0}\frac{2a_P~t}{c}, \\
\nu_{\tt model} = \nu_{0}\left[1 - \frac{2v_{\tt model}(t)}{c}\right]. 
\label{eq:delta_nu}
\end{eqnarray}
Here,  $\nu_{0}$ is the reference frequency, 
the factor $2$ is because we use two- and three-way data \cite{way}. 
$v_{\tt model}$ is the modeled  velocity of the
spacecraft due to the gravitational and other large forces 
discussed in Section \ref{navigate}.
(This velocity is outwards and hence produces a red shift.) 
We have already 
included the sign showing that $a_P$ is inward.  (Therefore, $a_P$  
produces a slight blue shift on top of the larger red shift.)  
By DSN convention \cite{drift}, the first of Eqs. (\ref{eq:delta_nu}) is  
$[\Delta \nu_{\tt obs} - \Delta \nu_{\tt model}]_{\tt usual} = 
- [\Delta \nu_{\tt obs} - \Delta \nu_{\tt model}]_{\tt DSN}$.  

\begin{figure}[!ht]
\begin{center}\noindent \vskip 0pt
\psfig{figure=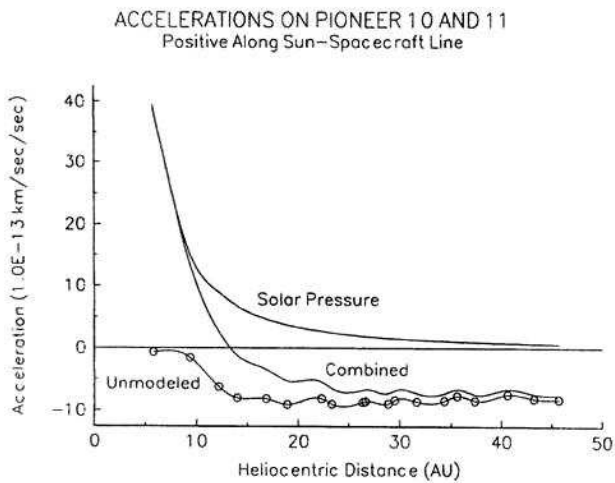,width=85mm}%,height=100mm}
\end{center} \vskip -20pt 
\caption
{ODP plots, as a function of distance from the Sun, of  accelerations on
Pioneers 10/11.  The accelerations are  a) the calculated  solar
radiation acceleration (top line), b) the unmodeled acceleration 
(bottom line), and c) the sum of the two above (middle line)
\cite{gleaned}.  
 \label{fig:forces}}
\end{figure}

Over the years the anomaly remained in the data of both Pioneer 10 and
Pioneer 11 \cite{bled}. (See Figure \ref{fig:forces}.)

%************

%************

\begin{figure}[!ht]
\begin{center}\noindent \vskip -10pt 
\psfig{figure=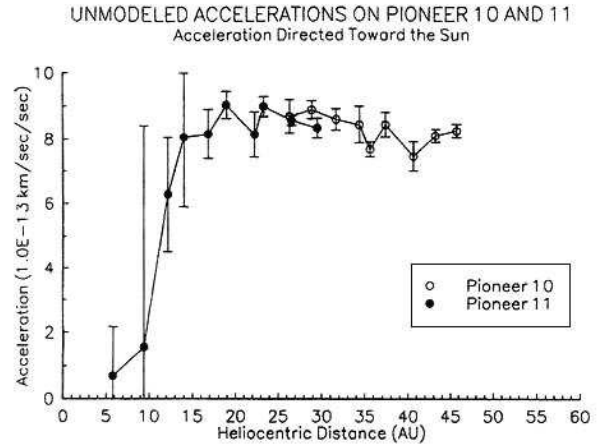,width=85mm}%,height=110mm}
\end{center}  \vskip -20pt 
  \caption
{An ODP plot of the early unmodeled 
accelerations of Pioneer 10 and Pioneer 11, from about 
1981 to 1989 and 1977 to 1989, respectively \cite{gleaned}.
\label{fig:correlation}}
\end{figure}

In order to model any unknown forces acting on Pioneer 10, the JPL group
introduced a stochastic acceleration, exponentially correlated in time,
with a time constant that can be varied. This stochastic variable is
sampled in ten-day batches of data. We found that a correlation time of
one year produces good results.    We did, however, experiment with
other time constants as well, including a zero correlation time (white
noise). The result of applying this technique to 6.5 years of Pioneer 10
and 11 data is shown in Figure \ref{fig:correlation}. 
The plotted points  represent our determination of the stochastic variable at
ten-day sample intervals.  We plot the stochastic variable as a function
of heliocentric distance, not time, because that is more fundamental in
searches for trans-Neptunian sources of gravitation.  

%************

%************

As possible  ``perturbative forces'' to explain this bias,  we considered
gravity  from the Kuiper belt, gravity from the galaxy, spacecraft ``gas
leaks,'' errors in the planetary ephemeris, and errors in the accepted
values of the Earth's orientation, precession,  and nutation.  We found
that none of these mechanisms could  explain the apparent acceleration, 
and some were three orders of magnitude or more too small.  [We also ruled
out a number of specific mechanisms involving heat radiation or  
``gas leaks,'' even though we feel these are  candidates for the cause
of the anomaly.  We will return to this in  Sections 
\ref{ext-systema} and \ref{int-systema}.]
 
We  concluded \cite{anderson}, from the JPL-ODP analysis,  that there is
an unmodeled  acceleration, $a_P$, towards the Sun  of
$(8.09\pm0.20)\times 10^{-8}$ cm/s$^2$ for Pioneer 10 and of $(8.56\pm
0.15) \times 10^{-8}$ cm/s$^2$ for Pioneer 11.  The error was determined
by use of a  five-day batch sequential filter with radial acceleration as
a stochastic parameter subject to white Gaussian noise ($\sim$ 500
independent five-day samples of radial acceleration) \cite{tap}. No
magnitude variation of $a_P$ with distance was found, within a  
sensitivity of $\sigma_0=2\times10^{-8}$ cm/s$^2$ over a  range of 40
to 60 AU. All our errors are taken from the covariance matrices associated
with the least--squares data  analysis. The assumed data errors are larger
than the standard error on the  post--fit residuals.  [For example, the
Pioneer S--band Doppler error  was set at 1 mm/s at a Doppler integration
time of 60 s,  as opposed to a characteristic $\chi^2$ value of 0.3
mm/s.]  Consequently, the quoted errors are  realistic, not formal, and
represent our attempt to include systematics  and a reddening of the
noise spectrum by solar plasma. Any spectral  peaks in the post-fit
Pioneer Doppler residuals were not  significant at a 90\%
confidence level
\cite{anderson}.

%*****************

\subsection{First Aerospace study of the apparent Pioneer acceleration}
\label{subsec:aero}

With no explanation of this data in hand, our attention focused on the
possibility that there was some error in  JPL's ODP.  To investigate
this, an analysis of the raw data was performed using an independent 
program, The Aerospace Corporation's Compact High Accuracy Satellite
Motion Program (CHASMP) \cite{chasmp} -- one of the standard Aerospace 
orbit analysis programs. CHASMP's orbit determination module is a
development  of a program called POEAS (Planetary Orbiter Error Analysis
Study program)  that was developed at JPL in the early 1970's
independently of JPL's ODP. As  far as we know, not a single line of code
is common to the two programs \cite{poeas}. 

Although, by necessity, both ODP and CHASMP use the same physical
principles,  planetary ephemeris, and timing and polar motion inputs, the
algorithms are  otherwise quite different. If there were an error in
either program, they would not agree.  

Aerospace analyzed a Pioneer 10 data arc that was initialized
on 1 January 1987 at  16 hr (the data itself started on 3 January)
and ended at 14 December 1994, 0 hr. The raw data set 
was averaged to 
7560 data points of which 6534 points were used.  This CHASMP analysis of
Pioneer 10 data also showed an unmodeled acceleration  in a direction
along the radial toward the  Sun \cite{aero}.  The value is $(8.65 \pm
0.03) \times 10^{-8}$ cm/s$^{2}$,  agreeing with JPL's result. The
smaller error here is because the CHASMP analysis  used a batch
least-squares fit over the whole orbit \cite{tap,chasmp}, not
looking  for a variation of the magnitude of $a_P$ with distance.

\begin{figure}
\begin{center}\noindent
\psfig{figure=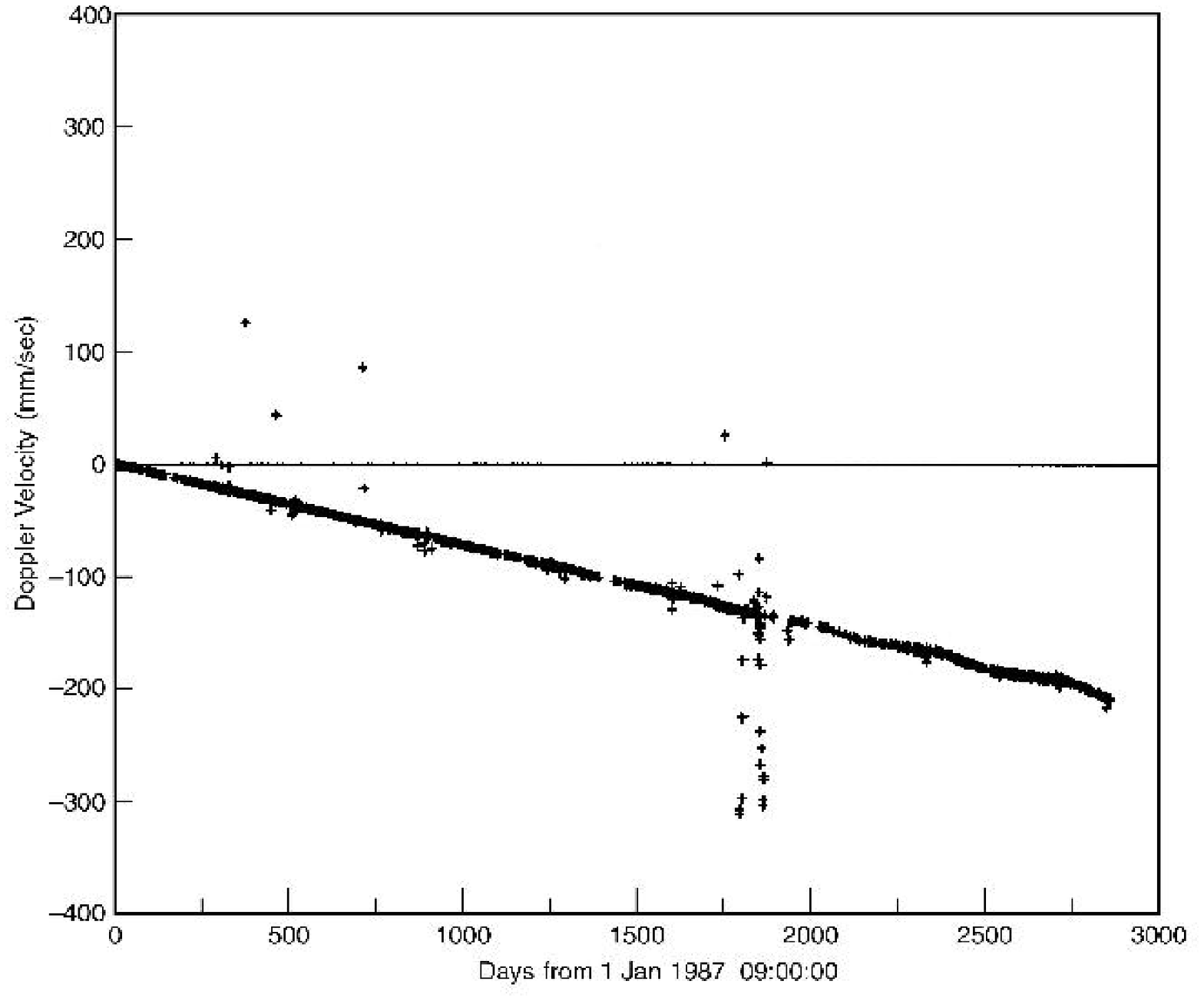,width=84mm}%,height=75mm}
\end{center} \vskip -10pt  
  \caption
{CHASMP two-way Doppler residuals (observed Doppler velocity 
minus model Doppler velocity) for Pioneer 10 vs. time. 1 Hz is equal
to 65 mm/s range change per second.  The model is fully-relativistic. 
The solar system's gravitational field is represented by the Sun and
its planetary systems \cite{Standish92}. 
\label{fig:aerospace}}

\begin{center}\noindent
\epsfig{figure=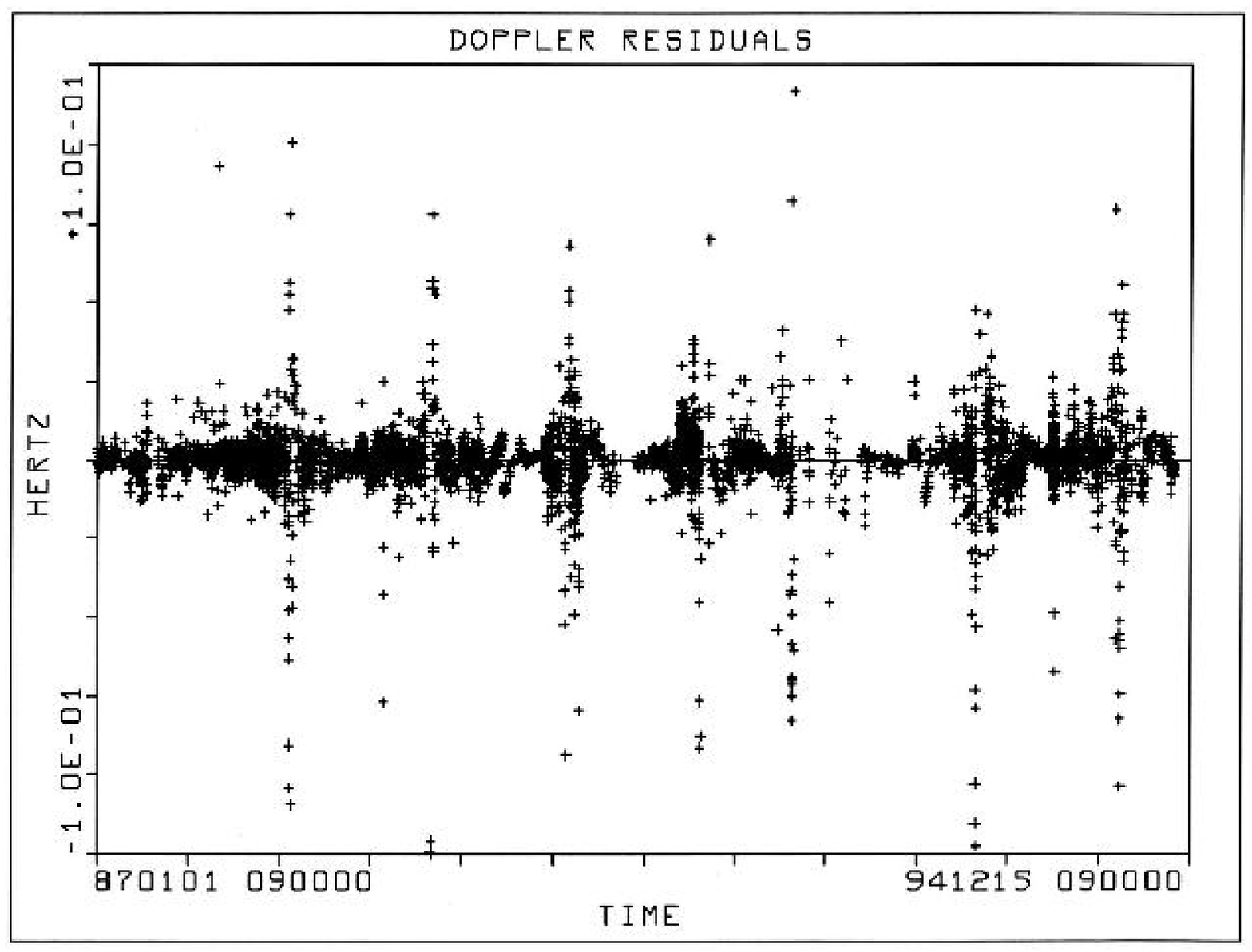,width=82mm}%,height=90mm}
\end{center}  \vskip -10pt 
   \caption
{CHASMP best fit for the Pioneer 10 Doppler residuals with the 
anomalous acceleration taken out.  
After adding one more parameter to the
model (a constant radial acceleration) the residuals are distributed 
about zero Doppler velocity with a systematic variation $\sim$ 3.0 mm/s
on a time scale of $\sim$ 3 months.  The outliers on the plot were
rejected from the fit. [The quality of the fit may be determined by the
ratio of residuals to the downlink carrier frequency, $\nu_0\approx$
2.29 GHz.]
 \label{fig:pio10best_fit}}
\end{figure}

Without using the apparent acceleration,  CHASMP shows a steady
frequency  drift \cite{drift} of about $-6 \times 10^{-9}$ Hz/s, or 1.5
Hz over 8 years  (one-way only).  (See Figure \ref{fig:aerospace}.)  
This equates to a  clock acceleration, $-a_t$, of
$-2.8\times 10^{-18}$ s/s$^{2}$. The identity with the apparent Pioneer
acceleration is 
\begin{equation}
a_t \equiv a_P/c.  \label{asubt}   
\end{equation}
The drift in the Doppler residuals
(observed minus computed data) is  seen in Figure 
\ref{fig:pio10best_fit}.

%************

%************

The drift is clear, definite, and  cannot be removed without either  the
added acceleration, $a_P$,   or the inclusion in the data itself  of a
frequency drift,  i.e., a ``clock acceleration'' $a_t$.  If there were a
systematic drift in the atomic clocks of the DSN  or in the
time-reference standard signals, this would appear   like a
non-uniformity of time; i.e., all clocks would be changing with a 
constant acceleration. We now have been able to rule out this
possibility. (See Section \ref{sec:timemodel}.) 

Continuing our search for an explanation, we considered the 
possibilities: i) that the Pioneer 10/11 spacecraft had internal 
systematic properties, undiscovered because they are of identical
design,  and ii) that the acceleration was due to some  not-understood
viscous drag force (proportional to the approximately  constant velocity
of the Pioneers). Both these possibilities could be investigated by
studying  spin-stabilized spacecraft whose spin axes  are not directed
towards the Sun, and whose orbital velocity vectors  are far from being
radially directed.  

Two candidates were Galileo in its  Earth-Jupiter mission phase and 
Ulysses  in Jupiter-perihelion  cruise out of the plane of the 
ecliptic.  As well as Doppler, these spacecraft also yielded a
considerable quantity of range data.   By having range data one can tell
if a spacecraft is accumulating a range effect due to a spacecraft
acceleration or if the orbit determination process is fooled by a Doppler
frequency rate bias. 

%*******************************

\subsection{Galileo measurement analysis}
\label{galileo}

We considered the dynamical behavior of Galileo's trajectory
during its  cruise   flight from second Earth encounter  (on 8 December
1992)  to arrival at Jupiter.   [This period ends just before the Galileo
probe release on 13 July 1995. The probe reached Jupiter on 7 December
1995.] During this time the spacecraft traversed a  distance of about 5
AU with an approximately constant  velocity of 7.19(4) km/s. 

A quick JPL look at limited Galileo data  (241 days from 8 January 1994
to 6 September 1994)  demonstrated that it was impossible to separate 
solar radiation effects from an anomalous constant acceleration.   The
Sun was simply too close and the radiation cross-section  too large.  The
nominal value obtained was $\sim 8 \times 10^{-8}$ cm/s$^2$.  

The Aerospace's analysis of the Galileo data covered the same arc as JPL 
and a  second arc from 2 December 1992 to 24 March 1993.   The analysis of
Doppler data from the first arc resulted in a determination for $a_P$ of  
$\sim (8 \pm 3) \times 10^{-8}$ cm/s$^2$,   a value similar to that from
Pioneer 10.  But the correlation with solar pressure was so high (0.99)
that it is impossible to decide whether solar pressure is a contributing
factor \cite{aT}. 

The second data arc was 113 days long, starting   six days prior to the
second Earth encounter. This solution  was also too highly correlated with
solar pressure, and the data analysis was  complicated by many mid-course
maneuvers in the orbit. The uncertainties in the maneuvers were so great, a
standard null result could not be ruled out.

%************
\begin{figure}[h]
\begin{center}\noindent
\psfig{figure=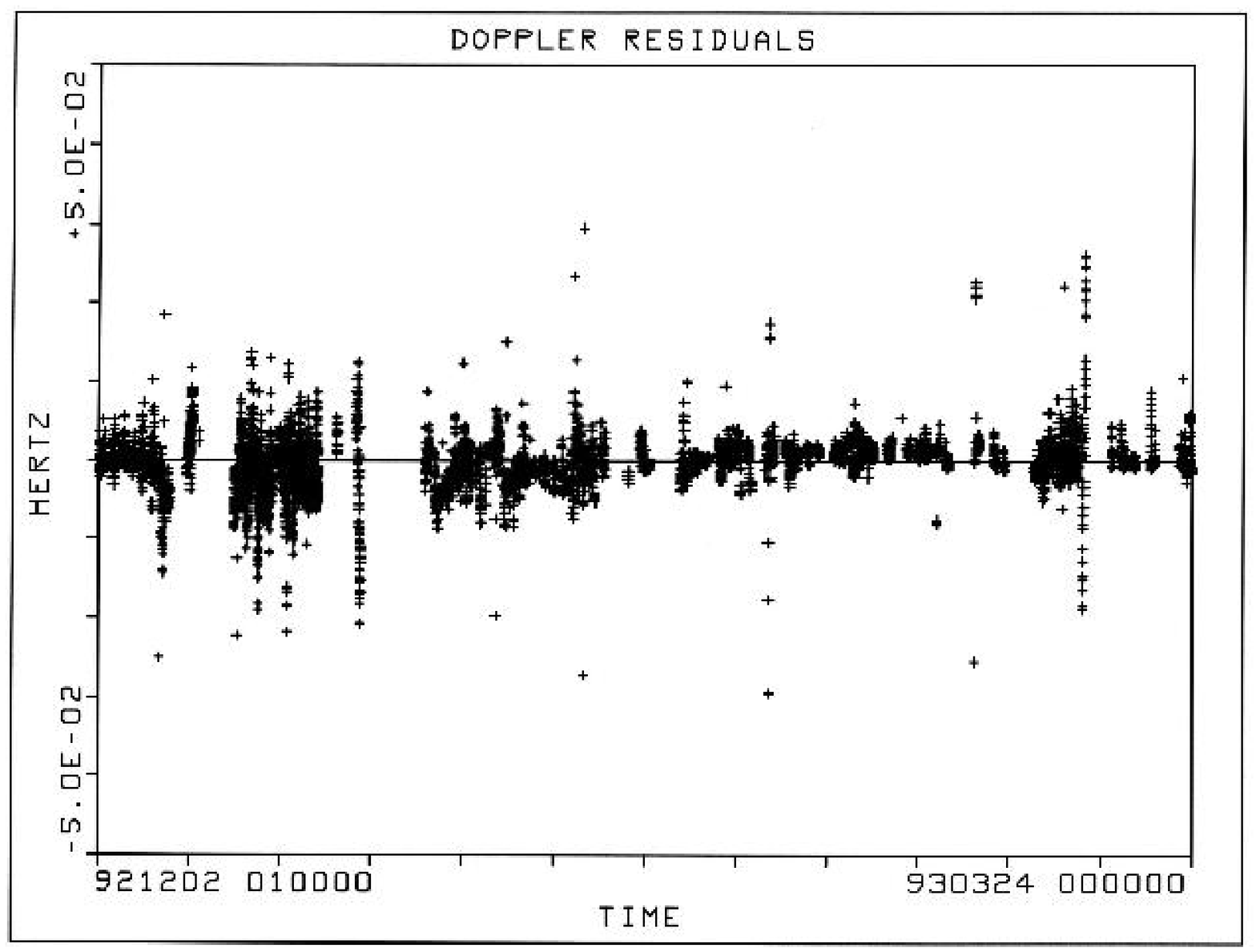,width=82mm}%,height=75mm}
\end{center}  
\begin{center}\noindent\vskip -10pt
\psfig{figure=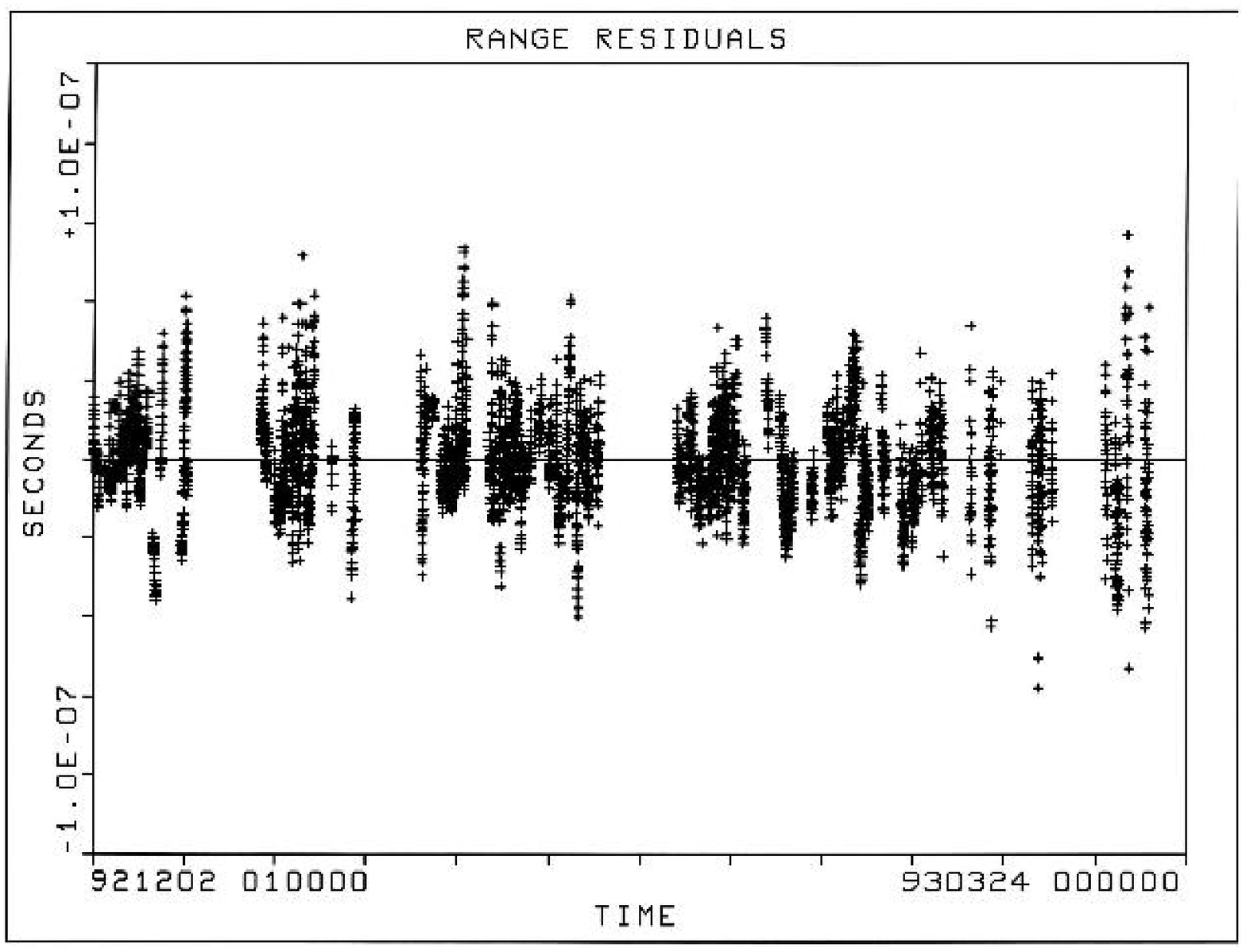,width=82mm}%,height=75mm}
\end{center} \vskip -10pt  
\caption{Galileo best fit Doppler and  range residuals using CHASMP. 
            \label{fig:galileo_range}}  
\end{figure}
%************

However, there was an additional result from the data of this second 
arc. This arc was chosen for study because of the  availability of
ranging data. It had 11596 Doppler points of which 10111 were
used and 5643 range points of which 4863 used. The two-way range change and
time integrated  Doppler are consistent  
(see Figure \ref{fig:galileo_range}) to $\sim 4$ m over  a time interval
of one day.   For comparison,
note that for a time of $t=1$ day, $(a_Pt^2/2)\sim 3$ m.   
For the apparent acceleration to be the result of 
hardware problems at the tracking stations, one would need a linear
frequency drift at all the DSN stations, a drift that is not observed.  

%****************

\subsection{Ulysses measurement analysis}
\label{ulysses}

%**************

\subsubsection{JPL's analysis}

An analysis of the radiation pressure on Ulysses,  
in its out-of-the-ecliptic journey from 5.4 AU near Jupiter in 
February 1992 to the perihelion at 1.3
AU in February 1995,  found a varying profile with distance \cite{uly}. 
The orbit solution requires a periodic updating of the solar radiation
pressure. The radio Doppler and ranging data can be fit to the noise level
with a time-varying solar constant in the fitting model
\cite{mcelrath}.   We obtained values for the time-varying solar
constant determined by Ulysses navigational data during this south polar
pass \cite{uly}.  The inferred solar constant is about 40 percent larger at
perihelion (1.3 AU) than at Jupiter (5.2 AU), a physical impossibility! 

We sought an alternative explanation. Using physical parameters of the
Ulysses spacecraft, we first converted the time-varying values of the solar
constant to a  positive (i.e., outward) radial spacecraft acceleration,
$a_r$, as a function of heliocentric radius. Then we fit the values of
$a_r$ with the following model:
\begin{equation} 
a_r = 
\frac{\mathcal{K}f_\odot A}{c M}\frac{\cos\theta(r)}{r^2} - a_{P(U)},
\label{Armodel_corr}
\end{equation}
where $r$ is the heliocentric distance in AU, $M$  is the total mass
of the spacecraft, $f_\odot=1367 ~{\rm W/m}^{2}$(AU)$^2$ is the 
(effective-temperature Stefan-Boltzmann) 
``solar radiation constant'' 
at 1 AU,  $A$ is the cross-sectional  area of the spacecraft and
$\theta(r)$ is the angle between the direction to the Sun at distance
$r$ and orientation of the antennae. [For the period analyzed 
$\theta(r)$ was almost a constant. Therefore its average value was
used which corresponded to $\langle{\cos\theta(r)}\rangle\approx 0.82$.]
Optical parameters defining the  reflectivity and emissivity of the
spacecraft's surface were taken to yield $\mathcal{K}\approx 1.8$. 
(See Section \ref{solarP} for a discussion on solar radiation pressure.) 
Finally, the parameter $a_{P(U)}$ was determined by linear least squares. 
The best--fit value was obtained
\begin{equation}
a_{P(U)} = (12 \pm 3)\times 10^{-8}~~{\rm cm/s}^2,
\label{SandA}
\end{equation}
where both random and systematic errors are included. 

So, by  interpreting this time variation 
as a true $r^{-2}$ solar pressure plus a constant radial
acceleration, we found that Ulysses was subjected to an
unmodeled acceleration  towards the Sun of  
(12 $\pm$ 3) $\times 10^{-8}$ cm/s$^{2}$.

Note, however, that the determined constant $a_{P(U)}$ is highly  
correlated  with solar radiation pressure (0.888). This shows that
the constant acceleration and the solar-radiation acceleration are not
independently determined, even over a heliocentric distance variation
from 5.4 to 1.3 AU.
 
%****************

\subsubsection{Aerospace's analysis}
\label{sec:AUlysses}

The next step was to perform  a detailed  calculation of the Ulysses orbit
from near Jupiter encounter to Sun  perihelion, using CHASMP to evaluate
Doppler and ranging data.   The data from 30 March  1992 to 11 August
1994 was processed. It consisted of 50213 Doppler points of which 46514
were used and 9851 range points of which 8465 were used.

Such a calculation would in principle allow a more precise and  believable
differentiation between an anomalous constant acceleration  towards the Sun
and systematics.  Solar radiation pressure and radiant  heat systematics
are both larger on Ulysses than on the Pioneers.

However, this calculation turned out to be a much more difficult than
imagined.   Because of a failed nutation damper,  an inordinate number of
spacecraft maneuvers were required (257).   Even so, the analysis was
completed.   But even though the Doppler and range residuals were 
consistent as for Galileo, the results were disheartening.  For an
unexpected  reason, any fit is not significant.   The anomaly is dominated
by  (what appear to be)  gas leaks \cite{ulygas}.   That is, after each
maneuver the measured anomaly changes.   The measured anomalies randomly
change sign and magnitude.  The values  go up to about an order of
magnitude larger than $a_P$. So, although the Ulysses data was useful for
range/Doppler checks  to test models (see Section \ref{sec:timemodel}),
like  Galileo it could not provide a good number to compare to $a_P$.

%***********************6) RECENT/LATEST RESULTS
%\newpage

\section{\label{recent_results}RECENT RESULTS}

Recent changes to our strategies and orbit determination programs, 
leading to new results, are threefold.   First, we have added a longer data
arc for Pioneer 10, 
extending the  data studied up to July 1998.   The entire data
set used (3 Jan. 1987 to 22 July 1998) 
covers a  heliocentric distance interval from 40 AU to 70.5 AU \cite{AU}. 
[Pioneer 11  was much closer in (22.42 to 31.7 AU) than Pioneer
10 during its data interval (5 January 1987 to 1 October 1990).]   
For later use in discussing systematics, we here note that 
in the ODP calculations, masses used for the Pioneers were  
$M_{Pio~10}=251.883$ kg and  $M_{Pio~11}=239.73$ kg.  CHASMP used 251.883 
kg for both \cite{gasuse}.  As the majority of our results are from Pioneer
10, we will make $M_0 = 251.883$ kg to be our nominal working mass.

Second, and as we discuss in the next subsection, we have studied the spin
histories of the craft.  In particular, the Pioneer 10 history exhibited
a very large anomaly in the period  1990.5 to 1992.5.  
This led us to take a closer look at any
possible   variation of $a_P$ among the three time intervals:  The JPL
analysis defined the intervals as I (3 Jan. 1987 to 17 July 1990);  II
(17 July 1990 to 12 July 1992) bounded by 49.5 to 54.8 AU;  and III (12
July 1992 to 22 July 1998).    (CHASMP used slightly different
intervals \cite{I/II})  
The total updated data set now
consists of  20,055 data points for Pioneer 10. (10,616 data points were
used for  Pioneer 11.)   This helped us to better 
understand the systematic due to   gas leaks, which is taken up in
Section \ref{sec:gleaks}.   

Third, in looking at the detailed measurements of $a_P$ as a function of
time using ODP, we found an anomalous oscillatory annual term,  smaller
in size than the anomalous acceleration \cite{moriond}.  As mentioned in
Section
\ref{sec:PE},  and as will be discussed in detail in Section
\ref{annualterm},    we wanted to make sure this annual term was not an
artifact of our  computational method.   For the latest results, JPL used
both the batch-sequential and the least-squares methods.  

All our recent results obtained with both the JPL and The Aerospace
Corporation software have given us a better understanding of systematic
error sources.  At the same time  they have increased our confidence in the
determination of the anomalous acceleration. We present a description and
summary of the new results in the rest of this section.      

%*************************************

\subsection{Analysis of the Pioneer spin history}
\label{spinhistory} 

\begin{figure}
\hskip -5pt
\epsfig{figure=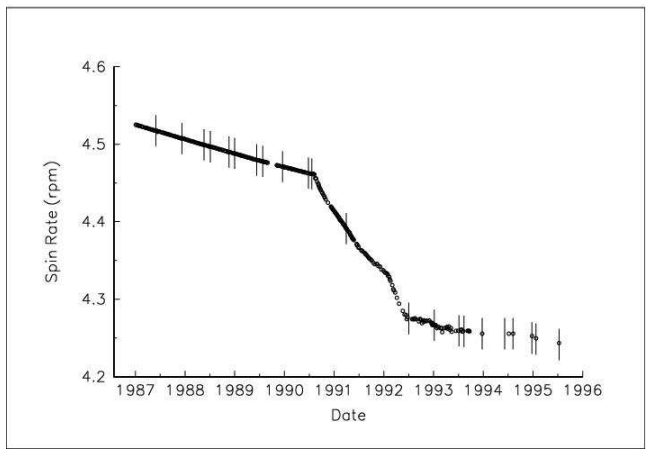,width=85mm}%,height=90mm}
    \caption
{The spin history of Pioneer 10. 
The vertical lines indicate the times when
precession  maneuvers were made.  How this spin data was obtained 
is described in   Section \ref{spincalibrate}.
The final data points were obtained at the 
times of maneuvers, the last being in 1995.  
      \label{fig:pioneer_spin}} 
\end{figure}

\begin{figure}
\epsfig{figure=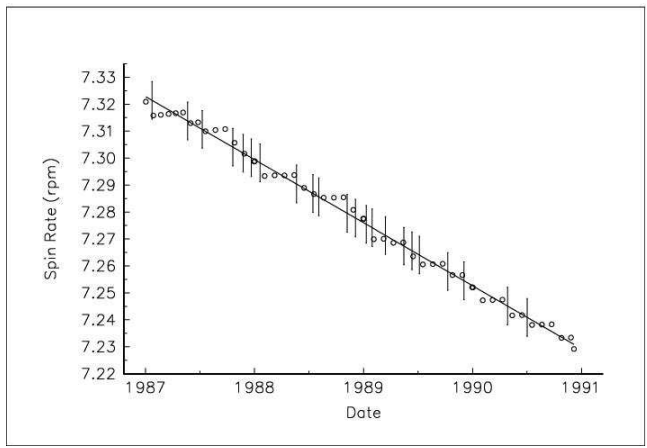,width=85mm}%,height=90mm}
    \caption
{The spin history of Pioneer 11 over the period of  analysis.  
The vertical lines indicate the times when
precession  maneuvers were made. This spin
calibration was done by the DSN until 17 July 1990.  At that time the DSN
ceased doing spin  calibrations. From 1990 until the loss of coherent
Doppler, orbit analysts made estimates of the spin rate. 
      \label{fig:pio11spin}}
\end{figure}

Both Pioneers 10 and 11 were spinning down during the respective data intervals
that determined their $a_P$ values.  Because any changes in spacecraft spin
must be associated with spacecraft torques (which for lack of a plausible
external mechanism we assume are internally generated), there is also a
possibility of a related internally generated translational force along the
spin axis.  Therefore, it is important to understand the effects of the spin
anomalies  on the anomalous acceleration. In Figures \ref{fig:pioneer_spin} 
and \ref{fig:pio11spin}  we show the spin histories of the two craft during the
periods of analysis.

%************
%************

%************
%************

Consider Pioneer 10 in detail.  In  time Interval I there is a slow spin
down   at an average  rate (slope) of   $\sim(-0.0181\pm 0.0001)$ rpm/yr.  
Indeed, a closer look at the curve (either by eye or from an expanded graph)
shows that the spin down is actually slowing with time (the curve is
flattening).  This last feature will be discussed in Sections \ref{subsec:katz}
and  \ref{subsec:mainbus}. 

Every time  thrusters are used, there tends to be a short-term leakage of gas
until the valves set (perhaps a few days later). But there can also be
long-term leakages due to some mechanism which does not quickly correct
itself.  The  major Pioneer 10 spin anomaly that marks the boundary of
Intervals I and II, is a case in point.  During this interval there was a major
factor of $\sim 4.5$ increase in  the average spin-rate change to
$\sim(-0.0861\pm0.0009)$ rpm/yr.  One also notices kinks during the 
interval.   

Few  values of the Pioneer 10 spin rate were obtained after 
mid-1993, so the long-term 
spin-rate change is not well-determined in Interval III.  
But from what was
measured, there was first a short-term transition region of about a year where
the spin-rate change was $\sim-0.0160$  rpm/yr.  Then things settled down
to a spin-rate change of about $\sim(-0.0073 \pm 0.0015)$ rpm/yr, which 
is small and less than that of interval I. 

The effects of the
maneuvers on the values of $a_P$ will allow an estimation of the gas leak
systematic in Section   \ref{sec:gleaks}.  Note, however, that in the time
periods studied, only orientation  maneuvers were made, not trajectory
maneuvers.  

Shortly after Pioneer 11 was launched 
on 5 April 1973, the spin period was
4.845 s. A spin precession maneuver on 18 May 1973 reduced the period
to 4.78 s and afterwards, because of a series of precession maneuvers,
the period lengthened until it reached 5.045 s at encounter with
Jupiter in December 1974. The period was fairly constant until 18
December 1976, when a mid-course maneuver placed the spacecraft on a
Saturn-encounter trajectory. Before the maneuver the period was 5.455
s, while after the maneuver it was 7.658 s. At Saturn encounter in
December 1979 the period was 7.644 s, little changed over the
three-year post maneuver cruise phase. At the start of our data
interval on 5 January 1987, the period was 7.321 s, while at the end of
the data interval in October 1990 it was 7.238 s. 

Although the linear fit to the Pioneer 11 spin rate shown in Figure 
\ref{fig:pio11spin} is similar to that for Pioneer 10 in Interval I, 
$\sim(-0.0234\pm 0.0003)$ rpm/yr, 
the causes appear to be very different.  
(Remember, although identical in design, Pioneers 10 and 11 were not
identical in quality \cite{design}.)  
Unlike Pioneer 10,
the spin period for Pioneer 11 was primarily affected at the time of spin
precession maneuvers.  One sees that at maneuvers the spin period decreases
very quickly, while in between maneuvers the spin rate actually tends to
{\it increase} at a rate of $\sim(+0.0073 \pm   0.0003)$   rpm/yr
(perhaps due to a gas leak in the opposite direction).   

All the above observations aid us in the interpretation of systematics in
the following three sections. 

%************************************

\subsection{Recent results using  JPL software}
\label{jplresults}

The latest results from JPL 
are based on an upgrade, \emph{Sigma}, to JPL's ODP
software \cite{sigma}. \emph{Sigma}, developed for NASA's Cassini
Mission to Saturn, eliminates   structural restrictions on memory  and
architecture that were imposed 30 years ago when JPL space navigation
depended solely on a Univac 1108 mainframe computer.  Five ODP programs and
their interconnecting files have been   replaced by the single program
\emph{Sigma} to support filtering, smoothing, and  mapping functions. 

%**************Begin Table**********************

\begin{table*}[ht]
\caption[Determinations of $a_P$ in units of $10^{-8}$ cm/s$^2$ from the
three time intervals of Pioneer 10 data and from Pioneer 11.]
{Determinations of $a_P$ in units of $10^{-8}$ cm/s$^2$ from the three
time intervals of Pioneer 10 data and from Pioneer 11.  As described  in
the text, results from various ODP/\emph{Sigma}  and CHASMP calculations
are listed.  For ODP/\emph{Sigma}, ``{\tt WLS}'' signifies  a weighted
least-squares calculation, which was used with i) no solar corona model
and   ii) the `Cassini' solar corona model.  Also for ODP/\emph{Sigma},
``{\tt BSF}'' signifies a batch-sequential filter calculation,  which
was done with iii)  the `Cassini' solar corona model.  Further (see
Section \ref{annualterm}),  a 1-day batch-sequential estimation for the
entire data interval  of 11.5 years for Pioneer 10 yielded a result 
$a_P= (7.77 \pm 0.16)\times 10^{-8}$ cm/s$^2$. The CHASMP calculations 
were all  {\tt WLS}.   These calculations were done with i) no solar
corona model,  ii) the `Cassini' solar corona model,  iii) the `Cassini'
solar corona model with corona data weighting and F10.7  time variation
calibration.   Note that the errors given are only formal calculational
errors. The much larger deviations of the results from each other
indicate the sizes of the systematics that are involved. 
\label{resulttable}}
\begin{center}
%\vskip 20pt 
\begin{tabular}{|l|c|c|c|c|} \hline\hline
Program/Estimation method & Pio 10 (I) & Pio 10 (II)&  Pio
10 (III) &  Pio 11
\\
\hline \hline
 \emph{Sigma}, {\tt WLS},& $$ & $$ & $$ & $$\\ 
no solar corona model  
& $8.02\pm0.01$ & $ 8.65\pm0.01$ & $7.83\pm0.01$ &
$8.46\pm0.04$\\ \hline
 \emph{Sigma}, {\tt WLS},  &$$ & $$ & $$ & $$\\ 
with solar corona model  
&$8.00\pm0.01$ & $8.66\pm0.01$ & 
$7.84\pm0.01$ & $8.44\pm0.04$\\\hline
 \emph{Sigma}, {\tt BSF}, 1-day batch,   &&&&\\
with solar corona model    
& $7.82\pm0.29$ & $8.16\pm0.40$ &
$7.59\pm0.22$ &  $8.49\pm0.33$ \\ 
\hline \hline
 CHASMP, {\tt WLS}, &&&&\\
no solar corona model  
& $8.25\pm0.02$ & $8.86\pm0.02$& $7.85\pm0.01$ & 
     $8.71\pm0.03$ \\ \hline
 CHASMP, {\tt WLS},   &&&&\\
 with solar corona model 
& $8.22 \pm 0.02 $ & $8.89\pm0.02$ & $7.92\pm0.01$ & 
     $8.69\pm0.03$ \\ \hline
 CHASMP, {\tt WLS}, with  &&&&\\
corona, weighting, and F10.7 & $8.25\pm0.03$ & $8.90\pm0.03$&
$7.91\pm0.01$ & 
     $8.91\pm0.04$ \\ 
\hline\hline
\end{tabular}
\end{center}
\end{table*}
%********************End Table*********************

We used \emph{Sigma} to reduce the Pioneer 10 (in  three time intervals)
and 11  Doppler  of the unmodeled acceleration, $a_P$, along the 
spacecraft spin axis.  As mentioned, the Pioneer 10  data interval was
extended to cover the   total time interval 3 January 1987 to 22 July
1998.  Of the total data set of  20,055 Pioneer 10 Doppler points, JPL
used  $\sim$19,403,  depending on the initial conditions and editing for
a particular run.   Of the available 10,616 (mainly shorter
time-averaged) Pioneer 11 data points, 10,252 were used (4919 two-way
and 5333 three-way).

We wanted  to produce independent (i.e., uncorrelated)  solutions for
$a_P$ in the three Pioneer 10 segments of data.    The word independent
solution in our approach means only the fact that data from any of the
three segments must not have any information (in any form) passed onto
it from the other two intervals while estimating the anomaly. We moved
the epoch from the beginning of one data interval to the next by
numerically integrating the equations of  motion and not iterating on
the data to obtain a better initial conditions for this consequent
segment. Note that this numerical iteration provided us only with an
\emph{a priori} estimate for the initial conditions for the data
interval in question. 

Other parameters included in the fitting model were the six spacecraft
heliocentric position and velocity coordinates at the  1987 epoch of 1
January 1987, 01:00:00 {\tt ET}, and 84 (i.e., $28\times 3$)
instantaneous velocity  increments along the three spacecraft axes for 28
spacecraft attitude (or spin orientation) maneuvers. If these orientation
maneuvers had been performed at exactly six month intervals, there would
have been 23 maneuvers over our 11.5 year data interval. But in fact,
five more maneuvers were performed than expected over  this 11.5 year 
interval giving a total of 28 maneuvers in all. 

As noted previously, in fitting the Pioneer 10  data over 11.5 years we
used the standard space-fixed J2000 coordinate system with planetary
ephemeris DE405,  referenced to ICRF.  The three-dimensional locations
of the tracking stations in the Earth's body-fixed coordinate system
(geocentric radius, latitude, longitude) were taken from a set
recommended  by ICRF for JPL's DE405.  The time-varying Earth
orientation in J2000 coordinates was defined by a 1998 version of JPL's
EOP file.  This  accounted for the geophysical motion of the Earth's
pole with respect to its spin axis  and the Earth's time varying spin
rate. 

JPL used both  the weighted least-squares ({\tt WLS}) and the 
batch-sequential filter ({\tt BSF}) algorithms  for the final
calculations.  In the first three rows of Table \ref{resulttable}  
are shown the ODP results for i) {\tt WLS} with no corona,  
ii)  {\tt WLS} with the Cassini  corona model, 
and iii) {\tt BSF} with the Cassini  corona model. 

Observe that the {\tt WLS} acceleration values for Pioneer 10 in
Intervals I, II,  and III  are larger or smaller, respectively,  just as
the   spin-rate changes in these intervals are larger or smaller,
respectively.   This indicates that the small  deviations may be due to a
correlation with the large gas leak/spin anomaly.  We will argue this
quantitatively in Section \ref{sec:gleaks}.   For now we just note that we
therefore   expect the number  from Interval III, $a_P= 7.83 \times
10^{-8}$cm/s$^2$, to be close to our  basic (least perturbed)  JPL result
for Pioneer 10.  We also note that the statistical errors and the effect
of the solar corona are both small for {\tt WLS}, and will be handled in
our error budget.

\begin{figure}[h!]
\epsfig{figure=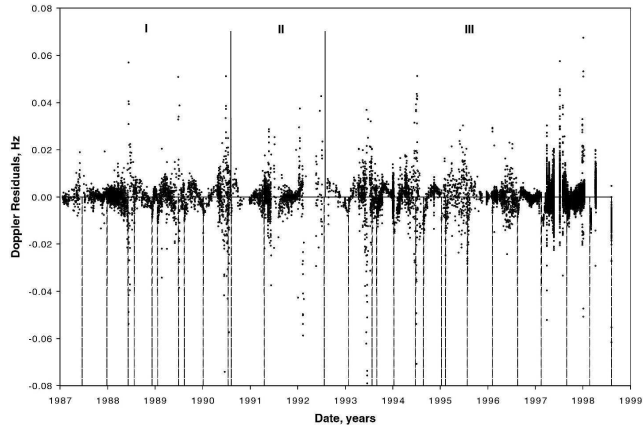,width=86mm}\vskip -10pt
    \caption
{ODP Doppler residuals 
in Hz for the entire Pioneer 10 data span.   
The two solid vertical lines in the upper part of the
plot indicate the boundaries between data Intervals  I/II and
II/III, respectively.   Maneuver times are indicated by the vertical
dashed lines in the  lower part of the plot.  
     \label{ODPall}}
\end{figure}

In Figure \ref{ODPall} we show  ODP/\emph{Sigma} {\tt WLS} 
Doppler residuals for the entire Pioneer 10 data set.   
The residuals were obtained by first solving for $a_P$ with no corona 
in each of the three Now look at the batch-sequential results in row 3 of Table
\ref{resulttable}. First, note that the statistical 
Intervals independently  and then subtracting
these solutions (given in Table \ref{resulttable}) from the 
fits within the corresponding data intervals.

%********************************Figure ODPallresid ********
%************
%************

One can easily see  
the very close agreement with the CHASMP residuals of
Figure \ref{fig:pio10best_fit}, which go up to 14 December 1994.

The Pioneer 11 number is significantly higher.   A deviation is not
totally unexpected since the data was relatively noisy, was from much
closer in to the Sun, and was taken during a period of high  solar
activity.  We also do not have the same handle on spin-rate change
effects as we did for Pioneer 10.  We must simply take the number 
for what it is, and give the basic JPL result for Pioneer 11 as 
$a_P= 8.46 \times 10^{-8}$ cm/s$^2$. 

Now look at the batch-sequential results in row 3 of Table
\ref{resulttable}. First, note that the statistical errors are an order of
magnitude larger than for {\tt WLS}.  This is not surprising since: i) the
process  noise significantly affects the precision, ii) {\tt BSF}
smoothes the data and the data from the various 
intervals is more correlated than in {\tt WLS}.  
The effects of all this are that all four numbers change 
so as to make them all closer to each other, but yet all
the numbers vary by less than $2
\sigma$ from their  {\tt WLS} counterparts.  

Finally, there is the annual term. It remains in the data (for both
Pioneers 10 and 11).   A representation of it can be seen in a 1-day
batch-sequential averaged over all 11.5 years.    It yielded a result
$a_P= (7.77 \pm 0.16) \times 10^{-8}$ cm/s$^2$, consistent with the 
other numbers/errors,   but with an added annual oscillation.   In the
following subsection we will compare JPL results showing the annual term
with the counterpart Aerospace results.  

We will argue in Section \ref{annualterm} that this annual term is  due
to the inability to model the angles of the Pioneers' orbits  accurately
enough.   [Note that this annual term is not to be confused with a  small
oscillation seen in Figure \ref{fig:aerospace} that can be  caused by
mispointing towards the spacecraft by the fit programs.]

%************************************

\subsection{Recent results using The Aerospace Corporation software}
\label{aerospaceresults}

As part of an ongoing upgrade to CHASMP's accuracy, Aerospace has used
Pioneer 10 and 11 as a test bed to confirm the revision's improvement.
In accordance with the JPL results of Section \ref{jplresults}, we  used
the  new version of CHASMP to  concentrate on the Pioneer 10 and 11 data.
The  physical models are basically the same ones that JPL used, but the
techniques  and methods used are largely different. (See Section
\ref{Ext_accuracy}.)  

The new results from the  Aerospace Corporation's
software  are based on first improving the  Planetary Ephemeris
and Earth orientation and spacecraft spin models required  by the program.
That is:  i) the spin data
file has been included with full detail; ii) a newer JPL Earth 
Orientation Parameters file was used;  iii) all IERS tidal terms were 
included; iv) plate tectonics were included; v) DE405 was used; vi)
no {\it a  priori} information on the solved for parameters was included 
in the fit; vii) Pioneer 11 was considered, 
viii) the Pioneer 10 data set used was extended to 14 Feb. 1998. 
Then the Doppler data was refitted.  

Beginning with this last point:   CHASMP uses the same original data file, but
it performs an additional data  compression.   This compression  combines  the
longest contiguous  data composed of adjacent data intervals or data spans with
duration  $\ge 600$ s (effectively it prefers 600 and 1980 second data
intervals).  It ignores short-time data points. Also,  Aerospace uses an
N-$\sigma$/fixed boundary rejection criteria that rejects all data in the fit
with a  residual greater  than $\pm 0.025$ Hz.  These rejection criteria
resulted in the loss of about 10 \% of the original data for both Pioneers 10
and 11.  In particular, the last five months of Pioneer 10 data, which was all
of data-lengths less than 600 s, was ignored.   Once these data
compression/cuts 
were made, CHASMP used 10,499 of its 11,610 data points for Pioneer 10  and
4,380 of its 5,137  data points for Pioneer 11. 

Because of the  spin-anomaly in the Pioneer 10 data, 
the data arc  was also divided into three time intervals (although the I/II 
boundary was taken as 31 August 1990 \cite{I/II}).  
In what was especially useful, 
the Aerospace analysis uses direct propagation of the trajectory data
and solves for the parameter of interest only for the data within a
particular data interval.  That means the three interval results were 
truly independent.  Pioneer 11 was fit as a single arc.  

Three types of runs are listed, with:  i) no corona;  
ii) with Cassini corona model of 
Sections \ref{corona+wt} and \ref{sec:corona}; and 
iii) with the Cassini corona model, but added are corona data weighting 
(Section  \ref{corona+wt}) and the time-variation called 
``F10.7''  \cite{F10-7}.  
(The number 10.7 labels the wavelength of solar radiation, 
$\lambda$=10.7 cm, that, in our analysis, is averaged over 81 days.)

The results are given in rows 4-6 of Table \ref{resulttable}.   The no
corona results (row 4) are in good   agreement with the 
\emph{Sigma} results of the first row.  This is especially true  
for the extended-time Interval III  values for Pioneer 10, which 
interval had clean data.   However
there is more disagreement with the values for Pioneer 10 in  Intervals I
and II and for Pioneer 11. These three data sets all were noisy and 
underwent more data-editing.  Therefore, it is significant that the
deviations between \emph{Sigma} 
and CHASMP in these arcs are all similar, but small, 
between $0.20$ to $0.25$ of our units.  
As before, the effect of the  solar corona is small, even with
the various model variations.   But most important, the   numbers
from \emph{Sigma} and CHASMP  for  Pioneer 10 Interval III 
are in excellent agreement.  

\begin{figure}[h]
\epsfig{figure=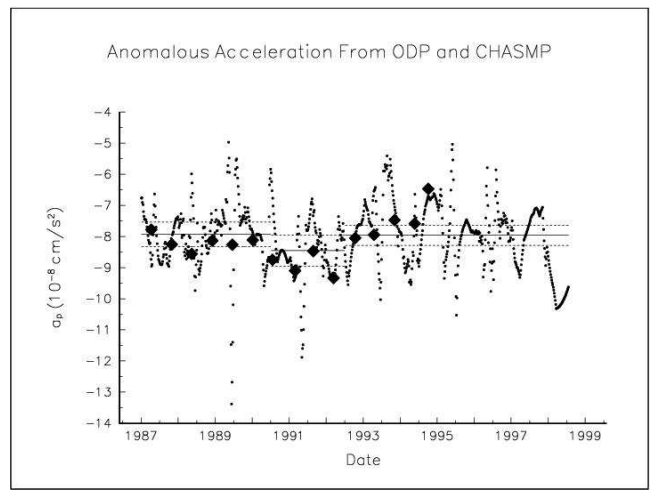,width=85mm}%,height=90mm}
    \caption
{Consistency of the ODP/{\em{Sigma}} and CHASMP time-variation signals. 
The dots show 5-day sample averages of the  anomalous acceleration of
Pioneer 10 from  ODP/{\em{Sigma}} using {\tt BSF} with a 200-day
correlation time. From  this data, the solid lines show the mean values
of $a_P$ in the three intervals corresponding to the three separate spin
down histories. The dashed lines represent the large batch-sequential 
computational error bounds on the three values of $a_P$.   The 200-day
acceleration values using CHASMP are the solid squares.  At the time
positions where there are CHASMP results, the agreement between the
CHASMP and the ODP/{\em{Sigma}} results  is clear. 
      \label{fig:rec_res_comb}}
\end{figure}

Further, CHASMP also found the annual term.  (Recall that CHASMP can also look
for a temporal variation by calculating short time averages.)
Results on the time variation in $a_P$ can be seen in Figure
\ref{fig:rec_res_comb}. Although there  could possibly be $a_P$ variations of
$\pm 2\times10^{-8}$ cm/s$^2$  on a 200-day time scale, a comparison of the
variations with the error  limits shown in Figure
\ref{fig:rec_res_comb} indicate that our measurements of these variations  are
not statistically significant.  The 5-day averages of $a_P$ from ODP (using the
batch-sequential method) are  not reliable at solar conjunction in the middle
(June) of each year, and  hence should be ignored there. The CHASMP 200-day
averages suppress the  solar conjunction bias inherent in the ODP 5-day
averages, and they  reliably indicate a constant value of $a_P$. Most
encouraging, these results clearly indicate that the obtained solution
is consistent, stable, and its mean value does not strongly depend  on the
estimation procedure used.  The presence of the small annual term on top 
of the obtained solution is apparent.

%************
%************

%*******************************************

\subsection{Our solution, before systematics, 
for the anomalous acceleration}
\label{final_sol}

\noindent From Table  \ref{resulttable}  we can 
intuitively draw a number of conclusions:  \\
\indent  A) The effect of the corona is small. 
This systematic will be analyzed in Section  \ref{solarwind}.  \\
\indent B) The numerical error is small.  This systematic will be analyzed 
in Section   \ref{leastsquares}. \\
\indent C) 
The differences between the \emph{Sigma} and CHASMP  Pioneer 10 
results for  Interval I and Interval II, respectively, 
we attribute to two main causes: especially  
i) the different data rejection techniques of the two analyses 
but also ii) the different maneuver simulations.
Both of these effects  
were especially significant in Interval II, where the data arc was 
small and  a large amount of noisy data was present. Also, to account for the
discontinuity in the spin data that occurred on 28 January 1992 (see Figure
\ref{fig:pioneer_spin}),  Aerospace introduced a  fictitious maneuver for
this interval.  Even so, the deviation in the two values of 
$a_P$ was relatively small, namely  $0.23$ and $0.21$, respectively, 
$\times 10^{-8}$ cm/s$^2$.    \\  
\indent D) The changes in $a_P$ in the different Intervals, correlated
with the changes in  spin-rate change, are likely (at least partially)
due to gas leakage.   This will be discussed in Section 
\ref{sec:gleaks}.  

But independent of the origin, this last correlation 
between shifts in $a_P$ and changes in spin rate actually allows us to 
calculate the best ``experimental'' base number for Pioneer 10.  
To do this, assume that the spin-rate change is directly contributing to
an anomalous acceleration offset.  Mathematically, this is saying
that in any interval
$i=\mathrm{I,II,III}$, for which the spin-rate change is 
an approximate constant, one has 
\begin{equation} a_{P}(\ddot{\theta}) = a_{P(0)} - \kappa~\ddot{\theta}, 
\label{gasleakeq}
\end{equation} 
where $\kappa$ is a constant with units of length  and $ a_{P(0)} \equiv
a_P(\ddot{\theta}=0)$ is the  Pioneer acceleration without any 
spin-rate change.  

One now can fit the data to Eq. (\ref{gasleakeq}) to obtain solutions for 
$\kappa$ and  $a_{P(0)}$.
The three intervals $i=\mathrm{I,II,III}$ provide three data combinations 
$\{a_{P(i)}(\ddot{\theta}),  \ddot{\theta}_i\}$. We take our base 
number, with which to reference systematics, to be the  weighted average 
of  the \emph{Sigma} and CHASMP results for $a_{P(0)}$
when no corona model was used. 
Start first with  the \emph{Sigma} Pioneer
10 solutions in row one of Table 
\ref{resulttable}  and the Pioneer 10 spin-down rates given in Section
\ref{jplresults}  and Figure \ref{fig:pioneer_spin}:
$a_{P(i)}^{\tt Sigma}=(8.02\pm 0.01,~ 8.65\pm 0.01,~7.83\pm 0.01)$ 
in units of $10^{-8}$ cm/s$^2$ and  
$\ddot{\theta}_{i}=-(0.0181\pm 0.0001,~0.0861\pm 0.0009,~0.0073\pm 0.0015)$ 
in units of rpm/yr, where   
\begin{eqnarray}
1~\mathrm{rpm/year}&=& 5.281\times10^{-10}~\mathrm{rev/s}^2 
\nonumber\\
        &=& 3.318\times10^{-9}~\mathrm{radians/s}^2.   
\end{eqnarray}

With 
these data we use the maximum likelihood and minimum variance approach 
to find the optimally weighted least-squares solution  for  
$a_{P(0)}$:    
% and $\kappa$: 
\begin{eqnarray} 
a^{\tt Sigma}_{P(0)}  &=& (7.82\pm 0.01) \times 10^{-8}~\mathrm{cm/s}^2,
\label{eq:sol_sigma_a} 
%\\ \kappa^{\tt Sigma}  &=& (29.2 \pm 0.7)~ \mathrm{cm}.  
%\label{eq:sol_sigma_k}  
\end{eqnarray} 
with solution for the parameter $\kappa$ obtained as $\kappa^{\tt Sigma} 
= (29.2 \pm 0.7)~ \mathrm{cm}$.  Similarly, for CHASMP one takes the
values for $a_P$ from  row four of Table \ref{resulttable}:
$a^{\tt CHASMP}_{P(i)}=(8.25\pm0.02,~ 8.86\pm0.02,~7.85\pm0.01)$ and uses 
them with the same $\ddot{\theta}_{i}$ as above.  The solution for 
$a_{P(0)}$ in this case is
\begin{eqnarray}
a^{\tt CHASMP}_{P(0)} & = &(7.89\pm 0.02) \times
10^{-8}~\mathrm{cm/s}^2   \label{eq:sol_chasmp},
%\\\kappa^{\tt CHASMP}  &= &(32.1 \pm 1.0)~ \mathrm{cm}.
\end{eqnarray}   
together with $\kappa^{\tt CHASMP} = (34.7 \pm 1.1)~ \mathrm{cm}$. The 
solutions for \emph{Sigma} and CHASMP are similar, 7.82 and 7.89 in our
units.   We take the weighted average of these two to yield our base line
``experimental'' number for $a_P$: 
\begin{eqnarray}
a_{P({\tt exper)}}^{\tt Pio10} &=& (7.84\pm
0.01)~\times~10^{-8}~\mathrm{cm/s}^2.   
\label{pio10lastresult}
\end{eqnarray}   
[The weighted average constant $\kappa$ is 
$\kappa_0  =(30.7\pm 0.6)$ cm.] 

For Pioneer 11, we only have the one  3$\frac{3}{4}$ year
data arc.  The weighted average of the two programs' 
no corona results is $(8.62\pm 0.02) \times 10^{-8}$  cm/s$^2$.  
We observed in Section \ref{spinhistory} that between maneuvers 
(which are accounted for - see Section \ref{model-maneuvers})
there is actually a spin rate {\it increase} of 
$\sim(+0.0073 \pm   0.0003)$   rpm/yr.  If one uses this spin-up rate 
and the Pioneer 10 value for  $\kappa_0=30.7$ cm given above, one 
obtains a spin-rate change corrected value for $a_P$.   We take this 
as the  experimental value for Pioneer 11:
\begin{equation} 
a_{P({\tt exper)}}^{\tt Pio 11}= (8.55\pm 0.02) 
\times 10^{-8} ~{\rm cm/s}^2.
\label{pio11lastresult}
\end{equation}

%***********7) SYSTEMATICS EXTERNAL**************
%\newpage

\section{\label{ext-systema}SOURCES OF SYSTEMATIC ERROR 
EXTERNAL TO THE SPACECRAFT}

We are concerned with possible systematic acceleration errors that could 
account for the unexplained anomalous acceleration directed toward the
Sun. There exist detailed publications describing analytic  recipes
developed to account for non-gravitational accelerations acting on
spacecraft.  (For a summary see Milani et al.
\cite{milani}.) With regard to the specific Pioneer spacecraft, possible
sources of systematic acceleration have been discussed before for Pioneer
10 and 11 at Jupiter \cite{null76} and Pioneer 11 at Saturn
\cite{null81}. 

External forces can produce three vector components of spacecraft
acceleration, unlike forces generated on board the spacecraft, where the
two non-radial components (i.e., those that are effectively perpendicular
to the spacecraft spin) are canceled out by spacecraft rotation. However, 
non-radial spacecraft accelerations are difficult to observe by the
Doppler technique, which measures spacecraft velocity along the
Earth-spacecraft line of sight.  But with several years of Doppler data,
it is  in principle possible to detect systematic non-radial acceleration
components \cite{sunearth}. 

With our present analysis  \cite{sunearth} we find that the Doppler data
yields only one significant component of unmodeled acceleration, and that
any acceleration components perpendicular to the spin axis are small. 
This is because in the fitting we tried including   three unmodeled
acceleration constants along the three spacecraft axes (spin axis and two
orthogonal axes perpendicular to the spin axis). The components
perpendicular  to the spin axis had values consistent with zero to a
1-$\sigma$ accuracy of 2 $\times$ 10$^{-8}$ cm/s$^{2}$ and  the radial
component was equal to the reported anomalous acceleration. Further, the
radial acceleration was not correlated with the other two unmodeled
acceleration components. 

Although one could in principle set up complicated engineering models to
predict all or each of the systematics, often the uncertainty of the models 
is too large to make them useful, despite the significant effort
required.  A different approach is to accept our ignorance about a 
non-gravitational acceleration and assess to what extent these can be
assumed a constant bias over the time scale of all or part of the mission. 
(In fact, a constant
acceleration produces a linear frequency drift that can be  accounted for in
the data analysis by a single unknown parameter.)  In fact, we will use
both approaches. 

In  most orbit determination programs some effects, like the
solar radiation pressure, are  included in the set of  routinely 
estimated parameters.  Nevertheless we  want to demonstrate their
influence on Pioneer's navigation from the general physics standpoint. 
This is not only to validate our results,  but also to be a model as to
how to study  the influence of the other physical phenomena that are not
yet included in the standard navigational packages  for future more
demanding missions.  Such missions will involve either spacecraft that
will be distant or spacecraft  at shorter distances where high-precision
spacecraft navigation will be required. 

In this section we will discuss possible systematics (including forces) 
generated  external to the spacecraft which might significantly affect our
results.  These start with true forces due to  (1) solar-radiation
pressure  and (2) solar wind pressure.  We go on to discuss (3) the
effect of the solar corona and its mismodeling,  (4) electro-magnetic
Lorentz forces,  (5) the influence of the Kuiper belt, (6) the phase
stability of the reference atomic clocks, and  (7) the mechanical and
phase stability of the DSN antennae, together with influence of the
station locations and troposphere and ionosphere contributions. 

%**************************

\subsection{Direct solar radiation pressure and mass}
\label{solarP}

There is an exchange of momentum when solar photons impact the
spacecraft and are either absorbed or reflected.  Models for 
this solar pressure effect were 
developed before either Pioneer 10 or 11 were launched
\cite{rad} and have been refined since then. The models take into account
various parts of the spacecraft exposed to solar radiation, primarily the
high-gain antenna.  It computes an acceleration directed away from the Sun
as a function of spacecraft orientation and solar distance.  

The models for the acceleration due to solar radiation can be formulated as  
\begin{equation}
a_{\tt s.p.}(r)=\frac{\mathcal{K} f_\odot A }{c~M}
\frac{ \cos\theta(r)}{r^2}.
 \label{eq:srp}
\end{equation}
$f_\odot=1367 ~{\rm W/m}^{2}$(AU)$^2$ is the 
(effective-temperature Stefan-Boltzmann) 
``solar   radiation
constant'' at 1 AU from the Sun and $A$ is the effective  size of the
craft as seen by the Sun \cite{solar_irr}. (For Pioneer the area was
taken to be the antenna dish of radius 1.73 m.)  $\theta$ is  the angle
between the axis of the antenna and the direction of the Sun, 
$c$ is the speed of light, $M$ is the mass of the spacecraft (taken to
be 251.883 for Pioneer 10), and $r$ is the distance from  the Sun to the
spacecraft in AU. $\mathcal{K}$ \cite{Lambda}  is the {\it effective}
\cite{effect} absorption/reflection coefficient.   For Pioneer 10 the
simplest approximately correct model yields $\mathcal{K}_{0}=1.71$  
\cite{effect}. 
Eq. (\ref{eq:srp}) provides a good model for analysis of the effect  of
solar radiation  pressure on the motion of distant spacecraft  and is
accounted for by most of the programs used for orbit determination. 

However, in reality the absorptivities, emissivities, and effective 
areas  of spacecraft parts parameters which, although modeled by design, 
are determined by calibration early in the mission \cite{sunparam}.  One
determines the magnitude of the solar-pressure acceleration at various
orientations using Doppler data. (The solar pressure effect can be
distinguished from gravity's $1/r^2$ law because $\cos\theta$ varies
\cite{massprog}.)  The complicated set of program input parameters that
yield the parameters  in Eq. (\ref{eq:srp})  are then set for later use
\cite{sunparam}.  Such a determination of the parameters for Pioneer 10
was done, soon after launch and later.  
When applied to the solar radiation acceleration in the  
region of Jupiter, this yields (from a 5 \%
uncertainty in $a_{\tt s.p.}$ \cite{null76}) 
\begin{eqnarray}
a_{\tt s.p.}(r={\tt 5.2 AU})&=& (70.0 \pm 3.5) 
\times 10^{-8}~{\rm cm/s}^2, \nonumber\\
 \mathcal{K}_{\tt 5.2} &=& 1.77. 
\label{aspS}
\end{eqnarray}
The second of Eqs. (\ref{aspS}) comes from putting the first into Eq. 
(\ref{eq:srp}).  Note, specifically, that in a fit a too high input mass 
will be compensated for by a higher effective $\mathcal{K}$.

Because of the $1/r^2$ law, by the time the craft reached 10 AU the solar
radiation acceleration was $18.9\times 10^{-8}$ cm/s$^2$ 
going down to 0.39 of those units by 70 AU.  
Since this systematic falls off as
$r^{-2}$, it can bias the Doppler determination of a constant
acceleration at some level, even though most of the systematic is 
correctly modeled by the program itself.  
By taking the average of the $r^{-2}$
acceleration curves over the Pioneer distance intervals, we estimate that
the systematic error from solar-radiation pressure in units of 10$^{-8}$
${\rm cm/s}^2$ is 0.001 for Pioneer 10 over an interval from 40 to 70
AU, and 0.006 for Pioneer 11 over an interval from 22 to 32 AU. 

However, this small uncertainty is not our main problem.  In actuality, 
since the parameters were fit 
the mass has decreased with the consumption of propellant.
Effectively, the $1/r^2$ systematic has changed its normalization with time.  
If not corrected for, the difference between the original  $1/r^2$ and the
corrected  $1/r^2$  will be interpreted as a bias in $a_P$.  
Unfortunately, exact information on gas usage is unavailable \cite{gasuse}.   
Therefore, in dealing with the effect of the temporal mass variation
during the entire data span (i.e. nominal input mass vs. actual mass
history \cite{mass,gasuse})  we have to address two effects on 
the solutions for the anomalous acceleration $a_P$.
They are  i) the effect of mass variation from gas consumption  
and ii) the effect of an incorrect input mass \cite{mass,gasuse}.

To resolve the issue of mass variation uncertainty we performed a
sensitivity analysis of our solutions to  different spacecraft input
masses.   We simply re-did the no-corona, WLS runs of Table
\ref{resulttable} with a range of different masses. 
The initial wet weight of the package was 259 kg with
about 36 kg of consumable propellant.  For Pioneer 10, the input mass 
in the program fit was 251.883 kg, roughly corresponding to the 
mass after spin-down. 
By our data period, roughly  half the fuel (18 kg) was gone so we take 
241 kg as our nominal Pioneer 10 mass.  Thus,  the effect of 
going from 251.883 kg to 241 kg we take to be  
our bias correction for Pioneer 10.  We take the uncertainty to be
given by one half the effect of going from plus to minus  9 kg 
(plus or minus a quarter tank) from the nominal mass of 241 kg.  

For the three intervals of Pioneer 10 data, using ODP/\emph{Sigma} 
yields the following changes in the accelerations:  
\begin{eqnarray}
\delta a^{\tt mass }_P &=& [(0.040 \pm 0.035),~(0.029 \pm 0.025), 
        ~~~~~~~~~~~~~~~~\nonumber \\
    &~& ~~~~~~~~~~~~~~~   
(0.020 \pm 0.017)]~\times  10^{-8}~  \mathrm{cm/s}^2.  
               \nonumber
\end{eqnarray} 
As expected,these results make $a_P$ larger. 
For our systematic bias we take the weighted
average of $\delta a^{\tt mass }_P$ for the three intervals of Pioneer
10.     The end result is  
\begin{equation} 
a_{\tt s.p.}= (0.03~ \pm~ 0.01) \times 10^{-8}~ \mathrm{cm/s}^2.
\end{equation}

For Pioneer 11 we did the same except our bias point was 3/4 of the fuel gone
(232 kg).  Therefore the  bias results by going from  the input mass of
239.73 to 232 kg.  The uncertainty is  again defined by $\pm$ 9 kg.   
The result for Pioneer 11 is  more sensitive to 
mass changes, and we find using ODP/\emph{Sigma}
\begin{equation} 
a_{\tt s.p.}= (0.09~ \pm~ 0.21) \times 10^{-8}~ \mathrm{cm/s}^2.
\end{equation}
The bias  number is three times larger than the  similar number 
for Pioneer 10, and the uncertainty much larger.   
We return to this difference in Section \ref{twospace}.

The previous analysis also allowed us to perform consistency checks on the
effective values of $\mathcal{K}$ which the programs were using. By taking 
$[r_{\mathrm{min}}r_{\mathrm{max}}]^{-1}= [\int(dr/r^2)/\int dr]$ for the
inverse distance squared of a data set, 
varying the masses, and determining the shifts in $a_P$ we could
determine the values of $\mathcal{K}$ implied,  We found: 
%\begin{eqnarray}
$\mathcal{K}_{\tt Pio-10(I)}^{\tt ODP} \approx 1.72$;  
%&~~~~~&
$\mathcal{K}_{\tt Pio-11}^{\tt ODP} \approx 1.82$;  
% \\
$\mathcal{K}_{\tt Pio-10(I)}^{\tt CHASMP} \approx 1.74$; and  
%&~~~~~&
$\hat{\mathcal{K}}_{\tt Pio-11)}^{\tt CHASMP} \approx 1.84$.
%  \label{Lequa}
%\end{eqnarray}
[The hat over the last $\mathcal{K}$ indicates it was multiplied by 
(237.73/251.883) because CHASMP uses 259.883 kg  instead of 239.73 kg 
for  the input mass.]  All these values of $\mathcal{K}$ are in the 
region expected and are clustered around the value $\mathcal{K}_{\tt 5.2}$
in  Eq. (\ref{aspS}). 

Finally, if you take the average values of $\mathcal{K}$ for Pioneers 10
and 11 (1.73, 1.83), multiply these numbers by the input masses
(251.883, 239.73) kg, and divide them by our nominal masses (241, 232)
kg, you obtain (1.87, 1.89), indicating our choice of
nominal masses was well motivated.

%************************************************************

%*********************

\subsection{The solar wind}
\label{solarwind}
  
The acceleration caused by the solar wind has the same form as Eq. 
(\ref{eq:srp}), with $f_\odot$ replaced by $m_pv^3n$, where $n
\approx5$  cm$^{-3}$ is the proton density at 1 AU and $v\approx400$ km/s is
the speed of the wind. Thus,   
\begin{eqnarray} 
\sigma_{\tt s.w.}(r)&=&\mathcal{K}_{\tt s.w.}\frac{m_pv^3\,n\,
A\cos\theta}{cM \,r^2}\nonumber\\
&\approx& 1.24\times10^{-13}
\left(\frac{20 ~\rm AU}{r}\right)^2~{\rm cm/s}^2.
\label{eq:sw}
\end{eqnarray}
\noindent Because the density can change by as much as 100\%, the exact 
acceleration is unpredictable. But there are measurements \cite{solar_irr}
showing  that it is about 10$^{-5}$ times smaller than
the direct solar radiation pressure. Even if we 
make the very conservative
assumption that the solar wind contributes only 100 times less force than 
the solar radiation, its smaller contribution is completely negligible.    

%*****************

\subsection{The effects of the solar corona and models of it}
\label{sec:corona}

As we saw in the previous Section \ref{solarwind}, the effect of the solar
wind pressure is negligible for distant spacecraft motion in the solar
system. However, the solar corona effect on  propagation of
radio waves between the Earth and the spacecraft needs to be analyzed in
more detail. 

Initially, to study the
sensitivity of $a_P$ to the solar corona model,  we were also solving for 
the solar corona parameters $A$, $B$, and $C$ 
of Eq. (\ref{corona_model_content}) in addition to $a_P$.  
However, we realized that the 
Pioneer Doppler data is not precise enough to produce credible results 
for these physical parameters.  
In particular, we found that solutions could yield a value of  $a_P$ which 
was changed by of order 10 \%  even though it gave unphysical values of 
the parameters (especially $B$, which previously had been poorly 
defined even by the Ulysses mission \cite{bird}).  
[By ``unphysical'' we mean electron densities that were either 
negative or positive with values that are vastly different from 
what would be expected.] 

Therefore, as noted in Section \ref{corona+wt},
we decided to use the newly obtained values for $A$, $B$, and  $C$ 
from the Cassini mission and use them as inputs for our analyses: 
$A= 6.0\times 10^3, B= 2.0\times 10^4, C= 0.6\times 10^6$, all in meters
\cite{Ekelund}.   This is the ``Cassini corona model.''  

The effect of the solar corona is expected to be small for Doppler and
large for range.  Indeed it is small for \emph{Sigma}.  
For ODP/\emph{Sigma}, the time-averaged  effect of the corona was 
small,  of order 
\begin{equation}
\sigma_{\tt corona} = \pm 0.02~ \times~ 10^{-8}~ \mathrm{cm/s}^2, 
\end{equation} as might be expected. We take this number to be the error 
due to the corona.  

What about the results from CHASMP.  
Both analyses use the same physical
model for the effect of the steady-state solar corona on radio-wave
propagation through the solar plasma  (that is given by
Eq. (\ref{corona_model})).  However, there is a slight difference in the 
actual implementation of the model in the two codes. 

ODP calculates the corona effect only when the Sun-spacecraft separation angle
as seen from the Earth  (or Sun-Earth-spacecraft angle) is less then $\pi/2$. 
It sets the corona contribution to zero in all other cases.   Earlier CHASMP
used the same model and got a small corona effect. Presently CHASMP  calculates
an approximate  corona contribution for all the trajectory.  Specific attention
is given to the region when the spacecraft is at   opposition  from the Sun and
the Sun-Earth-spacecraft angle $\sim \pi$.   
There CHASMP's implementation truncates the  code approximation to
the scaling factor $F$ in Eq. (\ref{corona_model}).  This is   specifically
done to remove the fictitious divergence in the region where ``impact
parameter'' is small,  $\rho \rightarrow 0$. 

However, both this and also the more complicated corona models (with
data-weighting and/or 
``F10.7'' time variation) used by CHASMP produce small deviations from
the no-corona results.  
Our decision was to  incorporate these small  deviations between the 
two results due to corona modeling into our overall error budget as  
a separate item:
\begin{equation}
\sigma_{\tt corona\_model} 
= \pm 0.02~\times~ 10^{-8}   ~~{\rm cm/s}^2.
\end{equation}
This number could be discussed  in Section   \ref{Int_accuracy}, 
on computational systematics.    
Indeed,  that is where it will be listed in our error budget.

%*****************

\subsection{Electro-magnetic Lorentz forces} 

The possibility that the spacecraft could hold a charge, and be deflected
in its trajectory by Lorentz forces, was a concern for the magnetic field
strengths at Jupiter and Saturn.  However, the magnetic field strength in
the outer solar system is on the  order of 
$<1~\gamma~(\gamma=10^{-5}$ Gauss).  This is about a factor of $10^5$ times
smaller than the magnetic field strengths measured by the  Pioneers   at
their nearest approaches to Jupiter:  0.185 Gauss for Pioneer 10  and 1.135
Gauss for the closer in Pioneer 11 \cite{edsmith}.

Also, there is an upper limit to the charge that a spacecraft can hold.  
For the Pioneers that limit produced an upper bound on the Lorentz
acceleration at closest approach to Jupiter of $20 \times 10^{-8}$
cm/s$^{2}$ \cite{null76}. With the interplanetary field being so much
lower than at Jupiter,  we conclude that the electro-magnetic force on the
Pioneer spacecraft in the outer solar system is at worst on the order of
$10^{-12}$ cm/s$^{2}$, completely negligible \cite{lorentz}. 

Similarly, the magnetic torques acting on the spacecraft were about a
factor of $10^{-5}$ times smaller than 
those acting on  Earth satellites, where they are  a concern. 
Therefore,  for the Pioneers any observed changes in spacecraft
spin cannot be caused by magnetic torques. 
 
%*****************

\subsection{The  Kuiper  belt's gravity}
\label{sec:kuiper}

\noindent From the study of the resonance effect of Neptune upon Pluto, 
two primary mass concentration resonances of 3:2 and 2:1 were discovered  
\cite{malhotra},
corresponding to 39.4 AU and 47.8 AU, respectively.    Previously,  Boss
and Peale had derived a model for a non-uniform density  distribution in
the form of an infinitesimally thin disc extending from 30 AU to 100 AU in
the ecliptic plane
\cite{liupeale}. We combined the results of Refs. \cite{malhotra}   and  
\cite{liupeale} to determine if the matter in the Kuiper 
belt could be the source of the anomalous 
acceleration of Pioneer 10 \cite{liudust}.

\begin{figure}[h]
\epsfig{figure=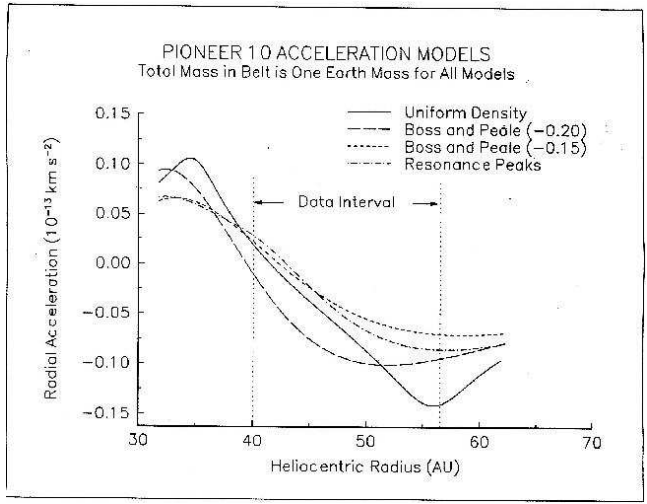,width=85mm}%,angle=270}
\caption{Possible acceleration caused by dust in the 
               Kuiper belt.
            \label{fig:pioneer_kb}}
\end{figure}

We specifically studied three distributions, namely: 
i) a uniform distribution, 
ii) a 2:1 resonance distribution with a peak at 47.8 AU, and 
iii) a 3:2 resonance distribution with a peak at 39.4 AU.  
Figure \ref{fig:pioneer_kb} exhibits the resulting acceleration
felt by Pioneer 10, from 30 to 65 AU which encompassed our data set
at the time.

%************
%************

We assumed a total mass of one Earth mass, 
which is significantly larger than standard estimates.  Even so,  
the accelerations are only on the order of $10^{-9}$ cm/s$^2$, 
which is two orders of magnitude smaller than the observed effect.  
(See Figure \ref{fig:pioneer_kb}.) 
Further, the accelerations are not constant across the data range.  
Rather, they show an increasing effect as Pioneer 10 approaches the belt
and a decreasing effect as Pioneer 10 recedes from the belt, even with a
uniform density model.  
For these two reasons, we excluded the dust belt as a source for the
Pioneer effect.   

More recent infrared observations have 
ruled out more than 0.3 Earth mass of  Kuiper Belt dust in the 
trans-Neptunian region \cite{backman,teplitzinfra}.  Therefore, we can 
now place a limit of  $\pm 3 \times 10^{-10}$ cm/s$^2$ for the 
contribution of the Kuiper belt.

Finally, we note that searches  for gravitational encounters 
of Pioneer with large Kuiper-belt  objects have so far not been successful
\cite{gio}.  

%***************

\subsection{Phase and frequency stability of clocks} 
\label{sec:clocs}

After traversing the mechanical components of
the antenna, the radio  signal enters the DSN antenna feed and passes
through a series of amplifiers,  filters, and cables.  Averaged over many
experiments, the net effect of this on the  calculated dynamical
parameters  of a  spacecraft should be very small. We
expect instrumental calibration instabilities to contribute
$0.2\times10^{-8} $ cm/s$^2$ to the anomalous acceleration on a
60 s time interval.  Thus, in order for the atomic clocks 
\cite{vessot_clocks} to have caused 
the Pioneer effect, all the atomic clocks used for signal referencing clocks
would have  had to have drifted in the same manner as the local DSN
clocks. 

In Section \ref{results} we observed that without using the apparent 
anomalous acceleration,  the CHASMP residuals show a steady
frequency  drift \cite{drift} of about $-6 \times 10^{-9}$ Hz/s, or 1.5 Hz
over 8 years  (one-way only).   This equates to a  clock acceleration,
$-a_t$, of $-2.8\times 10^{-18}$ s/s$^{2}$. 
(See Eq. (\ref{asubt}) and Figure \ref{fig:aerospace}.)
To verify that it is actually not  the clocks that are drifting,  
we analyzed the calibration of the frequency standards  used in the DSN
complex. 

The calibration system itself is referenced to  Hydrogen maser atomic
clocks. Instabilities in these clocks are another source of instrumental
error which needs to be addressed. 
The local reference is synchronized to the frequency standards generated
either at the National Institute of Standards and Technology (NIST),
located in Boulder, Colorado or at the U. S. Naval Observatory
(USNO), Washington, DC. These standards are presently distributed to local
stations by the Global Positioning System (GPS) satellites.  [During the
pre-GPS era, the station clocks used  signals from  WWV  to set the Cesium
or Hydrogen masers.  WWV, the radio station which broadcasts time and
frequency services, is located in Fort Collins, CO.] While on  a
track, the station is ``free-running,'' i.e., the  frequency and timing data
are generated locally at the station.  
The Allan variances are about $10^{-13}$ for Cesium and $10^{-15}$ for 
Hydrogen masers. Therefore, over the data-pass time interval,  the data
accuracy is on the order of one part in 1000 GHz or better. 

Long-term frequency stability tests are conducted with the 
exciter/transmitter subsystems and the DSN's radio-science open-loop
subsystem. An uplink signal generated by the exciter is translated at the
antenna by a test translator to a downlink frequency.  (See Section
\ref{Exp_tech}.) The downlink signal is then passed through the RF-IF
downconverter present at the antenna and into the radio science receiver
chain \cite{dsn}.    This technique allows  the processes to be
synchronized  in the DSN complex based on the frequency standards whose
Allan variances are of the order of  
$\sigma_y \sim 10^{-14}-10^{-15}$ for integration time in the range from 10 s
to 10$^3$ s. For the S-band frequencies of the Pioneers, 
the corresponding Allan variances are
1.3 $\times$ 10$^{-12}$ and 1.0 $\times$ 10$^{-12}$, 
respectively, for a 10$^3$ s Doppler integration time. 

Phase-stability testing characterizes stability over very short integration
times; that is, spurious signals whose frequencies  are very close to the
carrier (frequency). The phase noise region is defined to be frequencies within
100 kHz of the carrier.  Both amplitude and phase variations appear as phase
noise.  Phase noise is quoted in dB relative to the carrier,  in a 1 Hz band at
a specified deviation from the carrier; for example,  dBc-Hz at 10 Hz.  Thus,
for the frequency  1 Hz, the noise level is at $-51$ dBc and    10 Hz
corresponds to $-60$ dBc. This was not significant for our study.  

Finally, the influence of the clock stability on the   detected acceleration,
$a_P$, may be estimated based on the reported Allan variances for the clocks,
$\sigma_y$.  Thus, the standard `single measurement' error on acceleration as
derived by the time derivative of the Doppler frequency data is $(c
\sigma_y)/\tau$, where    the Allan variance, $\sigma_y$, is calculated for
1000 s Doppler integration time, and $\tau$ is the signal averaging time. This
formula provides a good rule of thumb when the Doppler power spectral density
function obeys a $1/f$ flicker-noise law, which is approximately the case when
plasma noise dominates the Doppler error budget. Assume a worst case
scenario, where only one clock was used for the whole 11 years study.   
(In reality
each DSN station has its own atomic clock.) To estimate 
the influence
of that one clock on the reported accuracy of the detected anomaly $a_P$, 
combine  $\sigma_y={\Delta\nu}/{\nu_0}$, the  fractional Doppler frequency 
shift  from the reference frequency of
$\nu_0=\sim 2.29$ GHz, with  the estimate for the Allan
variance,     
$\sigma_y =1.3 \times 10^{-12}$.  This yields
a number that characterizes the upper limit for a  frequency uncertainty
introduced in a single measurement by the instabilities in the atomic clock:  
$\sigma_\nu=\nu_0\sigma_y=2.98\times10^{-3}$ Hz for a 10$^3$ Doppler
integration time. 

In order to derive an estimate for the total effect, recall that the 
Doppler observation technique is essentially a continuous count of the
total number of complete frequency circles during observational time.
Within a year one can have as many as
$N\approx3.156\times10^3$ independent single measurements of the clock
with duration $10^3$ seconds.  This yields an upper limit for the 
contribution of atomic clock instability on the 
frequency drift of 
$\sigma_{\tt clock} = {\sigma_\nu}/{\sqrt{N}} \approx 
5.3\times 10^{-5}$ Hz/year.
But in Section \ref{subsec:aero} we noted that the observed $a_P$ corresponds
to a frequency drift of about 0.2 Hz/year, so the error in $a_P$ is about 
$0.0003 \times 10^{-8}$ cm/s$^2$.   Since all data is not
integrated over 1,000 seconds and  is data is not available for  all time,
we increase the numerical factor to $0.001$, which is still  negligible to
us.   [But further, this upper limit
for the error becomes even smaller if one accounts for the number of
DSN stations and corresponding atomic clocks that were used
for the study.]   

Therefore, we conclude that the clocks are not a contributing
factor to the anomalous acceleration at a meaningfully level.
We will return to this issue in  Section
\ref{sec:timemodel} where we will discuss a number of phenomenological time
models that were used to fit the data.

%************************

\subsection{DSN antennae complex}
\label{sec:dsn_complex}

The mechanical structures which support the reflecting surfaces of the
antenna  are not perfectly stable.  Among the numerous effects influencing
the DSN antennae performance,  we are only interested in those whose
behavior might contribute to  the estimated solutions for $a_P$.  The
largest systematic instability over a long period is   due to gravity loads
and the aging of the structure. As discussed in   
\cite{SoversJacobs96}, antenna deformations due to gravity loads should be
absorbed almost entirely  into biases of the estimated station locations
and clock offsets.  Therefore, they will have little effect on  the
derived solutions for the purposes of spacecraft navigation.
 
One can also consider ocean loading, wind loading, thermal expansion, and
aging of the structure. We found none of these can produce the constant
drift in the Doppler frequency on a time scale comparable to the Pioneer
data. Also,  routine tests are performed by  DSN personnel on a
regular basis to access all the effects that may contribute to the
overall   performance of the DSN complex. 
The information is available and it shows all parameters 
are in the required ranges.
Detailed assessments of all these effect on the astrometric VLBI
solutions were published in \cite{SFJ98,SoversJacobs96}. The results for the
astrometric errors introduced by the above factors may be directly
translated to the error budget for the Pioneers, scaled by the number of
years.  It yields  a negligible contribution.

Our analyses also estimated  errors  introduced by a number of 
station-specific parameters.  These include  
the error due to imperfect knowledge in
a DSN station location,  errors due to troposphere and
ionosphere models for different stations, 
and errors due to the Faraday
rotation effects in the Earth's atmosphere.    
Our analysis indicates that at
most these effects would  produce a distance- and/or 
time-dependent drifts
that would be easily noticeable in the radio Doppler data. What is more
important is that none of the effects would be able to produce a constant
drift in the Doppler residuals of Pioneers over such a long time scale. 
The updated version of the ODP, {\it Sigma},  routinely accounts for
these error factors.  Thus, we run covariance analysis for the whole set of
these parameters using both {\it Sigma} and CHASMP.   Based on these
studies we   conclude that mechanical and phase stability of the DSN
antennae together with geographical locations of the antennae, geophysical 
and atmospheric conditions on the antennae site  have negligible effects on
our solutions for $a_P$. At most their contributions are  at the level
of $\sigma_{\tt DSN}\leq10^{-5}a_P$.

%*******************8) INTERNAL SYSTEMATICS
%\newpage

\section{\label{int-systema}SOURCES OF SYSTEMATIC  ERROR INTERNAL TO
THE SPACECRAFT}

In this section we will discuss the forces that   may be 
generated by spacecraft systems. The mechanisms we consider that may
contribute to the found constant acceleration, $a_P$, and that may be
caused by the on-board mechanisms include: (1) the radio beam reaction 
force, (2) RTG heat reflecting off the spacecraft, (3) differential
emissivity of the RTGs, (4) non-isotropic radiative cooling of the
spacecraft, (5) expelled Helium  produced  within the RTG, (6)
thruster gas leakage, and (7) the difference in experimental results 
from the two spacecraft.

%***********************************

\subsection{Radio beam reaction force}
\label{radioantbeam}

The Pioneer navigation does not require that the  spacecraft   constantly
beam its radio signal, but instead it does so only when it is requested to
do so from the ground control. Nevertheless, the recoil force due to the
emitted radio-power must also be analyzed. 

The Pioneers have a total nominal emitted radio power of eight Watts.  
It is parameterized as 
\begin{equation}
P_{\tt rp}~=\int_0^{\theta_{\tt max}} d\theta~ \sin\theta~ {\cal
P}(\theta),
\end{equation}
${\cal P}(\theta)$ being the antenna power distribution.
The radiated power has been kept constant in time, independent of  the
coverage from ground stations.  That is, the  radio transmitter is always
on, even when not received by a ground station.  

The recoil from this emitted  radiation produces an acceleration  
bias, $b_{\tt rp}$, on the spacecraft away from the Earth of
\begin{equation}
b_{\tt rp}= \frac{\beta \,P_{\tt rp}}{Mc}. 
\label{eq:rp}
\end{equation}
$M$ is taken to be the Pioneer mass when half the fuel is gone
\cite{mass}. $\beta$ is the fractional component of the radiation
momentum  that is going in a direction opposite to $a_P$:
\begin{equation}
\beta =\frac{1}{P_{\tt rp}}
 {\int_0^{\theta_{\tt max}} d\theta~ \sin\theta~\cos\theta~
       {\cal P}(\theta)}.
\label{radiopower}
\end{equation}

Ref \cite{piodoc} describes the HGA
and shows its downlink antenna pattern in Fig. 3.6-13. 
(Thermal antenna expansion  mismodeling is thought to be negligible.)
The gain is given as $(33.3 \pm 0.4)$ dB at zero (peak) degrees. 
The intensity  is down by a factor of two ($-3$ dB) at 1.8 degrees. 
It is down a factor of 10 ($-10$ dB) at 2.7 degrees
and down by a factor of 100 ($-20$ dB) at 3.75 degrees.  
[The first diffraction minimum is at a little over four degrees.]
Therefore, the pattern is a very good conical beam.  
Further, since $\cos [3.75^\circ] = 0.9978$, we can  take 
$\beta = (0.99 \pm 0.01)$, yielding $b_{\tt rp}=1.10$. 
 
Finally, taking the error for  the nominal 8 Watts power to be given by 
the 0.4 dB antenna error ($0.10$) and the 
error due to the uncertainty in our nominal mass ($0.04$),  we arrive at the
result  
\begin{equation}
a_{\tt rp} =b_{\tt rp} \pm \sigma_{\tt rp}
=(1.10 \pm 0.11)\times 10^{-8} ~{\rm cm/s}^2.
\end{equation}

%******** katz ***********

\subsection{RTG heat reflecting off the spacecraft}
\label{subsec:katz} 

It has been argued that the anomalous acceleration seen in the  Pioneer
spacecraft is due to  anisotropic heat reflection off of the back   of the
spacecraft high-gain antennae, the heat coming from the RTGs \cite{katz}.
Before launch, the four RTGs had a
total thermal fuel inventory of 2580 W (now $\sim$ 2070 W). They produced 
a total electrical power  of 160 W (now $\sim$ 65 W).  
Presently  $\sim 2000$ W of RTG heat must be dissipated.   
Only $\sim63$ W  of directed power could explain the  anomaly. 
Therefore, in principle there is enough  power to explain the
anomaly this way.  However,  there are two reasons that preclude such a
mechanism, namely:

i) {\sf The spacecraft geometry:}  The RTGs are located at the end of
booms, and  rotate about the spacecraft in a plane that contains the
approximate base of  the antenna.  
From the closest axial center point of the RTGs,  
the antenna is seen nearly ``edge on'' (the longitudinal angular width is
24.5$^o$).  The total solid angle subtended is  $\sim$ 1-2\%  of $4\pi$
steradians \cite{s}.   
Even though a more detailed calculation yields a value of 
1.5\% \cite{ss}, even taking the higher bound of 2\% means  this 
proposal could provide at most $\sim 40$ W.  But there is more 
\cite{heatreflect}. 

ii) {\sf The RTGs' radiation pattern:}   The above estimate was based on
the assumption that   the RTGs are spherical black bodies.  But they  are
not.   The main bodies of the RTGs are cylinders and they  are grouped in
two packages of two.  Each  package has the two cylinders end to end
extending away from  the antenna.  Every RTG has six fins 
separated by equal angles of 60 degrees 
that go radially out from the cylinder.   
Presumably this results in a symmetrical 
radiation of thermal power into space. 

Thus, the fins are ``edge on'' to the antenna (the
fins point  perpendicular to the cylinder axes).  
The largest opening angle of the fins is seen only by the narrow-angle
parts of the antenna's outer edges.   
Ignoring these edge  effects, 
only $\sim$2.5\% of the surface area of the RTGs is facing the antenna.  
This is a factor 10 less than that from integrating
the directional intensity from a hemisphere: 
$[(\int^{\tt h.sph.}d\Omega\cos\theta)/(4\pi)]=1/4$.  
So, one has only 4 W of directed power. 
This suggests a systematic bias of  
$\sim0.55 \times 10^{-8}$ cm/s$^2$. Even adding an uncertainty of 
the same size yields a systematic for heat reflection of 
\begin{equation}
a_{\tt h.r.}= (-0.55 \pm 0.55) \times 10^{-8}~\mathrm{ cm/s}^2. 
\end{equation}

But there are reasons to consider this an upper bound.  The Pioneer SNAP
19 RTGs have larger fins than the earlier test models and the packages
were insulated so that the end caps have lower temperatures.  
This results in lower radiation from the end caps than 
from the cylinder/fins \cite{tele,Rconf}. As a result, 
even though this is not exact, we can argue that 
the vast majority of the  heat radiated by the RTGs is
symmetrically directed to space unobscured by the antenna.  Further, 
for this mechanism to work 
one still has to assume that the energy hitting the antenna is
completely reradiated in the direction of the spin axis 
\cite{heatreflect}.  

Finally, if this mechanism were the cause, ultimately an unambiguous
decrease in 
the size of $a_P$ should be seen  because the RTGs' 
radioactively produced radiant heat is
decreasing.   As noted previously, the heat produced is now about 80\% of
the original magnitude.   In fact, one would similarly 
expect  a decrease of about 
$0.75\times 10^{-8}$ cm/s$^2$ in $a_P$ over the 11.5 year Pioneer 10 data
interval if this mechanism were the origin of $a_P$.

So, even though a complete thermal/physical model of the spacecraft 
might be able to ascertain if there are any other 
unsuspected heat systematics, we conclude that this particular 
mechanism does not provide enough power to explain 
the Pioneer anomaly \cite{uskatz}. 

In addition to the observed constancy of the
anomalous acceleration,  any explanation involving thermal radiation must
also discuss the absence of a  disturbance  to the spin of the
spacecraft.   There may be a small correlation of the spin angular
acceleration with the anomalous linear acceleration.  However,   as
described in Section \ref{recent_results}, the linear acceleration
is much more constant than the spin. This suggests that most of the linear
acceleration is not caused by whatever disturbs the spin, thermal or not. 

However,   a careful look at the Interval I results of Figure 
\ref{fig:pioneer_spin} shows that the nearly steady, background spin-rate
change of about   $6 \times 10^{-5}$ rpm/day   is slowly decreasing.

In principle this could be caused by heat.  

The spin-rate change produced by the torque of 
radiant power directed against the rotation with a lever arm $d$ is 
\begin{equation}
\ddot{\theta} = \frac {P~d}{c~{\cal I}_{\tt z}},  \label{tdd}
\end{equation}
where ${\cal I}_{\tt z}$ is the moment of inertia,  588.3 kg m$^2$
\cite{vanallen}.   We take a base unit of $\ddot{\theta}_0$ for a power of
one Watt and a  lever arm of one meter.  This is 
\begin{eqnarray}
\ddot{\theta}_0 &=& 5.63 \times 10^{-12}~\mathrm{rad/s^2}
=4.65 \times 10^{-6}~\mathrm{ rpm/day} = \nonumber\\
&=&  1.71 \times 10^{-3}~\mathrm{ rpm/yr}. 
\end{eqnarray}
So, about 13 Watt-meters of directed power could cause the base spin-rate
change.  

It turns out that  such sources could, in principle,  
be available.  There are  $3\times 3 = 9$ 
radioisotope heater units (RHUs) with one Watt power  to heat the
Thruster Cluster Assembly (TCA). (See pages  3.4-4 and  3.8-1--3.8-17 
of Ref. \cite{piodoc}.)
The units are on the edge of the antenna 
of radius 1.37 m,  in the housings of
the TCAs which are  approximately 180$^\circ$ apart
from each other. At one position there are  six RHUs 
and at the other position there are three.  An additional RHU 
is near the sun sensor which is located near the second assembly. 
The final RHU is located  at the magnetometer, 6.6 meters out from the
center of the spacecraft.   

The placement  gives an ``ideal'' rotational asymmetry  of two Watts. 
But note, the real asymmetry should be less, since these RHUs do not 
radiate only in one direction.  
Even one Watt unidirected at the magnetometer, is not enough to cause the 
baseline spin rate decrease. Further,  
since the base line is decreasing faster than what would come from
the change cause by radioactive decay decrease, one cannot look for this
effect or some complicated RTG source as the entire origin of the baseline
change.  One would suspect a very small gas leak or a combination 
of this and heat from the powered bus.   
(See Section    \ref{subsec:mainbus}.)  
Indeed, the factor $1/c$ in Eq. (\ref{tdd}) is a manifestation of the 
energy-momentum conservation power needed to produce $\ddot{\theta}$ by
heat vs. massive particles.  

But in any event, this baseline spin-rate change  
is not  significantly correlated with
the anomalous acceleration, so we do not have to pursue it further.

%********************DIFFERENTIAL EMISSIVITY 

\subsection{Differential emissivity of the RTGs}
\label{differemit}

Another suggestion related to the RTGs is the following \cite{slusher}: 
during the early parts of the missions, there might have been a differential
change of the radiant emissivity of the solar-pointing sides of the RTGs
with respect to the deep-space facing sides.   Note that, especially
closer in the Sun, the inner sides were subjected to
the solar wind.  Contrariwise, the outer sides were  sweeping through the
solar-system dust cloud.   Therefore, it can be  argued that these two
processes could have caused the effect.  However, other information seems 
to make it difficult for this explanation to work.  

The six fins of each RTG, designed to ``provide the bulk of the heat 
rejection capacity,'' were fabricated of HM21A-T8 magnesium alloy plate
\cite{tele}.  The metal, after being specially prepared, was coated with 
two to three mils of zirconia in a sodium silicate binder to provide a high 
emissivity $(\sim 0.9)$ and low absorptivity $(\sim 0.2)$.  
Depending on how symmetrically fore-and-aft they radiated,
the  relative fore-and-aft emissivity of the alloy would have had to have 
changed by $\sim10$\% to  account for $a_P$ (see below).  Given our
knowledge of the 
solar  wind and the interplanetary dust (see Section \ref{sec:know}), 
we find that this  amount of a radiant change would be 
difficult to explain, even if it were of the right sign. 
(In fact, even the brace bars holding the RTGs were built such that 
radiation is roughly fore/aft symmetric,) 

We also have ``visual'' evidence from the Voyager spacecraft. 
As mentioned, the Voyagers are not spin-stabilized.  They have imaging
video cameras attached 
\cite{camera}.  The cameras are mounted on a scan platform
that is pointed under  both celestial and inertial attitude control modes
\cite{plate}. The cameras {\it do not} have lens covers \cite{hansen}.  
During the outward cruise calibrations, the cameras were sometimes pointed
towards an imaging target plate mounted at the rear of the spacecraft. But
most often they were pointed all over the sky at specific star fields in
support of ultraviolet spectrometer observations. Meanwhile, the 
spacecraft antennae were pointed towards Earth.  Therefore, at an  angle,
the lenses were sometimes hit by the solar wind and sometimes by the 
interplanetary dust.  Even so, there was no noticeable   deterioration of
the images received, even when Voyager 2 reached  Neptune \cite{neptune}. 
We infer, therefore, that this mechanism can not  explain the Pioneer
effect.   

It turned out that the greatest radiation damage occurred during
the flybys.  The peak Pioneer 10 radiation flux near Jupiter was about 
10,000 times that of Earth for electrons (1,000 times for protons).  
Pioneer 11 experienced an even  higher radiation flux and also went by 
Saturn \cite{piopr2}.  (We return to this in Section \ref{twospace}.)  
Therefore, if radiation damage was a
problem, one should have seen an approximately uniform change in
emissivity during flyby.   Since the total heat flux, $\cal{F}$,  
from the RTGs was a constant over a flyby, 
there would have been a change in the RTG surface
temperature manifested by the radiation formula 
${\cal{F}} \propto \epsilon_1T_1^4 = \epsilon_2T_2^4$, the $\epsilon_i$ 
being the emissivities of the fin material.  There are several
temperature sensors mounted at RTG fin bases. They measured average 
temperatures of approximately 330 F, roughly 440 K.  Therefore, a 10\%
change in the {\it total average} emissivity 
would have produced a temperature change of $\sim$12.2 K $=$ 22 F. 
Such a change would have been noticed. 
(Measurements would be compared from, say, 30 days before and after 
flyby to eliminate the flyby power/thermal distortions.)
Since (see below) a 10\% 
{\it differential} fore/aft emissivity could cause the Pioneer effect, 
the lack of observation of a 10\% {\it total average} emissivity change 
limits the size of the differential emissivity systematic. 

To obtain a reasonable estimate of the uncertainty, consider if 
one side (fore or aft) of the
RTGs had its emissivity changed by 1\% with respect to the other side.  
In a  simple cylindrical model of the RTGs, with  2000 W power  
(here we presume only radial emission with no loss out the sides),   
the ratio of power emitted by the two sides would be 0.99 = 995/1005, 
or a differential emission between the half cylinders of 10 W.
Therefore, the fore/aft asymmetry towards the normal would be 
$[10~{\mathrm{W}}] \times  \int_0^\pi [\sin 
\phi]d\phi/\pi \approx 6.37$ W. If one does a more sophisticated fin
model, with 4 of the 12 fins facing the normal (two flat and two at 
30$^\circ$), one gets a number of 6.12 W. We take this to yield our
uncertainty,  
\begin{equation}
\sigma_{\tt d.e.} = 0.85 \times 10^{-8} ~ {\mathrm{cm/s}}^2.
\end{equation}
Note that $10~ \sigma_{\tt d.e.}$ almost equals our final $a_P$.  This is
the origin of our previous statement that $\sim 10$\% differential 
emissivity (in the correct direction) would be needed to explain $a_P$.   

Finally, we want to comment on the significance of radioactive decay
for this mechanism.  Even acknowledging the
Interval jumps due to gas leaks (see below),   we reported a one-day
batch-sequential value (before systematics) for $a_P$,  averaged over the
entire  11.5 year interval, of $a_P = (7.77 \pm 0.16)\times 10^{-8}$  
cm/s$^2$.   From radioactive decay, the
value of $a_P$  should have decreased by $0.75$ of these units over 11.5
years.  This is 5 times the above variance, which  is very large with batch
sequential.  Even more stringently, this bound is good for {\it all}
radioactive heat sources.  So,  what if one were to argue that 
emissivity changes occurring before 1987 were the cause
of the Pioneer effect?   There still should have been a decrease in
$a_P$ with time since then, which has not been observed.    
 
We will return to these points in Section \ref{twospace}.

%************ murphy *************************************

\subsection{Non-isotropic radiative cooling of the spacecraft}
\label{subsec:mainbus}

It has also been suggested that the  anomalous acceleration seen in the
Pioneer 10/11 spacecraft can be, ``explained, at least in part, by
non-isotropic  radiative cooling of the spacecraft \cite{murphy}.''  So,
the question is, does  ``at least in part'' mean this effect  comes  near
to explaining the anomaly?  We argue it does not \cite{usmurphy}. 

Consider radiation of the main-bus electrical systems power from 
the spacecraft rear.    For the Pioneers, the aft has a louver system,
and  ``the louver system acts to control the heat  rejection of the
radiating platform.  A bimetallic spring, thermally coupled radiatively to
the platform, provides the motive force for  altering the angle of each
blade. In a closed position (below 40 F) the heat rejection of the 
platform is minimized by virtue of the blockage of the  blades while  
open fin louvers provide the platform with a nearly  unobstructed view of
space \cite{piodoc}.''

If these louvers were open (above $\sim$ 88 F) and all the
diminishing  electrical-power heat was radiated only out of the louvers,  
this mechanism could produce a significant effect. However, by nine AU the
actuator spring  temperature had already reached $\sim$40  F
\cite{piodoc}.  This means the  louver doors were   closed  (i.e.,  the
louver angle was zero)  from where we obtained our data. 
Thus, from that time on of the radiation properties,  the contribution
of the thermal radiation to the Pioneer anomalous acceleration should be 
small.  Although one might speculate that a louver stuck, there are 
30 louvers on each craft.  They clearly worked as designed, 
or else the temperature of the crafts' interiors would have
fallen to disastrous levels. 

\begin{figure}[h]
\begin{center}
      \epsfig{figure=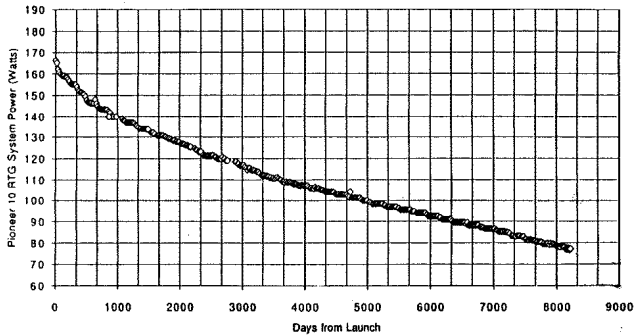, width=3.4in}
\end{center}
\caption{
The Pioneer 10 electrical power generated at the RTGs as a function of
time from launch to near the end of 1994.   By 1998.5, only 
$\sim$68 W was generated.
\label{epower}}
\end{figure}

As shown in Figure \ref{epower}, 
in 1984 Pioneer  10 was at about 33 AU and the power was about 105 W. 
(Always reduce  the total power numbers by 8 W to account for the radio
beam power.) In (1987, 1992, 1996) the spacecraft was at $\sim$(41, 55, 65) 
AU and the power was $\sim$(95, 82, 73) W. The louvers were inactive, and 
no decrease in $a_P$ was  seen.
  
%********************
%**********************

In fact, during the entire 11.5 year period from 1987 to 1998 the
electrical 
power  decreased from around 95 W to around 68 W, a change of 27 W. 
Since we already have noted that about $\sim 65$ W is needed to cause our
effect, such  a large decrease in the ``source'' of the acceleration
would have been seen.   But as shown in Section \ref{recent_results}, it
was not.  Even the small  differences in the three intervals are most
likely to be from gas leaks (as will be demonstrated in Section 
\ref{sec:gleaks}).  

Later a double modification of this idea was given.  It was first suggested 
that  ``most, if not all, of the unmodeled
acceleration''  of  Pioneer 10 and 11 is due to an  essentially constant
supply of heat coming from the central compartment, directed out the
front of the  craft through the closed louvers \cite{scheffer}-(a). 
However, when one studies the electrical power history in both parts
(instruments and experimental) of the central compartment, there is no
constancy of heat.  (See the details in \cite{usscheffer}.)  Indeed during
our data period the heat from this compartment decreased from about 73 W to
about 57 Watts, or a factor of 1.26.  This is inconsistent with the
constancy of our result.  Further, if one looks at the earlier, very roughly
analyzed \cite{earlydata} data in Figure \ref{fig:correlation} one sees
nothing close to the internal
power change of 93 to 57 W (a factor of 1.6) \cite{usscheffer}.

To address this inconsistency a second modification \cite{scheffer}-b,c 
was made.  It was arbitrarily 
argued that there was an incorrect determination of the
reflection/absorption coefficients by a large factor.  But 
these coefficients are known to 5\%.  If they were as poorly determined 
as speculated, the mission would have failed early on.  (Further  
discussion is in \cite{usscheffer}.)  

We conclude that neither the original proposal \cite{murphy} nor the
modification \cite{scheffer} can explain the anomalous 
Pioneer acceleration \cite{usmurphy,usscheffer}.  A bound on the 
constancy of $a_P$ comes from first noting the 11.5 year 
1-day batch-sequential result, sensitive to time variation: 
$a_P = (7.77 \pm 0.16)\times 10^{-8}$  cm/s$^2$.  Also given the 
constancy of the earlier imprecise date, it is conservative to 
take three times this error 
to be our systematic uncertainty for radiative cooling of the 
craft, $\sigma_{\tt r.c.}= \pm 0.48 \times 10^{-8}$  cm/s$^2$.

Although doubtful, one can also speculate that some  mechanism 
like this might be involved with the baseline spin-rate
change discussed in Section \ref{subsec:katz}.  In 1986-7, Pioneer 10 power 
was about 97  W, decreasing at about 2.5-3.0  W/yr.  If you take a
lever arm of 0.71 meters (the hexagonal bus size), this is more than enough
to provide the 13 W-meters necessary to produce the baseline spin-rate
change of Figure \ref{fig:pioneer_spin}.  Further for the first three
years the decrease about matches the bus power loss rate.  Then after the 
complex changes associated with the end of 1989 to 1990, there is a 
decrease in the base rate with a continued similar slope.  

Perhaps the ``baseline" rate is indeed from the heat of the bus being vented 
to the side. But the much larger gas leaks would be on top of the baseline.

%***************** helium ************************************* 

\subsection{Expelled Helium  produced  within the RTGs}
\label{subsec:helium}

Another possible on-board systematic is from the expulsion of the  He being
created in the RTGs from  the $\alpha$-decay of $^{238}$Pu.  To make this
mechanism work, one would need that the He leakage from the RTGs  be 
preferentially directed away from the Sun,   with a velocity large enough
to cause the acceleration. 

The SNAP-19 Pioneer RTGs were designed in a such a way that the He pressure
has not been totally  contained within the Pioneer heat source  over the
life of RTGs \cite{tele}. Instead, the Pioneer heat source contains a
pressure relief device which allows the generated He to vent out of the
heat source and into the thermoelectric converter.  (The strength member
and the capsule clad contain small holes to permit He to escape into the
thermoelectric converter.) The thermoelectric converter housing-to-power
output receptacle interface is sealed with a viton O-ring. 
The O-ring allows the helium gas within the converter to
be released by permeation 
to the space environment throughout the mission life of the Pioneer RTGs. 

Information on the fuel pucks \cite{puck} shows that they each have
heights of 0.212 inches with diameters of 2.145 inches. With 18  in each
RTG and four RTGs per mission, this gives a  total volume of fuel of about
904 cm$^3$.  The fuel is   PMC Pu conglomerate.    The amount of
$^{238}$Pu in this fuel is about 5.8 kg. With a half life of 87.74
years, that means the rate of He production (from Pu decay) is about 0.77
gm/year, assuming it all leaves the cermet.  Taking on operational
temperature on  the RTG surface  of 320 F = 433 K, implies a $3kT/2$
helium velocity of  1.22 km/s. (The possible energy loss coming out of
the viton is neglected  for helium.)   Using this in the rocket
equation,  
\begin{equation} 
a(t) = -v(t) \frac{d}{dt} \Big[\ln M(t)\Big]
\end{equation} 
with our nominal  Pioneer mass with half the fuel gone {\it and the
assumption}  that the gas is all unidirected, yields  a maximal bound on
the possible acceleration of $1.16 \times 10^{-8}$ cm/s$^2$.  So, we can
rule out helium permeating through the O-rings as the cause of $a_P$
although it is a systematic  to be dealt with.   

Of course, the gas is not totally unidirected.   As one can see by
looking at Figures \ref{fig:trusters} and  III-2 of \cite{tele}:   the
connectors  with the O-rings  are  on the RTG cylinder surfaces,  on the
ends of the cylinders where the fins are notched. They are equidistant
(30 degrees)  from two of the fins.  The placement is exactly at the
``rear'' direction of the RTG  cylinders, i.e., at the position closest
to the Sun/Earth.   The axis through the O-rings is parallel to the
spin-axis.   
The O-rings, sandwiched by the receptacle and connector plates, ``see''
the outside  world through an angle of about 90$^\circ$ in latitude
\cite{hefromrtg}.  (Overhead of the O-rings is towards the Sun.)  
In longitude the O-rings see the direction of the bus
and space through about 90$^\circ$, and  ``see'' the fins through most of
the rest of the longitudinal angle.  

If one assumes a single elastic reflection, one can estimate  the
fraction of the bias away from the Sun.   (Indeed, multiple and back
reflections will produce an even greater bias.  Therefore, we  feel this
approximation is justified.)  This estimate is $(3/4) \sin30^\circ$ 
times the average of the heat momentum component  parallel to the
shortest distance to the RTG fin.   Using this, we find the bias would be
$0.31 \times 10^{-8}$ cm/s$^2$.  This bias effectively increases the
value of our solution for $a_P$, which we  hesitate to accept given all
the true complications of the real system.   Therefore we take the
systematic expulsion  to be $a_{\tt He} = (0.15 \pm 0.16)
\times 10^{-8}$ cm/s$^2$.

%******************8.6) Propulsive mass expulsion due to gas leakage *****

\subsection{Propulsive mass expulsion due to gas leakage}
\label{sec:gleaks}
 
The effect of propulsive mass expulsion due to gas leakage has to be
assessed. Although this effect is largely unpredictable, many  
spacecraft  have experienced gas leaks producing accelerations on the
order of $10^{-7} ~{\rm cm/s^2}$.  [The reader will recall the even
higher figure for Ulysses found in Section \ref{sec:AUlysses}.]  As noted
previously,  gas leaks  generally behave differently after each
maneuver.  The leakage often decreases with  time and  becomes  
negligibly small.  

Gas leaks can originate from Pioneer's propulsion system, which is used
for mid-course trajectory maneuvers, for spinning-up or -down the
spacecraft, and for orientation of the spinning spacecraft.  The
Pioneers  are equipped with three pairs of hydrazine thrusters which are
mounted on the circumference of the Earth-pointing high gain antenna. 
Each pair of thrusters forms a Thruster Cluster Assembly (TCA) with two
nozzles aligned in opposition to each other. For attitude control,   two
pairs of  thrusters can be fired forward or aft and are used to precess
the spinning antenna  (See Section \ref{sec:prop}.) The other pair of
thrusters is aligned  parallel to the rim of the antenna with nozzles
oriented in co- and contra-rotation directions for spin/despin
maneuvers.  

During both observing intervals for the two Pioneers, there  were no
trajectory or spin/despin maneuvers.  So, in this analysis   we are
mainly concerned with precession  (i.e., orientation or attitude control)
maneuvers only.   (See Section \ref{sec:prop}.) Since the valve seals
in the thrusters can never be perfect, one can ask if the leakages
through the hydrazine thrusters could be the  cause of the  anomalous
acceleration, $a_P$. 

However, when we investigate the total computational accuracy of
our solution in Section \ref{Ext_accuracy}, we will show that the
currently implemented models of propulsion maneuvers may be
responsible for an uncertainty in $a_P$ only  at the level of  
$\pm0.01\times 10^{-8}$ cm/s$^2$. Therefore, the maneuvers themselves are
the main contributors neither to the total error budget 
nor to the gas leak uncertainty, as we now detail    

The serious uncertainty comes from the possibility 
of undetected gas leaks.  We will address this issue in some detail.
First consider the possible action of gas leaks  originating
from the spin/despin  TCA.  Each nozzle from this pair of thrusters  is
subject to a certain amount of gas leakage. But 
only a differential leakage from the two nozzles would produce an observable
effect causing the spacecraft to  either spin-down or spin-up
\cite{leaks}.  So, to obtain a gas leak uncertainty (and we emphasize
``uncertainty'' vs. ``error'' because  we have no other evidence) let us 
ask how large  a differential  force is needed to cause the 
spin-down or spin-up effects observed?  

Using the moment of inertia about the spin
axis, ${\cal I}_{\tt z}=\sim 588.3$ kg$\cdot$m${^2}$
\cite{vanallen}, and the antenna radius, ${\cal{R}}=1.37$ m, 
as the lever arm,  one can calculate that the differential 
force needed to torque the spin-rate change, 
$\ddot{\theta}_i$, in  Intervals $i=$I,II,III   is 
\begin{eqnarray}
F_{\ddot{\theta}_i} &=& 
\frac{{\cal I}_{\tt z}{\ddot{\theta}_i}}{{\cal{R}}}
=\big(2.57, ~12.24, ~1.03\big) \times 10^{-3}~~{\rm
dynes}.\hskip 20pt 
\label{FthetaI} 
\end{eqnarray}

It is possible that a similar mechanism of undetected gas leakage could
be responsible for the net differential force acting in the direction
along  the line of sight. In other words, what if there were some
undetected gas leakage from the thrusters oriented  along the spin axis
of the spacecraft that is causing
$a_P$?  How large would this have to be?  A force of ($M= 241$ kg)
\begin{equation}
F_{a_P}=  M \, a_P
=21.11\times10^{-3}~{\rm dynes}
\end{equation}
would be needed to produce our final unbiased value of $a_P$.   (See  
Section \ref{budget}.
That is, one would need even more force than is needed to produce 
the anomalously high rotational gas leak of Interval II.   
Furthermore,  the differential leakage 
to produce this $a_P$ would have had to have been constant
over many years and in the same direction for both spacecraft, without
being detected as a spin-rate change.   That is possible, but certainly
not demonstrated.
Furthermore if the gas leaks hypothesis were true, one  would expect to see a
dramatic difference in   $a_P$ during the three Intervals of Pioneer
10 data. Instead an almost 500 \% spin-down rate change between
Intervals I and II resulted only in a less than 8\% change in $a_P$.

Given the small amount of information, we propose to {\it conservatively} 
take as our gas leak uncertainties the acceleration values that would be 
produced by  differential forces equal to 
\begin{eqnarray}
F_{a_P(i)\tt g.l.}&\simeq & \pm \sqrt{2}F_{\ddot{\theta}_i} =
\\
&=&  \big(\pm 3.64,~\pm 17.31,~\pm 1.46\big)
\times 10^{-3}~~{\rm dynes}.\nonumber
\label{eq:diff}
\end{eqnarray}
The argument for this is that, in the root sum of squares 
sense,  one is accounting for
the  differential  leakages from the two pairs of thrusters 
with their nozzles oriented along the line of sight direction. 
This directly translates into the acceleration errors introduced by the
leakage during the three intervals of Pioneer 10 data,
\begin{eqnarray}
\sigma(a_{P(i) \tt g.l.})&=& \pm F_{a_P(i)\tt g.l.}/M =\\ 
&=&\big(\pm 1.51,~\pm 7.18,~\pm 0.61\big)\times10^{-8}~{\rm cm/s}^2.
\nonumber
\end{eqnarray}
Assuming
that these errors are uncorrelated and  are normally distributed around
zero mean,  we find the gas leak uncertainty for the entire Pioneer 10
data span to be 
\begin{equation}
\sigma_{\tt g.l.} =  \pm 0.56 \times
10^{-8}~~{\rm cm/s}^2. 
\label{gluncertC}
\end{equation}  
This is one of our largest uncertainties. 

The data set from Pioneer 11 is over a much smaller time span,  taken
when Pioneer 11 was much closer to the Sun (off the plane of the 
ecliptic), and during a maximum of solar activity.   For Pioneer 11 the
main  effects of gas leaks occurred at the maneuvers,  when there were 
impulsive lowerings of  the spin-down rate.  These dominated the over-all
spin rate change of   $\ddot{\theta}_{11}= -0.0234$ rpm/yr.  (See Figure 
\ref{fig:pio11spin}.)   But in between maneuvers the spin rate was
actually  {\it increasing}. One can argue that this explains the higher
value for 
$a_{P(11)}$ in Table \ref{resulttable} as compared to  $a_{P(10)}$. 
Unfortunately, one has  no {\it a priori} way of  predicting the effect
here.  We do not know that the same specific gas leak mechanism applied
here as did in the case of Pioneer 10 and there is no well-defined 
interval set as there is for Pioneer 10.  Therefore, although we feel this
``spin up'' may be part of the explanation of the higher value of $a_P$ 
for Pioneer 11, we leave the different numbers as a separate systematic
for the next subsection. 

At this point, we must conclude that the gas leak mechanism for explaining
the anomalous acceleration seems very unlikely, because  it
is hard to understand why it would affect Pioneer 10 and 11 at the same
level (given that both spacecraft had different quality of propulsion
systems, see Section \ref{sec:prop}).  One also expects a gas
leak would obey the rules of a Poisson distribution. 
That clearly is not true.  Instead, our analyses of different data sets
indicate that $a_P$ behaves as a constant bias rather than as a random
variable.  (This is clearly seen in the time history of $a_P$ obtained
with batch-sequential estimation.)

%******************* 2 SPACECRAFT **************

\subsection{Variation between determinations from the two spacecraft}
\label{twospace}

Finally there is the important point that we have  two ``experimental'' 
results from  the two spacecraft, given in Eqs. (\ref{pio10lastresult})  
and (\ref{pio11lastresult}): 7.84 and 8.55, respectively, in units of
$10^{-8}$ cm/s$^2$.  If the Pioneer effect is real, and not a 
systematic, these numbers should be approximately equal. 

The first number, 7.84, is for Pioneer 10.  In Section \ref{final_sol} 
we obtained this number by correlating the values of $a_P$ in
the three data Intervals with the different spin-down rates in these
Intervals.  The weighted correlation between a shift in $a_P$ and 
the spin-down rate is  $\kappa_0  =(30.7\pm 0.6)$ cm.
(We argued in  the previous Section \ref{sec:gleaks} that this 
correlation is the manifestation of the rotational gas leak systematic.)
Therefore, this number
represents the entire 11.5 year data arc of Pioneer 10.   Similarly, 
Pioneer 11's number, 8.55, represents a 3$\frac{3}{4}$ year data arc.  

Even though the Pioneer 11 number may be less reliable since the 
craft was so much closer to the Sun, we calculate 
the time-weighted average of the experimental results from the two craft: 
$[(11.5)(7.84) + (3.75)(8.55)]/(15.25)
= 8.01$  in units of $10^{-8}$ cm/s$^2$.  This implies a bias of
$b_{\tt 2\_craft}=+0.17\times10^{-8}$ cm/s$^2$ with respect to the Pioneer
10 experimental result $a_{P({\tt exper})}$. We also take this
number to be our two spacecraft uncertainty.  This means   
\begin{eqnarray}
a_{\tt 2-craft}&=&b_{\tt 2-craft}\pm \sigma_{\tt 2\_craft} =
\nonumber\\
&=&
(0.17 \pm 0.17)~\times~10^{-8}~\mathrm{cm/s}^2.
\end{eqnarray}

The difference between the two craft could be due to  different gas
leakage.  But it also could be due to  heat emitted from the RTGs.  In
particular,  the two sets of RTGs have had different histories and so
might have  different emissivities.  Pioneer 11 spent more time in the
inner solar system (absorbing radiation).  Pioneer 10 has  swept out
more dust in deep space.  Further,  Pioneer 11 experienced  about twice
as much Jupiter/Saturn radiation as Pioneer 10.

Further, note that 
$[a_{P({\tt exper)}}^{\tt Pio11} - a_{P({\tt exper)}}^{\tt Pio10}]$  and
the uncertainty from differential emissivity of the RTGs, $\sigma_{\tt
d.e.}$, are of the same size:  0.71 and 0.85 $\times10^{-8}$ cm/s$^2$.
It could therefore be argued that Pioneer 11's offset from Pioneer 10 
comes from Pioneer 11 having obtained twice as large a  differential
emissivity bias as Pioneer 10. Then our final value of $a_P$, given in  
Section \ref{budget}, would be reduced by about $0.7$ of our units 
since  $\sigma_{\tt d.e.}$ would have become mainly a negative bias, 
$b_{\tt d.e.}$.  This would make the final number closer to 
$8 \times10^{-8}$ cm/s$^2$.  Because this model and our final number are 
consistent, we present this observation only for completeness and as a
possible reason for the different results of the two spacecraft.

%***********************9) DATA/SOFTWARE TESTS************************
%*******************************************************************
%\newpage

\section{\label{Int_accuracy}COMPUTATIONAL SYSTEMATICS}

Given the very large number of observations for the same spacecraft, the
error contribution from observational noise is very small and not a
meaningful measure of uncertainty. It is therefore necessary to consider
several other effects in order to assign realistic errors. 
Our first consideration is the statistical and 
numerical stability of of the calculations.   We then go on to 
the cumulative influence of all modeling errors and editing decisions. 
Finally we discuss 
the  reasons for and significance of the annual term.  

Besides the factors mentioned above,  we will discuss in this section
errors that may be attributed to the specific hardware used to run the
orbit determination computer codes, together with computational algorithms
and statistical methods used to derive the solution.

%*************NUMERICAL STABILITY

\subsection{Numerical stability of least-squares estimation}
\label{leastsquares}

Having presented estimated solutions along with their formal statistics,
we should now attempt to characterize the true accuracy of these results. 
Of course, the 
significance of the results must be assessed on the  basis of the
expected measurement errors. These expected errors are used to  weight a
least-squares adjustment to parameters which describe the  theoretical
model. [Examination of experimental systematics from sources both external
to and also internal to the spacecraft was covered in Sections 
\ref{ext-systema}-\ref{int-systema}.]

First we look at the numerical stability of the least squares 
estimation algorithm and  the derived solution. The leading computational
error source turns out to be subtraction of similar  numbers. Due to the
nature of floating point arithmetic, two numbers with  high order digits
the same are subtracted one from the other results in the  low order digits
being lost. 
% For example, if 123456.666666667 is subtracted  from
% 123457.666666667 the result is 0.99999999746269 or 1.00000000000000 
% depending  on the rounding of the last bit. 
This situation occurs with
time  tags on the data. Time tags are referenced to some epoch, 
such as say 1 January 1 1950 which is used by CHASMP.  
As more than one billion seconds have passed since 1950,
time tags on  the Doppler data have a start and end time that have five or 
six common leading  digits. Doppler signal is computed by a differenced
range formulation (see Section \ref{Dopp_tech}). This noise in  the time
tags causes noise in the computed Doppler at the 0.0006 Hz level for both
Pioneers. This noise can be reduced by shifting the reference epoch  closer
to the data or increasing the word length of the computation, however,  it
is not a significant error source for this analysis. 
 
In order to guard against possible computer compiler and/or hardware
errors we ran orbit determination programs on  different
computer platforms. JPL's ODP  resides on an HP workstation. The
Aerospace  Corporation ran  the analysis on three different computer
architectures: (i) Aerospace's DEC 64-bit RISC architecture workstation 
(Alphastation 500/266), (ii) Aerospace's DEC 32-bit CISC architecture 
workstation (VAX 4000/60), and (iii) Pentium Pro PC. Comparisons of
computations performed for CHASMP in the three machine show consistency to 15 
digits which is just sufficient to represent the data. While this comparison
does not eliminate the possibility of systematic errors that are common to 
both systems, it does test the numerical stability of the  analysis on
three very different computer architectures.  

The results of the individual programs 
were given in  Sections \ref{results}and
\ref{recent_results}. In a  test we took the JPL results for a 
batch-sequential {\it Sigma} run with  50-day averages of the 
anomalous acceleration of Pioneer 10, 
$a_P$.  The data interval was from January 1987 to July 1998.  We compared
this to an Aerospace determination using CHASMP, where the was split 
into 200 day intervals, over a shorter data interval ending in 1994. 
As seen in Figure \ref{fig:rec_res_comb}, the results basically agree.   

Given the excellent agreement in these
implementations of  the modeling software,  
we conclude that differences in analyst choices (parameterization
of  clocks, data editing, modeling options, etc.) give rise to 
coordinate discrepancies  only at the level of  $0.3$ cm. This
number corresponds to an uncertainty in estimating the anomalous
acceleration on the order of   $8\times 10^{-12}$ cm/s$^2$.

But there is a slightly larger error to contend with.  In principle 
the STRIPPER  can give output to 16 significant figures.  From the
beginning the output was-rounded off to
15 and later to 14 significant figures.  
When Block 5 came on near the beginning of 1995, the output was rounded off 
to 13 significant figures. Since the Doppler residuals are 1.12 mm/s
this last truncation  means an error of order 0.01 mm/s. 
If  we divide this number by 2 for an average round off, 
this translates to $\pm 0.04\times10^{-8}$ cm/s$^2$. The roundoff occurred 
in approximately all the data we added for this paper.  
This is the cleanest  1/3 of the Pioneer 10 data.  
Considering this we take the uncertainty to be 
\begin{equation}
\sigma_{\tt num} \pm 0.02 \times 10^{-8}   ~~{\rm cm/s}^2.
\label{eq:num_st}
\end{equation}

It needs to be stressed that such tests examine only the accuracy of
implementing a given set of model codes, without consideration of the 
inherent accuracy of the models themselves.  Numerous external tests, 
which we have been discussing in the previous three sections, 
are possible for assessing the accuracy of the 
solutions. Comparisons between the two software packages enabled us to 
evaluate the implementations of the theoretical models within  a
particular software. Likewise, the results of independent radio tracking 
observations obtained for the different spacecraft and analysis programs 
have enabled us to 
compare our results to infer  realistic error levels from
differences in data sets and  analysis methods.  Our analysis of the
Galileo and Ulysses missions (reported in Sections \ref{galileo} and 
\ref{ulysses}) was done partially for this purpose.

%*************** CONSISTENCY/MODEL TESTS ************

\subsection{Accuracy of consistency/model tests}
\label{Ext_accuracy}

\paragraph{Consistency of solutions:}
A code that models the motion of solar system bodies and spacecraft 
includes numerous lengthy calculations.  Therefore, the software used 
to obtain solutions from the Doppler data is, of necessity, very complex. 
To guard  against potential errors in the
implementation of these models, we used two software packages;  
JPL's ODP/\emph{Sigma} modeling software \cite{Moyer71,Moyer81} and  
The  Aerospace Corporation's
POEAS/CHASMP software package \cite{chasmp,poeas}. The differences between
the JPL and Aerospace orbit determination program  results are now 
examined. 

As discussed in Section \ref{sec:OD}, in estimating parameters
the CHASMP code uses a standard variation of parameters method whereas ODP
uses the Cowell method to integrate the equations of motion and the
variational equations. In other words, CHASMP integrates six first-order
differential equation, using the Adams-Moulton predictor-corrector method 
in the orbital elements. 
Contrariwise,  ODP integrates three second-order
differential equations for the accelerations using the Gauss-Jackson
method. (For more details on these methods see Ref. \cite{herrick}.)  

As seen in our results of Sections \ref{results}and
\ref{recent_results}, agreement was
good; especially considering that each program uses independent methods, 
models, and constants. Internal consistency tests indicate that a
solution is consistent at the level of one part in $10^{15}$. This implies
an acceleration error on the order of no more then one part in  
$10^{4}$ in $a_P$.

\paragraph{Earth orientation parameters:} 
In order to check for possible problems with Earth orientation, CHASMP
was modified to accept Earth orientation information from three different 
sources. (1) JPL's STOIC program that outputs {\tt UT1R-UTC}, (2) JPL's
Earth  Orientation Parameter files ({\tt UT1-UTC}), and (3) The
International Earth  Rotation Service's Earth Orientation Parameter file
({\tt UT1-UTC}). We found that all three
sources gave virtually  identical results and changed the value of $a_P$
only in the 4th digit \cite{Folkner}. 

\paragraph{Planetary ephemeris:}  Another possible source of problems 
is the planetary ephemeris. To explore this a fit was first done with
CHASMP that used DE200. The solution of that  fit was then used in a fit
where DE405 was substituted for DE200. The result  produced a  small 
annual signature before  the fit. After the fit, the maneuver solutions
changed a small amount (less then 10\%) but the value of the anomalous
acceleration remained the same to seven  digits. The post-fit residuals to
DE405 were virtually unchanged from those  using DE200. This showed that
the anomalous acceleration was unaffected by  changes in the planetary
ephemeris.  

This is pertinent to note for the following subsection.  
To reemphasize the above,  a small ``annual term'' can be 
introduced by changing the planetary ephemerides.  
This annual term can then be totally taken up by changing the 
maneuver estimations.   Therefore,  in principle,  any possible 
mismodeling in the planetary ephemeris could be at least partially 
masked by the maneuver estimations.

\paragraph{Differences in  the codes' model implementations:}  
The impact of an analyst's choices is difficult to address, largely
because of the time and expense required to  process a large data set
using complex models.   This is especially important when it comes to
data editing.   It should be understood that small differences are to be
expected as  models differ in levels of detail and accuracy.   The
analysts' methods, experience, and judgment differ. The independence of
the analysis of JPL and Aerospace has been consistently and  strictly
maintained in order to provide confidence on the validity of the analyses.
Acknowledging such  difficulties, we still feel that using the very
limited tests given above is preferable to an implicit assumption that
all analysts' choices were  optimally made. 

Another source for differences in  the results presented in Table
\ref{resulttable} is the two codes' modeling of  spacecraft
re-orientation maneuvers. ODP uses a model that  solves for the resulted
change in the Doppler observable $\Delta v$ (instantaneous burn model). 
This is a more convenient model for  Doppler velocity measurements. 
CHASMP  models the change in acceleration,  solves for $\Delta a$ (finite
burn model), and only then produces a solution for
$\Delta v$. Historically, this was done  in order  to incorporate   range
observations (for Galileo and Ulysses) into the analysis.

Our best handle on this is the no-corona results, especially given that
the two  critical Pioneer 10 Interval III results differed by very little,
$0.02 \times 10^{-8}$ cm/s$^2$.  This data is least affected by 
maneuver modeling, data editing, corona modeling, and spin calibration.  
Contrariwise, for the other data, the differences were larger. 
The Pioneer Interval I and II results and the Pioneer 11 results differed, 
respectively, by  (0.21, 0.23, 0.25) in units of $10^{-8}$ cm/s$^2$.  
In these intervals models of maneuvers and data editing were crucial.    
Assuming that these errors are uncorrelated, we compute
their combined effect on  anomalous acceleration $a_P$ as 
\begin{equation}
\sigma_{\tt consist/model} = \pm 0.13 \times 10^{-8}~\textrm{cm/s}^2.  
\end{equation}

\paragraph{Mismodeling of maneuvers:} 
A small contribution to the error comes from a possible mismodeling of
the propulsion maneuvers.   In  Section \ref{model-maneuvers} we found 
that for a typical maneuver the standard error in the residuals is
$\sigma_0\sim0.095$ mm/s. 

Then we would expect that in the period  between
two maneuvers,  which on average is  $\tau=$ 11.5/28 year,  the effect
of the mismodeling would produce a contribution to the acceleration
solution with a magnitude on the order of  
$\delta a_{\tt man}= {\sigma_0}/{\tau} = 0.07 \times 10^{-8}$
cm/s$^2$. 
Now let us assume that the errors in the
Pioneer Doppler residuals  are normally distributed around zero mean with
the standard deviation of $\delta a_{\tt man}$ that constitute a single
measurement accuracy.  Then, since there are $N=28$ maneuvers in the
data set, the total error due to maneuver  mismodeling is  
\begin{equation}
\sigma_{\tt man} = \frac{\delta a_{\tt man}}{\sqrt{N}} 
= 0.01 \times 10^{-8} ~~{\rm cm/s}^2.
\label{gluncertA}
\end{equation}

\paragraph{Mismodeling of the solar corona:}  
Finally, recall that our number for mismodeling of the solar corona, 
$ \pm 0.02 \times 10^{-8}$ cm/s$^2$, was already explained in Section 
\ref{sec:corona}.

%******************** ANNUAL TERM *********************

\subsection{Apparent annual/diurnal periodicities in the solution}
\label{annualterm}

In Ref. \cite{moriond} we reported,  in addition to the constant anomalous
acceleration term, a possible annual sinusoid. If approximated by a
simple sine wave, the amplitude of this oscillatory term is about $1.6
\times 10^{-8}$ cm/s$^2$. The  integral of a sine wave in the
acceleration, $a_P$, with angular velocity  
$\omega$ and amplitude $A_0$ yields the following first-order Doppler 
amplitude in two-way fractional frequency:
\begin{equation}
\frac{\Delta \nu}{\nu} =  \frac{2A_0}{c~ \omega}.
\label{lasttwo}
\end{equation}
\noindent
The resulting Doppler amplitude for the  annual angular velocity  
$\sim 2 \times 10^{-7}$ rad/s is $\Delta \nu/\nu$ = 5.3 $\times$
10$^{-12}$.    At the Pioneer downlink S-band carrier frequency of  $\sim
2.29$ GHz, the  corresponding Doppler amplitude is 0.012 Hz 
(i.e. 0.795 mm/s).    

This term was first seen in ODP using the {\tt BSF} method.  As
we discussed in Section \ref{sec:PE}, treating $a_P$ as a stochastic
parameter in  JPL's batch--sequential analysis allows one to search for a
possible temporal variation in this parameter. Moreover, when many  short
interval times  were used with least-squares CHASMP, the effect was also
observed.  (See Figure \ref{fig:rec_res_comb} in Section
\ref{recent_results}.)

The residuals obtained from both programs  are of the same
magnitude.   In particular, the  Doppler residuals  are distributed about
zero Doppler velocity with a systematic variation $\sim$ 3.0  mm/s on a
time scale of $\sim$ 3 months. More precisely, the least-squares
estimation residuals from both  ODP/\emph{Sigma} and  CHASMP are
distributed well within a half-width taken to be  0.012 Hz.  (See, for
example,  Figure \ref{fig:pio10best_fit}.)  Even the general structures
of the two sets of residuals are similar. The fact that both programs
independently were able to produce similar post-fit residuals gives us
confidence in the  solutions.

\begin{figure}
\epsfig{figure=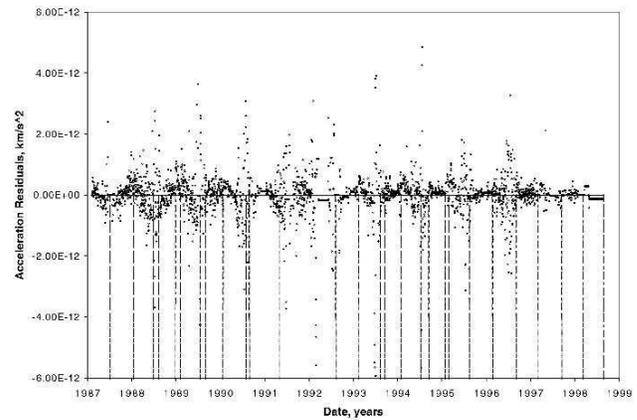,width=86mm}%,height=85mm}
    \caption
{ODP 1-day batch-sequential acceleration
residuals using the  entire Pioneer 10  data set.  
Maneuver times are indicated by the vertical dashed lines. 
      \label{annualresiduals}}
\end{figure}

With this confidence, we next looked in greater detail at the  acceleration
residuals from  solutions for $a_P$.   Consider Figure
\ref{annualresiduals}, which shows the $a_P$  residuals from a value
for $a_P$ of $(7.77\pm 0.16) \times 10^{-8}$ cm/s$^2$. The data was
processed using   ODP/\emph{Sigma} with a batch-sequential filter and
smoothing algorithm. The solution for $a_P$ was obtained using  1-day
batch sizes. Also shown  are the maneuver times. At early times the
annual term is largest.  During Interval II, the interval of the large
spin-rate change anomaly, coherent oscillation is lost.   
During Interval III the oscillation is smaller and begins to die out. 

%************
%************

In attempts to understand the nature of this annual term,  
we first examined a number of possible sources, 
including effects introduced by imprecise modeling of maneuvers, 
the  solar corona, and the Earth's troposphere. We also
looked at the  influence of the data editing strategies that were used. 
We concluded that these effects could not account for the annual term. 

Then, given that the effect is particularly large in the
out-of-the-ecliptic  voyage of Pioneer 11 \cite{moriond},  we focused
on the possibility that inaccuracies in solar system
modeling are the cause of the annual term in the Pioneer solutions. In
particular, we looked at the modeling of the Earth orbital orientation and
the accuracy of the planetary ephemeris.

%****************************

\paragraph{Earth's orientation:}

We specifically modeled the  Earth orbital elements $\Delta p$ and
$\Delta q$ as stochastic parameters.  ($\Delta p$ and $\Delta q$ are  two
of the Set III elements defined by Brouwer and Clemence \cite{bc}.)  
\emph{Sigma} was applied to the entire Pioneer 10 data set  with $a_P$,
$\Delta p$, and $\Delta q$ determined as  stochastic parameters sampled
at an interval of five days and exponentially correlated with a
correlation time of 200 days.   Each interval was fit independently, but
with information on the spacecraft  state (position and velocity) carried
forward from one  interval to the next.   Various correlation times,
0-day, 30-day, 200-day, and 400-day, were investigated.   The {\it a
priori} error and process noise on  $\Delta p$ and $\Delta q$ were set
equal to 0, 5, and 10 $\mu$rad in separate runs, but only the 10
$\mu$rad case removed the annual term.  This value is at least 
three orders of magnitude too  large a deviation when compared to 
the present accuracy of the Earth orbital elements. It is most 
unlikely that such a deviation is  causing the annual term.  Furthermore,
changing to the latest  set of EOP has very little effect on the
residuals.   [We also looked at  variations of the other four Set III
orbital elements, essentially defining the Earth's orbital shape, size,
and longitudinal phase angle.  They had little or no effect on the annual
term.]

%*******************************************************

\paragraph{Solar system modeling:}

We concentrated on Interval III, 
where the spin anomaly is at a minimum and where
$a_P$ is presumably best determined. 
Further, this data was partially taken after the DSN's Block 5 hardware 
implementation from September 1994 to August 1995.  
As a result of this implementation the data is less noisy than before.  
Over Interval III  the annual term is roughly in the form of a sine wave. 
(In fact, the modeling error is not strictly a sine wave. 
But it is close enough to a sine wave for purposes of our error analysis.)
The peaks of the sinusoid are 
centered on conjunction, where the Doppler noise is at a maximum. 
Looking at a CHASMP set of residuals for Interval III,  
we found a 4-parameter, nonlinear,
weighted, least-squares fit to an annual sine wave  with  the parameters  
amplitude $v_{\tt a.t.}=(0.1053\pm 0.0107)$ mm/s, phase $(-5.3^\circ \pm
7.2^\circ$),  angular velocity 
$\omega_{\tt a.t}=(0.0177 \pm 0.0001$) rad/day, and bias ($0.0720 \pm
0.0082$) mm/s.  The weights eliminate data taken inside of  solar
quadrature, and also account for different Doppler integration times
$T_c$ according to
$\sigma = (0.765 {\rm ~mm/s})\,[(60$ s$)/T_c]^{1/2}$.  This rule yields 
post-fit weighted RMS residuals of  0.1 mm/s.     

The amplitude, 
$v_{\tt a.t.}$, and angular velocity, $\omega_{\tt a.t.}$, of  the
annual term results in a small acceleration amplitude of 
$a_{\tt a.t.}=v_{\tt a.t.}\omega_{\tt a.t.} = (0.215 \pm 0.022) \times
10^{-8}$ cm/s$^2$.  We will argue below that the cause is  most likely 
due to errors in the navigation programs' determinations of the 
direction of the  spacecraft's orbital inclination to the ecliptic. 

\begin{figure}[h]
\psfig{figure=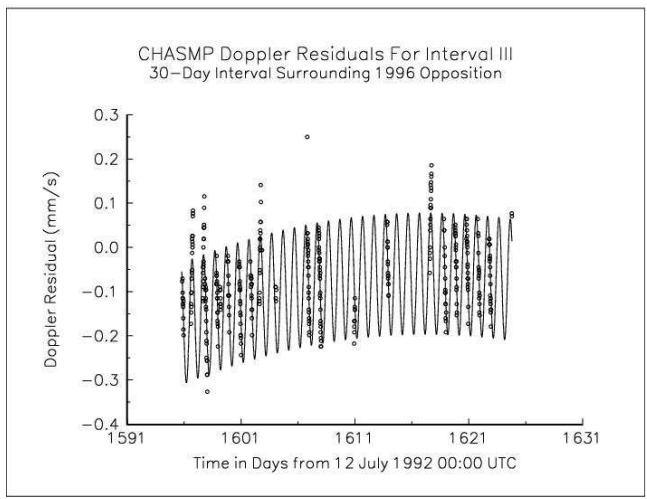,width=85mm}%,height=4.5in}
     \caption
{CHASMP acceleration residuals from 
23 November 1996 to 23 December 1996.  
A clear modeling error is
represented by the solid diurnal curve.  
(An annual term maximum is also seen as a background.) 
   \label{opp96}}
\end{figure}

A similar troubling modeling error exists on a much shorter time scale
that is most likely an error in the spacecraft's orbital inclination to
the Earth's equator.  We looked at CHASMP acceleration residuals over a
limited data interval, from 23 November 1996 to 23 December 1996, 
centered on opposition where the data is least affected by 
solar plasma.  As seen in Figure \ref{opp96}, 
there is a significant diurnal term in the Doppler residuals,
with period approximately equal to  the Earth's sidereal rotation period 
($23^{\rm h}56^{\rm m}04^{\rm s}$.0989 mean solar time). 

%************
%************

After the removal of this diurnal term, the RMS Doppler residuals are
reduced to  amplitude 0.054 mm/s for $T_c = 660$ s  ($\sigma_\nu/\nu = 2.9
\times 10^{-13}$ at $T_c = 1000$ s).   The amplitude of the  diurnal
oscillation  in the fundamental Doppler observable, $v_{\tt d.t.}$,  is
comparable to that in the annual oscillation, $v_{\tt a.t.}$, but the
angular velocity, $\omega_{\tt d.t.}$, is much larger than 
$\omega_{\tt a.t.}$.  This means the  magnitude of the  apparent angular
acceleration, $a_{\tt d.t.}=v_{\tt d.t.}\omega_{\tt d.t.} = (100.1 \pm
7.9) \times 10^{-8}$ cm/s$^2$, is large compared to $a_P$. Because of the
short integration times, $T_c=660$ s, and long observing intervals, $T\sim
1$ yr, the high  frequency, diurnal, oscillation signal  averages out to
less than $0.03\times 10^{-8}$ cm/s$^2$ over a year.  This intuitively
helps to explain why the apparently noisy  acceleration residuals still
yield a  precise value of $a_P$.

Further, all the residuals from CHASMP and  ODP/\emph{Sigma} are
essentially the same. Since ODP and CHASMP both use the same Earth
ephemeris and the same Earth orientation models, this is not surprising. 
This is another check that neither program introduces serious modeling
errors of its own making. 

Due to the long distances from the Sun,  the spin-stabilized attitude
control, the long  continuous  Doppler data history, and the fact that the
spacecraft communication systems utilize coherent  radio-tracking,  the
Pioneers allow for a very sensitive and precise positioning on the sky. 
For some cases, the Pioneer 10 coherent Doppler data provides accuracy
which is even better than that achieved with  VLBI observing natural
sources.  In summary, the Pioneers are simply  much more sensitive
detectors of  a number of solar system modeling errors than other
spacecraft. 

The annual and diurnal terms are very likely different manifestations of
the same modeling problem. The magnitude of the Pioneer 10 post-fit
weighted RMS residuals of  $\approx 0.1$ mm/s, implies that   the
spacecraft angular position on the sky  is known to $\le 1.0$
milliarcseconds (mas). (Pioneer 11, with $\approx 0.18$ mm/s, yields the 
result $\approx 1.75$ mas.)  At their great distances,   the trajectories
of the Pioneers are not gravitationally affected by the Earth. (The 
round-trip light time is now  $\sim 24 $ hours for Pioneer  10.)  This
suggests that the sources of the annual and diurnal terms are both Earth
related. 

Such a modeling problem arises when there are errors in any of the
parameters of the spacecraft orientation with respect to the chosen
reference frame. Because of these errors, the system of equations 
that describes the spacecraft's motion in this reference frame is 
under-determined and its solution  requires non-linear estimation 
techniques.
In addition, the whole estimation process is subject to Kalman filtering
and smoothing methods. Therefore, if there are modeling errors in the
Earth's ephemeris, the orientation of the Earth's spin axis (precession
and nutation), or in the station coordinates (polar motion and length of
day variations), the least-squares process (which  determines best-fit
values of the three direction cosines) will leave small diurnal and
annual components in the Doppler residuals, like those  seen in Figures 
\ref{annualresiduals}-\ref{opp96}.

Orbit determination programs are particularly sensitive to an error in a 
poorly observed direction  \cite{melbourne}. If not corrected for, 
such an error could in principle  significantly affect the overall
navigational accuracy.  In 
the case of the Pioneer spacecraft, navigation was performed using only
Doppler tracking, or line-of-sight observations. The other directions,
perpendicular to the line-of-sight or in the plane of the sky, are poorly
constrained by the data available.  At present, it is infeasible to
precisely parameterize the systematic errors with a physical model. 
That  would have   allowed one to reduce the errors to a level 
below those from the best available ephemeris and Earth orientation
models. A local empirical parameterization is possible, but not a
parameterization over many months.  

We conclude that for both Pioneer 10 and 11, there are  small periodic
errors in solar system modeling that are largely masked by maneuvers and
by the overall plasma noise.  But because these sinusoids are essentially
uncorrelated with the constant $a_P$, they do not present important
sources of systematic error. The characteristic signature of $a_P$ is a
linear drift in the Doppler, not annual/diurnal  signatures  
\cite{myles}.  

%************************************** 
\paragraph{Annual/diurnal mismodeling uncertainty:} 

We now  estimate the annual term contribution to the error budget for 
$a_P$.  First observe that the standard errors for radial velocity, $v_r$, 
and acceleration, $a_r$, are essentially what one would expect for a
linear regression. The caveat is that they are scaled by the root sum of
squares (RSS) of the Doppler error and unmodeled sinusoidal errors, rather
than just the Doppler error. Further, because the error is systematic, it
is unrealistic to assume that the errors for $v_r$ and $a_r$ can be reduced
by a factor 1/$\sqrt{N}$, where $N$ is the number of data points.
Instead,  averaging their correlation matrix over the data interval, $T$,
results in the estimated systematic error of 
\begin{eqnarray}
\sigma_{a_r}^2  = \frac{12}{T^2}~\sigma_{v_r}^2 =
\frac{12}{T^2}~\Big(\sigma_{T}^2 + 
\sigma_{v_{\tt a.t.}}^2+\sigma_{v_{\tt d.t.}}^2\Big). 
\label{syserror}
\end{eqnarray}
$\sigma_{T}=0.1$ mm/s is the Doppler error averaged over $T$ (not the
standard error on a single Doppler measurement).   
$\sigma_{v_{\tt a.t.}}$ and $\sigma_{v_{\tt d.t.}}$ are equal to the
amplitudes of corresponding unmodeled annual and diurnal sine waves
divided by $\sqrt{2}$. The resulting RSS error in radial velocity
determination is about 
$\sigma_{v_r}= (\sigma_{T}^2 + \sigma_{v_{\tt a.t.}}^2+
\sigma_{v_{\tt d.t.}}^2)^{1/2}=0.15$ mm/s for both Pioneer 10 and 11. 
Our four interval values of $a_P$ were determined over
time intervals of longer than a year.  At the same time,  to detect
an annual signature in the residuals, one needs at least half of the
Earth's orbit complete. Therefore, with $T = 1/2$ yr,  Eq.
(\ref{syserror}) results in an acceleration error of
\begin{equation} 
\sigma_{{\tt a/d}} = \frac{0.50~~{\rm mm/s}}{T}
      = 0.32~\times 10^{-8}~{\mathrm{cm/s}}^2.
\label{aderror}
\end{equation} 
We use this number for the systematic error from 
the annual/diurnal term.

%************************10) ERROR BUDGET
%\newpage

\section{\label{budget}ERROR BUDGET AND FINAL RESULT}

It is important to realize  that our experimental observable is  a
Doppler frequency shift, i.e., $\Delta \nu (t)$. [See Figure
\ref{fig:aerospace} and Eq. (\ref{eq:delta_nu}).]  In actual fact it is
a  cycle count.   We \emph{interpret} this as an apparent acceleration
experienced by the spacecraft.  However, it is possible that the
Pioneer  effect  is not  due to a real acceleration. (See
Section~\ref{newphys}.) Therefore, the question arises ``In what units
should we report our errors?''  The best choice is not clear at this
point.  For reasons of clarity we chose units of acceleration.

%\newpage
%====================================

\begin{table*}[ht]
\begin{center}
\caption{Error Budget: A Summary of Biases and Uncertainties.
\label{error_budget}} \vskip 20pt
\begin{tabular}{rlll} \hline\hline
Item   & Description of error budget constituents & 
Bias~~~~~& Uncertainty    \\     &       &
           $10^{-8} ~\rm cm/s^2$ &  $10^{-8} ~\rm cm/s^2$  \\\hline
   &                 &             &         \\
 1 & {\sf Systematics generated external  to the spacecraft:}  && \\
   & a) Solar radiation pressure and mass   &  $+0.03$   & $\pm 0.01$\\
   & b) Solar wind       && $ \pm < 10^{-5}$ \\
   & c) Solar corona      & & $ \pm 0.02$ \\
   & d) Electro-magnetic Lorentz forces     &&  $\pm < 10^{-4}$  \\
   & e) Influence of the Kuiper belt's gravity     && $\pm 0.03$\\
   & f) Influence of the Earth orientation       && 
        $\pm 0.001$ \\
   & g) Mechanical and phase stability of DSN antennae    
        &&  $\pm < 0.001$ \\
   & h) Phase stability and clocks      &&  $\pm <0.001$ \\
   & i) DSN station location      &&  $\pm < 10^{-5}$ \\
   & j) Troposphere and ionosphere      &&  $\pm < 0.001$ \\[10pt]
 2 & {\sf On-board generated systematics:} &&         \\
   & a) Radio beam reaction force         & $+1.10$&$\pm 0.11$ \\
   & b) RTG heat reflected off the craft      &  $-0.55$&$\pm 0.55$ \\
   & c) Differential emissivity of the RTGs         & &  $\pm 0.85$ \\
   & d) Non-isotropic radiative cooling of the spacecraft && $\pm 0.48$\\
   & e) Expelled Helium  produced  within the RTGs 
           &$+0.15$ & $\pm 0.16$   \\
   & f) Gas leakage           &         &  $\pm 0.56$   \\
   & g) Variation between spacecraft determinations  
           &   $+0.17$  & $\pm 0.17$  \\[10pt]
 3 & {\sf Computational systematics:}          &&         \\
   & a) Numerical stability of least-squares estimation & &$\pm0.02$ \\
   & b) Accuracy of consistency/model tests          & &$\pm0.13$ \\
   & c) Mismodeling of maneuvers   &        &  $\pm 0.01$          \\
   & d) Mismodeling of the solar corona      & &$\pm 0.02$ \\
   & e) Annual/diurnal terms                 & & $\pm 0.32$  \\[10pt]  
\hline
   &                              &&         \\
   & Estimate of total bias/error     & $+0.90$& $\pm 1.33$       \\
   &                              &&         \\
\hline\hline
\end{tabular} 
\end{center} 
\end{table*}

%==========================================================  

The tests documented in the preceding sections  have considered various 
potential sources of systematic error. The results of these tests are 
summarized in Table \ref{error_budget}, which serves as a systematic
``error  budget.'' This budget is useful both for evaluating the accuracy
of our solution for $a_P$ and also for guiding  possible future efforts
with other spacecraft.   In  our case it actually is hard to totally
distinguish ``experimental'' error from ``systematic error.''  (What
should a drift in the atomic clocks be called?)  Further, there is the 
intractable mathematical problem of how to handle combined experimental
and systematic errors.  In the end we have decided to treat them all in
a least squares {\it uncorrelated} manner.    

The results of our analyses  are  summarized in Table
\ref{error_budget}.  There are two columns of results.  The first gives
a bias, $b_P$, and the second gives an uncertainty, $\pm \sigma_P$. The
constituents of the error budget are listed separately in three
different categories: 1) systematics generated external  to the
spacecraft; 2) on-board generated systematics, and 3)  computational
systematics. Our final result then will become some average
\begin{equation} a_P = a_{P({\tt exper)}}~ +   b_P ~\pm \sigma_P,
\end{equation} where,  from Eq. (\ref{pio10lastresult}),
$a_{P({\tt exper)}} = (7.84\pm 0.01)  \times 10^{-8}$ cm/s$^2$.

The least significant factors of our error budget are in the first group
of effects, those external to the spacecraft.  From the table one sees
that some are near  the limit of contributing.   But in totality,  they
are insignificant.  

As was expected,  the on-board generated systematics are the 
largest contributors to our total error budget. All the important
constituents are listed in the second group of effects in Table
\ref{error_budget}.  Among these effects, the radio beam reaction force 
produces the largest bias to our result, $1.10\times 10^{-8}$ cm/s$^2$.   
It makes the Pioneer effect larger.  The largest bias/uncertainty 
is from RTG heat reflecting off the spacecraft.  We argued for 
an effect as large as $(-0.55 \pm 0.55) \times 10^{-8}$ cm/s$^2$.  
Large uncertainties  also come from differential emissivity of 
the RTGs, radiative cooling, and gas leaks, $\pm 0.85$, $\pm 0.48$, and 
$\pm 0.56$, respectively, $\times 10^{-8}$ cm/s$^2$.
The computational systematics  
are listed in the third group of  Table \ref{error_budget}. 

Therefore, our final 
value for $a_P$ is 
\begin{eqnarray}
a_P &=& (8.74  \pm 1.33)  \times 10^{-8}~{\rm cm/s}^2
\nonumber\\
    &\sim& (8.7  \pm  1.3)  \times 10^{-8}~{\rm cm/s}^2.
\end{eqnarray}
The effect is clearly significant  and remains to be explained.  

%************** POSSIBLE PHYSICAL ORIGINS ***************** 
%\newpage

\section{\label{newphys}POSSIBLE PHYSICAL ORIGINS OF THE SIGNAL}

%****************

\subsection{A new manifestation of known physics?}     
\label{sec:know}

With the anomaly still not accounted for, possible 
effects from applications of known physics have been advanced.  
In particular, Crawford \cite{crawford} suggested a novel new effect: 
a gravitational frequency shift of the radio signals that is 
proportional to the distance to the spacecraft and the density of 
dust in the intermediate medium.  
In particular, he has argued that the gravitational interaction of
the S-band radio signals with the interplanetary dust  may be responsible
for  producing an  anomalous acceleration similar to that seen by the
Pioneer spacecraft. The effect of this interaction is a frequency shift
that is proportional to the distance and the square root of the density of
the medium in which it travels. 
Similarly,   Didon, Perchoux, and Courtens \cite{courtens} 
proposed that the effect comes from resistance of the spacecraft 
antennae as they transverse the interplanetary dust.  This is 
related to more general ideas that an asteroid or comet belt, with its
associated dust, might cause the effect by gravitational interactions 
(see Section \ref{sec:kuiper}) or resistance to dust particles. 

However, these ideas have problems with known properties of the 
interplanetary medium that were outlined in Section \ref{sec:kuiper}.     
In particular, infrared observations  rule out more than 0.3 Earth mass
from Kuiper Belt dust in the  trans-Neptunian region
\cite{backman,teplitzinfra}.   Ulysses and Galileo measurements in the
inner solar system find  very few dust grains in the $10^{-18}-10^{-12}$
kg range \cite{dust}.   The density varies greatly, up and down, within
the belt  (which precludes a constant force) and, in any event, the density 
is  not large enough to produce a gravitational acceleration on the order 
of $a_P$ \cite{malhotra}-\cite{liudust}. 

One can also speculate that there is some unknown  interaction of the radio
signals with the solar wind.  An experimental answer  could be given with
two different transmission frequencies.   Although the main communication
link on the Ulysses mission  is S-up/X-down mode, a small fraction of the
data is S-up/S-down.  We had hoped to utilize this option in  further
analysis. However, using them in our attempt to study a possible frequency
dependent  nature of the anomaly, did not provide any useful results.  This
was in part due to the fact that X-band data (about 1.5~\% of the whole
data available) were  taken only in the close proximity to the Sun, thus
prohibiting the study  of a possible frequency dependence of the anomalous
acceleration.

%***************

\subsection{Dark matter or modified gravity?}
\label{sec:DMgr}

It is interesting to speculate on the unlikely 
possibility that the origin of the anomalous signal 
is new physics \cite{photon}.  This is true 
even though the  probability is that some ``standard physics''
or some as-yet-unknown systematic will be found to explain this  
``acceleration.''  The first paradigm is obvious. 
``Is it dark matter or a modification of gravity?''  
Unfortunately, neither easily works. 

If the cause is dark matter, it is hard to understand.   A
spherically-symmetric distribution of matter which goes as
$\rho \sim r^{-1}$ produces a constant acceleration {\it inside} 
the distribution.  To produce our anomalous acceleration even  only out to
50 AU would require the total dark matter to be greater than $3 \times
10^{-4}  M_\odot$. But this  is in conflict with the
accuracy of the ephemeris, which allows only  of order a few times
$10^{-6} M_\odot$ of dark matter even within the  orbit of Uranus
\cite{ephem}. (A 3-cloud neutrino model also did not solve 
the problem \cite{jgscold}.)

Contrariwise, the most commonly studied 
possible modification of gravity (at various scales) is an 
added Yukawa force \cite{physrep}. Then the 
gravitational potential is 
\begin{equation}
V(r) = -\frac{GMm}{(1+\alpha)r}\left[1 +\alpha e^{-r/\lambda}\right],
  \label{V}
\end{equation}
where $\alpha$ is the new coupling strength relative to
Newtonian gravity, and $\lambda$ is the new force's range.
Since the radial force is $F_r = -d_r V(r) =ma$,
the power series for the acceleration 
yields an inverse-square term, no inverse-$r$ term, then a constant
term.   Identifying this last term as the Pioneer acceleration yields 
\begin{equation}
a_P = -\frac{\alpha {a_1}}{2(1+\alpha)}
        \frac{r_1^2}{\lambda^2},   \label{solution}
\end{equation}
where $a_1$ is the Newtonian
acceleration at distance $r_1 =1$ AU.  (Out to 65 AU there
is no observational evidence of an $r$ term in the acceleration.) 
Eq. (\ref{solution}) is the solution curve; for example, 
$\alpha = -1 \times 10^{-3}$ for $\lambda = 200$ AU.  

It is also of interest to consider some of the recent proposals to 
modify gravity, as alternatives to dark matter \cite{milgrom}-\cite{mil}. 
Consider Milgrom's proposed modification of gravity \cite{mil}, 
where the gravitational acceleration of a massive body is 
$a \propto 1/r^2$ for some constant $a_0 \ll a$ and 
$a \propto 1/r$ for $a_0 \gg a$.  Depending on the 
value of $H$, the Hubble constant, $a_0 \approx a_P$!  Indeed, as a 
number of people have noted, 
\begin{equation}
    a_H = cH \rightarrow 8 \times 10^{-8}~ {\rm cm/s}^2,  \label{hubble}
\end{equation}
if $H = 82$ km/s/Mpc. 

Of course, there are (fundamental and deep) theoretical problems if 
one has a new force of  the phenomenological types of those above.
Even so, the deep space data piques our curiosity.  In fact,  Capozziello 
et al. \cite{Capozzielloetal} note  the  Pioneer anomaly 
in their discussion of astrophysical structures as manifestations of 
Yukawa coupling scales.  This ties into the above discussion.  

However, any universal gravitational explanation 
for the Pioneer effect comes up against a hard experimental wall. 
The anomalous acceleration  is too
large to have gone undetected in planetary orbits, particularly
for Earth and Mars.  NASA's Viking mission
provided radio-ranging measurements 
to an accuracy of about 12 m \cite{reasenberg,mg6}.
If a planet experiences a small, anomalous, radial acceleration,
$a_A$,  its orbital radius $r$ is
perturbed  by  
\begin{equation}
\Delta r =-\frac{{\it l}^6 a_A}{(GM_\odot)^4} 
         \rightarrow  - \frac{r~ a_A}{a_N} , 
  \label{deltar}
\end{equation}
where {\it l} \, is the orbital angular momentum per unit mass and $a_N$
is the Newtonian acceleration at $r$. 
(The right value in Eq. (\ref{deltar}) holds in the circular orbit limit.)

\indent For  Earth and Mars, $\Delta r$ is about 
$-21$  km and $-76$ km.  However, the Viking data determines the
difference  between the Mars and Earth orbital radii to about a 100 m
accuracy, and  their sum  to an accuracy of about 150 m.  The Pioneer
effect is not seen.  

Further, a perturbation in $r$ produces a perturbation to the
orbital angular velocity of 
\begin{equation}
\Delta \omega =  \frac{2{\it l}a_A}{GM_\odot}
   \rightarrow \frac{2 \dot{\theta}~ a_A}{a_N}.
\end{equation} 
The determination of the
synodic angular velocity $(\omega_E - \omega_M)$ is accurate to 7 parts
in 10$^{11}$, or to about 5 ms accuracy in synodic period. The
only parameter that could possibly mask the spacecraft-determined 
$a_R$ is  $(GM_\odot)$. But a large error here 
would cause inconsistencies with the overall planetary ephemeris
\cite{ephem,Standish92}.  [Also, there would be a problem with the 
advance of the perihelion of Icarus \cite{sanmil}.] 

We conclude that the Viking ranging data limit any unmodeled radial
acceleration acting on Earth and Mars to no more than
$0.1 \times 10^{-8}$ cm/s$^2$. Consequently, if the anomalous radial
acceleration acting on spinning spacecraft is gravitational in origin, it
is {\it not} universal.  That is, it must affect bodies in the 1000 kg
range more than bodies of planetary size by a factor of 100 or more.  
This would be a strange violation of the Principle of Equivalence 
\cite{pe}. 
(Similarly, the $\Delta \omega$ results rule out the universality of 
the $a_t$ time-acceleration model.  In the age of the universe, $T$,  
one would have $a_t T^2/2 \sim 0.7~T$.)

A new dark matter model was recently proposed  by Munyaneza and Viollier
\cite{MunyanezaViollier} to explain the Pioneer anomaly. The dark matter is
assumed to be gravitationally clustered around the Sun in the form of a
spherical halo of a degenerate gas of heavy neutrinos. However,  although
the resulting mass distribution is consistent with constraints on the mass
excess within the orbits of the outer planets previously mentioned,   it
turns out that the model fails to produce a viable mechanism for the
detected anomalous acceleration.

%********

\subsection{New suggestions stimulated by the Pioneer effect}
\label{sec:neww}

Due to the fact that the size of the anomalous acceleration is of 
order $cH$, where $H$ is the Hubble constant
(see Eq. (\ref{hubble})), the Pioneer results have stimulated 
a number of new physics suggestions.  
For example, Rosales and S\'anchez-Gomez \cite{rosales} 
propose that $a_P$ is due to a local curvature 
in light geodesics in the expanding spacetime universe.  
They argue that the Pioneer effect represents a new cosmological
Foucault experiment, since the solar system coordinates are not
true inertial coordinates with respect to the expansion of the universe. 
Therefore, the Pioneers are  
mimicking the role that the  rotating Earth plays in Foucault's
experiment. Therefore, in  this picture the 
effect is not a ``true physical effect'' and a
coordinate transformation to the  co-moving cosmological coordinate frame
would entirely remove the Pioneer effect. 

\indent From a  similar viewpoint, Guruprasad \cite{guru} finds 
accommodation for the constant term while trying to explain the annual term
as a tidal  effect on the physical structure of the spacecraft itself. 
In particular, he suggests that the deformations of the physical
structure of the spacecraft (due to external factors such as the 
effective solar and galactic tidal forces) combined
with the spin of the spacecraft are directly responsible for the detected
annual anomaly. Moreover,  he proposes a   hypothesis of the planetary
Hubble's flow and  suggests that Pioneer's anomaly does not contradict the
existing planetary data, but supports his new theory of relativistically
elastic space-time. 

{\O}stvang \cite{ostvang} further exploits the fact that the gravitational
field of the solar system is not static with respect to the cosmic
expansion.  He does  note, however,  that in order to
be acceptable, any non-standard explanation of the effect 
should follow from a general
theoretical framework.  
Even so,   {\O}stvang still presents quite a radical  model.
This model advocates the use of an expanded PPN-framework that includes a 
direct effect on local scales due to the cosmic space-time expansion. 

Belayev \cite{belayev} considers a Kaluza-Klein  model in 5 
dimensions  with a
time-varying  scale factor for the compactified fifth dimension. His
comprehensive analysis led to the conclusion that a variation of the physical
constants on a cosmic time scale is responsible for the appearance of the
anomalous acceleration observed in the Pioneer 10/11 tracking data. 

Modanese  \cite{modan} considers the effect 
of a scale-dependent cosmological term in the gravitational action.
It turns out that, even in the case of a static spherically-symmetric 
source,  the external solution of his modified gravitational field
equations  contains  a non-Schwartzschild-like component that depends on
the size of the test particles. He  argues that this additional term may
be  relevant to the observed  anomaly. 

Mansouri, Nasseri and Khorrami
\cite{MansouriNasseriKhorrami}  argue that there is an effective time
variation in the Newtonian gravitational constant that in turn may be
related to the anomaly. In particular, they consider the time evolution
of $G$ in a model universe with variable space dimensions. When analyzed
in the low energy limit, this theory produces a result that may be
relevant to the long-range acceleration discussed here. A similar
analysis  was  performed  by Sidharth
\cite{Sidharth}, who also  discussed cosmological models with a
time-varying Newtonian gravitational constant.

Inavov \cite{ivanov} suggests that the Pioneer anomaly is 
possibly the
manifestation of a superstrong interaction of photons with single
gravitons that form a dynamical background   in the solar
system.  Every gravitating body would experience a deceleration
effect from such a background  with a magnitude proportional to Hubble's
constant. Such a deceleration would  produce an observable effect on a 
solar system scale. 
 
All these ideas produce predictions that are close to Eq. (\ref{hubble}),
but they certainly must be judged against discussions in  the following
two  subsections.    

%********************  

In a different framework, Foot and Volkas \cite{foot}, suggest the anomaly
can be explained if there is mirror matter of mirror dust in the solar
system.   this could produce a drag force and not violate solar-system mass
constraints.  

Several scalar-field ideas have also appeared.  Mbelek and Lachi\`eze-Rey 
\cite{rey}  have a model based on a long-range scalar field,
which also  predicts an oscillatory decline in $a_P$ beyond about 100
AU.   This model does explain the fact that $a_P$ stays  approximately
constant for a long period  (recall that Pioneer 10 is now past  
70 AU).  From a similar standpoint Calchi Novati et al. \cite{novati}
discuss a weak-limit, scalar-tensor extension to the standard gravitational
model.  However,  before any of these proposals can be seriously considered
they must explain  the  precise timing data for millisecond binary pulsars,
i.e.,  the gravitational radiation indirectly observed in PSR 1913+16 by 
Hulse and Taylor \cite{millipulsar}.  Furthermore, there should be 
evidence of a distance-dependent scalar field  if it is uniformly coupled
to  ordinary matter.

Consoli and Siringo  \cite{consoli_siringo} and Consoli \cite{consoli}
consider the Newtonian regime of  gravity to be the  long wavelength
excitation of a  scalar  condensate from electroweak symmetry breaking.
They speculate that  the self-interactions of the condensate could be the
origins of both  Milgrom's inertia modification 
\cite{milgrom,mil} and also of the Pioneer effect.  

Capozziello and Lambiase \cite{CapozzielloLambiase} argue that 
flavor oscillations of neutrinos in the Brans-Dicke theory of gravity may
produce a quantum mechanical phase shift of 
neutrinos. Such a  shift would 
produce observable effects on astrophysical/cosmological length and  
time scales. 
In particular, it results in a variation of the
Newtonian gravitational constant and, in the 
low energy limit, might be relevant to our study.

Motivated by the work of Mannheim \cite{mann,mann2},
Wood and Moreau \cite{moreau} investigated the theory of conformal
gravity with dynamical mass generation.  They argue that the Higgs scalar 
is a feature of the theory that cannot be ignored.  In particular, within 
this framework  they find one can reproduce the standard gravitational
dynamics and  tests within the solar system, and yet the Higgs fields 
may leave room for the Pioneer effect on small bodies.  

In summary,  as highly speculative as all these ideas are, it can be seen 
that at the least the  Pioneer anomaly  is influencing the 
phenomenological discussion of modern gravitational physics 
and  quantum cosmology \cite{BertolamiNunes}.

%*************************************** 

\subsection{Phenomenological time models}
\label{sec:timemodel}

Having noted the relationships 
$a_P = c~ a_t$ of Eq. (\ref{asubt})  
and that of Eq. (\ref{hubble}), we were motivated to try
to  think of any (purely phenomenological) ``time'' distortions  that
might fortuitously fit   the CHASMP Pioneer results shown in Figure 
\ref{fig:aerospace}.    In other words, are Eqs. (\ref{hubble}) and/or
(\ref{asubt}) indicating something?  Is there any evidence that some kind
of ``time acceleration''  is being seen?  

The Galileo and Ulysses spacecraft radio tracking data 
was especially useful.  We examined numerous ``time'' models searching 
for any (possibly radical) solution.   It was thought that these  
models would contribute to the definition of the different time scales
constructed on the basis of Eq. (\ref{eq:time}) and discussed in the
Section \ref{sec:time_scales}. The nomenclature of the standard time
scales  \cite{Moyer81}-\cite{exp_cat} was phenomenologically extended in our
hope to find a desirable quality of the trajectory solution for the
Pioneers.

In particular we considered:  

i) {\sf Drifting Clocks.}
This model adds a constant  acceleration term to the Station Time 
({\tt ST}) clocks, i.e., in the {\tt ST-UTC} (Universal Time
Coordinates)  time transformation. The model may be given as follows:
\begin{equation}
\Delta{\tt ST}={\tt ST}_{\tt received}-{\tt ST}_{\tt sent}
~~\rightarrow ~~\Delta{\tt ST}+\frac{1}{2}a_{\tt clocks}
\cdot\Delta{\tt ST}^2
\end{equation}
\noindent where ${\tt ST}_{\tt received}$ and ${\tt ST}_{\tt sent}$
are the atomic proper times of sending and receiving the signal by a DSN
antenna. The model fit Doppler well for Pioneer 10, Galileo,  and Ulysses
but failed to model range data for  Galileo and Ulysses.  

ii)  {\sf Quadratic Time Augmentation.} 
This model adds a quadratic-in-time augmentation
to the {\tt TAI-ET} (International Atomic Time -- Ephemeris Time) 
time transformation, as follows
\begin{equation}
{\tt ET} 
~~\rightarrow ~~ {\tt ET}+\frac{1}{2}a_{\tt ET}\cdot{\tt ET}^2.
\end{equation} 
The model fits Doppler fairly well but range very badly.  

iii)   {\sf Frequency Drift.}  
This model adds a constant frequency drift to the 
reference S-band carrier frequency:
\begin{equation}
\nu_{\tt S-band}(t)=\nu_{0}
\Big(1+\frac{a_{\tt fr.drift}\cdot{\tt TAI}}{c} \Big).
\end{equation}
 The model also fits Doppler well but again fits 
range poorly. 

iv) {\sf Speed of Gravity}. This model adds a ``light time'' delay  
to the actions of the Sun and planets upon the spacecraft:
\begin{equation}
v_{\tt grav}=c
\Big(1+\frac{a_{\tt sp.grav}
\cdot|\vec{r}_{\tt body}-\vec{r}_{\tt Pioneer}|}{c^2}
\Big).
\end{equation}
The model fits
Pioneer 10 and Ulysses well.  But the Earth flyby of Galileo  fit was
terrible, with Doppler  residuals as high as 20 Hz. 

All these models were rejected due either to poor fits or to inconsistent 
solutions among spacecraft.

%*****************

\subsection{Quadratic in time model}

There was one model of the above type that was especially
fascinating.  This model adds a quadratic in time term to the 
light time as seen by the DSN station.  Take any labeled time 
${\tt T}_a$ to be 
\begin{equation}
{\tt T}_a = t_a - t_0 \rightarrow  t_a - t_0
  + \frac{1}{2}a_t\left(t_a^2 - t_0^2\right).
\end{equation}
Then the light time is 
\begin{eqnarray}
\Delta{\tt TAI}&=&{\tt TAI}_{\tt received}-{\tt TAI}_{\tt sent} 
~~\rightarrow \nonumber\\
&&\hskip -30pt \rightarrow~~
\Delta{\tt TAI}+\frac{1}{2}a_{\tt quad}\cdot
\Big( {\tt TAI}_{\tt received}^2-{\tt TAI}_{\tt sent}^2\Big).
\label{eq:aqt}
\end{eqnarray}
\noindent It mimics a line of sight acceleration of the spacecraft, and
could be thought of as an {\it expanding space}  model.  Note that $a_{\tt
quad}$ affects only the data. This is in contrast to the $a_t$ of
Eq. (\ref{asubt}) that affects both the data and the trajectory.

This model fit both Doppler and range very well. Pioneers 10 and
11, and Galileo have similar solutions although Galileo solution is 
highly correlated with solar pressure; however, the  range coefficient 
of the quadratic is negative for the Pioneers and Galileo while positive
for Ulysses.  Therefore we originally rejected the model because of the
opposite  signs of the coefficients.  But when we later appreciated  that
the Ulysses anomalous acceleration is dominated by gas leaks  (see
Section \ref{sec:AUlysses}), which makes the  different-sign coefficient of
Ulysses meaningless, we reconsidered it.    

The fact that the Pioneer 10 and 11, Galileo, and  Ulysses are
spinning spacecraft whose spin axis are periodically adjusted so  as to
point towards Earth turns out to make  the quadratic in time model and the
constant spacecraft acceleration model highly correlated and therefore very
difficult to separate.  The quadratic in time model produces residuals only
slightly ($\sim20\%$) larger than the constant spacecraft acceleration
model. However, when estimated together with  no {\it a priori} input
{ i.e.}, based only the tracking data,  even though the correlation
between the two models is 0.97,  the  value $a_{\tt quad}$ determined for
the quadratic in time model is zero while the value for the constant
acceleration model $a_P$ remains the same as before.

The orbit determination process clearly prefers the constant 
acceleration model, $a_P$, over that the quadratic in time model, 
$a_{\tt quad}$ of Eq.~(\ref{eq:aqt}). This implies that a real
acceleration is being observed and not a pseudo acceleration. 
We have not rejected this model as it may be too simple in that 
the motions of the spacecraft and the Earth may need to be included to
produce a true expanding space model. Even so, the numerical relationship
between the Hubble constant and $a_P$, which many people have observed (cf.
Section \ref{sec:neww}), remains an interesting conjecture.

%****************************12) CONCLUSIONS
%\newpage

\section{\label{disc}CONCLUSIONS}

In this paper we have discussed the equipment, theoretical models, and 
data analysis techniques involved in obtaining the anomalous Pioneer 
acceleration $a_P$.  We have also reviewed the possible systematic  errors
that could explain this effect.  These included computational errors as
well as experimental systematics, from systems both  external to and
internal to the spacecraft.  Thus, based on further data for the Pioneer
10 orbit determination   (the extended data spans 3 January 1987 to 22
July 1998)  and more detailed studies of  all the systematics,  we can
now give  a total error  budget for our analysis and a  latest result of  
$a_P = (8.74 \pm 1.33) \times 10^{-8}$ cm/s$^2$.

This investigation was possible because modern radio tracking techniques  
have provided us with the means  to investigate gravitational interactions
to  an accuracy  never before possible.   With these techniques,
relativistic  solar-system  celestial mechanical experiments using the
planets and  interplanetary spacecraft provide critical new information. 

Our investigation has emphasized that effects that previously   thought  to
be insignificant, such as rejected thermal radiation or mass 
expulsion, are now within (or near) one order of magnitude of possible 
mission requirements. This has unexpectedly emphasized the need to
carefully  understand all systematics to this level.  

In projects proposed for the near future, such as a Doppler measurement of 
the solar gravitational deflection  
using the  Cassini spacecraft \cite{GAB} and
the  Space Interferometry  Mission \cite{sim}, navigation
requirements  are more stringent than those for current spacecraft. 
Therefore, all the  effects we have discussed will  have to be well-modeled
in  order to obtain  sufficiently good trajectory solutions. That is, a
better understanding of the nature of these extra small forces  will be
needed to achieve the stringent navigation requirements for these  
missions.  

Currently, we find no mechanism or theory that explains the  anomalous
acceleration. What we can say with some confidence is that the  anomalous
acceleration is a line of sight  constant acceleration  of the  spacecraft
toward the Sun \cite{sunearth}.  Even though  fits to the Pioneers appear
to match     the noise level of the data, in reality the fit levels are as
much as  50 times above the fundamental noise limit of the data. 
Until more is known, we must admit that the most  likely cause of this
effect is an  unknown systematic.  (We ourselves are divided as to whether
``gas leaks'' or ``heat'' is this  ``most likely cause.'')  

The arguments for ``gas leaks'' are:  i) All spacecraft experience a gas
leakage at some level.  ii) There is  enough gas available to cause the
effect.  iii) Gas leaks require no new physics.  However, iv) it is 
unlikely that the two Pioneer spacecraft would have gas  leaks at 
similar rates,  over the entire data interval, 
especially when the valves have  been used for so many maneuvers. 
[Recall also that one of the Pioneer 11 thrusters
became inoperative soon after launch.  (See Section \ref{sec:prop}.)]
v) Most importantly, it would require that these gas leaks be precisely
pointed towards the front \cite{rearfront} of the spacecraft so as  not
to cause a large spin-rate changes.  But vi) it could still be true anyway.  

The main arguments for ``heat'' are: i)  There is so much heat available
that a small amount of the total could cause the effect. ii) In deep space
the spacecraft will be in approximate thermal equilibrium. The heat should
then be emitted at an approximately constant rate,  deviating from a
constant only because of   the slow exponential decay of the Plutonium heat
source.  It is hard to resist the notion that this heat somehow must be the
origin of the effect.  However, iii) there is no solid explanation in hand
as to how a specific heat mechanism could work.  Further, iv)  the 
decrease in the heat supply over time should have been seen by now.  

Further experiment and analysis is obviously needed to resolve this
problem.  

On the  Pioneer 10  experimental front, 
there now exists data up to July 2000.  Further, there exists archived
high-rate data from 1978 to the beginning of our data arc in Jan 1987 
that was not used in this analysis.  Because this early data originated
when the Pioneers were much closer in to the Sun, greater effort would be
needed to perform the data analyses and to model the systematics.
 
As Pioneer 10 continues to recede into
interstellar space, its signal is becoming dimmer.  Even now, the return
signal is hard to detect with the largest DSN antenna. However, with
appropriate instrumentation, the 305-meter antenna of the  Arecibo
Observatory in Puerto Rico will be able to detect Pioneer's signal for a
longer time. If contact with Pioneer 10 can be maintained with conscan
maneuvers,  such further extended data would be very useful, since the 
spacecraft is now so far from the Sun.  

Other spacecraft can also  be used in the study of $a_P$.   
The radio Doppler and range data from the Cassini mission could offer a
potential contribution. This mission was launched on 15 October 1997. 
The potential data arc will be the cruise phase from after the Jupiter flyby 
(30 December 2000) to the vicinity of 
Saturn (just before the Huygens probe release) in July 2004.  
Even though the Cassini spacecraft is in
three-axis-stabilization  mode, using on-board  active thrusters, it was
built with very sophisticated radio-tracking capabilities, with  
X-band being the main navigation frequency.  (There will also be  
S- and K-band links.)   Further, during much of the 
cruise phase, reaction wheels will be used for
stabilization instead of thrusters.  Their use will  aid 
relativity experiments at solar conjunction 
and gravitational wave experiments at solar opposition.  (Observe, however, 
that the relatively large systematic from the close in Cassini RTGs will 
have to be accounted for.)

Therefore,  Cassini  could yield important orbit data, 
independent of the Pioneer hyperbolic-orbit data.    
A similar opportunity may exist, out of the plane of the
ecliptic,  from the proposed Solar Probe mission.  Under consideration
is  a low-mass  module to be ejected  during solar flyby. 
On a longer time scale, 
the reconsidered Pluto/Kuiper  mission (with arrival at Pluto 
around  2029) could  eventually provide 
high-quality data from very deep space. 

All these missions might help test our current 
models of precision navigation and also provide a new test for the
anomalous $a_P$.  In particular, 
we anticipate that, given our analysis of the Pioneers, in the 
future precision orbital analysis may concentrate more on systematics. 
That is, data/systematic modeling analysis may be assigned more 
importance relative to the astronomical modeling techniques people have  
concentrated on for the past 40 years \cite{tuck}-\cite{bartlett}.

Finally, we  observe that if no convincing explanation is  to
be obtained, the possibility remains that the effect is real.  
It could even be related to cosmological quantities, as 
ha]s been intimated.
[See Eq. (\ref{asubt}) and  Sec. \ref{newphys}, especially the 
text around Eq. (\ref{hubble}).] 
This possibility necessitates a cautionary note on phenomenology:  At this
point in time, with the limited results available, there is a
phenomenological equivalence between the $a_P$ and $a_t$ points of view. 
But somehow, the choice one makes affects one's outlook and 
direction of attack.  If
one has to consider new physics one should be open to both points of
view.  In the unlikely event that there is  new physics, one does not
want to miss it because one  had the wrong mind set.

%************************%\section*{ACKNOWLEDGEMENTS}

\begin{acknowledgments}

Firstly we  must acknowledge the many people who have helped us with
suggestions,  comments, and constructive criticisms.   Invaluable
information on the history and status of Pioneer 10  came from Ed Batka,
Robert Jackson, Larry Kellogg, 
Larry Lasher, David Lozier,  and Robert Ryan.  E. Myles
Standish critically reviewed  the manuscript and provided a number of
important insights, especially on time scales, solar system  dynamics
and planetary data analysis.   We also thank John E. Ekelund, Jordan
Ellis,  William Folkner, Gene L. Goltz, William E. Kirhofer,  Kyong J.
Lee, Margaret Medina,  Miguel Medina, Neil Mottinger, George W. Null,
William L. Sjogren,  S. Kuen Wong, and Tung-Han You of JPL for their aid 
in obtaining and understanding  DSN Tracking Data. Ralph McConahy
provided us with very useful information on the history and current
state of the DSN complex at Goldstone.  R. Rathbun and A. Parker of TRW
provided information  on the mass  of the Pioneers.   S. T. Christenbury
of Teledyne-Brown, to whom  we are very grateful, supplied us with
critical  information on the RTGs.  Information was also supplied by  G.
Reinhart of LANL, on the RTG fuel pucks, and by C. J. Hansen of JPL,  
on the operating characteristics of the Voyager image cameras.  We thank
Christopher S. Jacobs of JPL for encouragement and  stimulating
discussions on present VLBI capabilities.    Further guidance and
information were provided by  John W. Dyer, Alfred S. Goldhaber, Jack G.
Hills, Timothy P. McElrath,  Irwin I. Shapiro, Edward J. Smith, and
Richard J. Terrile.  Edward L. Wright sent useful observations on the
RTG emissivity analysis. We also thank Henry S. Fliegel, Gary B. Green, 
and Paul Massatt  of The Aerospace Corporation  for suggestions and
critical reviews of the present manuscript. 

This work was supported by the 
Pioneer Project, NASA/Ames Research Center, and was performed at the Jet
Propulsion Laboratory, California Institute of Technology, under contract
with the  National Aeronautics and Space Administration.
P.A.L. and A.S.L. were supported by a grant from NASA through the
Ultraviolet, Visible, and  Gravitational Astrophysics Program. M.M.N.
acknowledges support  by the U.S. DOE.  

Finally, the collaboration especially acknowledges the contributions of 
our friend and colleague, Tony Liu, who passed away while the manuscript
was nearing completion. 

\end{acknowledgments}

%****************************  APPENDIX ***********

\newpage

%*************************************************

\appendix
\section{APPENDIX}
\label{appendix}

In Table \ref{poeas00479} we give the 
hyperbolic orbital parameters for Pioneer 10 and Pioneer 11 at
epoch 1 January 1987, 01:00:00 {\tt UTC}. 
The semi-major axis is $a$, 
$e$ is the eccentricity, 
$I$ is the inclination, 
$\Omega$ is the longitude of the ascending node, 
$\omega$ is the argument of the perihelion, 
$M_0$ is the mean anomaly, 
$f_0$ is the true anomaly at epoch,   
and $r_0$ is the heliocentric radius at the epoch. 
The direction cosines of the spacecraft
position for the axes used are 
$(\alpha, \,\beta, \,\gamma)$.   
These direction cosines and angles 
are referred to the mean equator and equinox of J2000. The ecliptic
longitude $\ell_0$ and latitude $b_0$ are also listed for an obliquity of
23$^\circ$26$^{'}$21.$^{''}$4119. The numbers in parentheses denote
realistic standard errors in the last digits.  

%************************Table  ************

\begin{table}[h]
\begin{center}
\caption{Orbital parameters for Pioneer 10 and Pioneer 11 at
epoch 1 January 1987, 01:00:00 {\tt UTC}. 
\label{poeas00479}} 
\vskip 20pt
\begin{tabular}{|l|r|r|}\hline\hline
Parameter & Pioneer 10   & Pioneer 11 \\   \hline
$a$ [km]    & $-1033394633(4)$ & $-1218489295(133)$  \\[1pt]
$e$  &  $1.733593601(88)$  &  $2.147933251(282)$ \\[1pt]
$I$ [Deg] &  $26.2488696(24)$  &  $9.4685573(140)$ \\[1pt]
$\Omega$ [Deg]  &  $-3.3757430(256)$  &  $35.5703012(799)$  \\[1pt]
$\omega$ [Deg] &  $-38.1163776(231)$  &  $-221.2840619(773)$  \\[1pt]
$M_0$  [Deg]  &  $259.2519477(12)$  &  $109.8717438(231)$  \\[1pt]
$f_0$  [Deg]  &  $112.1548376(3)$  &  $81.5877236(50)$  \\[1pt]
$r_0$ [km] &  $5985144906(22)$ &  $3350363070(598)$ \\ [1pt]
$\alpha$  &  $0.3252905546(4)$  &  $-0.2491819783(41)$ \\[1pt]
$\beta$  &  $0.8446147582(66)$  &  $-0.9625930916(22)$  \\[1pt]
$\gamma$  &  $0.4252199023(133)$  &  $-0.1064090300(223)$ \\[1pt]
$\ell_0$ [Deg]  &  $70.98784378(2)$  &  $-105.06917250(31)$ \\[1pt]
$b_0$ [Deg]  &  $3.10485024(85)$  &  $16.57492890(127)$ \\[1pt]
\hline 
\end{tabular} 
\end{center} 
\end{table}

%************* End Appendix Table *******************

%*************************************BIBLIOGRAPHY

%*****************************************************************

%*****************************************************************


\begin{thebibliography}{99}

\bibitem{science} See the special issue of Science {\bf 183},
No. 4122, 25 January 1974;
specifically, J. D. Anderson, G. W. Null, and S. K. Wong, 
Science {\bf 183}, 322 (1974).

\bibitem{piopr} R. O. Fimmel, W. Swindell, and E. Burgess, 
{\it Pioneer Odyssey: Encounter with a Giant}, 
NASA document No. SP-349 (NASA, Washington 
D.C., 1974).

\bibitem{piopr2} R. O. Fimmel, J. Van Allen, and E. Burgess, 
{\it Pioneer: First to Jupiter, Saturn, and beyond}, 
NASA report NASA--SP-446 (NASA, Washington D.C., 1980).  

\bibitem{piodoc} 
{\it Pioneer F/G Project: Spacecraft Operational Characteristics},
Pioneer Project NASA/ARC document No. PC-202  
(NASA, Washington, D.C., 1971).   

\bibitem{extended} {\it Pioneer Extended Mission Plan}, Revised,
NASA/ARC document No. PC-1001 (NASA, Washington, D.C., 1994).

\bibitem{pioweb} For web summaries of Pioneer, go to: 
{\tt http://quest.arc.nasa.gov/pioneer10},     \\
{\tt http://spaceprojects.arc.nasa.gov/    \\
Space\_Projects/ pioneer/PNhome.html}

\bibitem{jdakuiper}  J. D. Anderson, E. L. Lau, K. Scherer, D. C. 
Rosenbaum, and V. L. Teplitz, 
%``Kuiper Belt Constant from Pioneer 10'', 
{Icarus} {\bf 131}, 167  (1998). 

\bibitem{ephem} J. D. Anderson, E. L. Lau, 
T. P. Krisher,  D. A. Dicus, D. C. Rosenbaum,
and V. L. Teplitz, {Astrophys. J.} {\bf 448}, 885 (1995).

\bibitem{pulsar} K. Scherer, H. Fichtner, J.~D. Anderson, 
and E.~L. Lau, 
%``A Pulsar, the heliosphere, and Pioneer 10: probable Mimicking 
%of a Planet of PSR B1257+12 by Solar Rotation'', 
{ Science} {\bf 278}, 1919 (1997).

\bibitem{anderson85} J.~D. Anderson and  B. Mashoon,
%``Pioneer 10 Search for Gravitational waves -- Limits on a Possible
%Isotopic Cosmic Background of Radiation in the Microhertz Region,'' 
{Astrophys. J.} {\bf 290}, 445 (1985).

\bibitem{anderson93} J.~D. Anderson, J.~W. Armstrong, and E.~L.Lau,
%``Upper Limits for Gravitational Radiation from Supermassive Binaries,''
{Astrophys. J.} {\bf 408}, 287  (1993). 

\bibitem{anderson} J. D. Anderson, P. A. Laing, E. L. Lau, A. S. Liu,  
M. M. Nieto, and S. G. Turyshev, Phys. Rev. Lett. {\bf 81}, 2858 (1998).
Eprint gr-qc/9808081.

\bibitem{moriond} S. G. Turyshev, J. D. Anderson, P. A. Laing, 
E. L. Lau, A. S. Liu, and  M. M. Nieto, 
% ``The Apparent Anomalous, 
% Weak, Long-Range Acceleration  of Pioneer 10 and 11,'' 
in: {\it Gravitational Waves and Experimental Gravity, 
Proceedings of the {XVIIIth} Moriond Workshop of the  
Rencontres de Moriond}, ed. by J. Dumarchez and J. Tran Thanh Van
(World Pub., Hanoi,  2000), pp. 481-486. 
Eprint gr-qc/9903024.

%#% {\bf End of references for Section \ref{intro}.***************}




\bibitem{design}  There were four Pioneers built of this particular
design.  After testing, the best components were placed in Pioneer 10.
(This is probably why Pioneer 10 has lasted so long.)
The next best were placed in Pioneer 11.  The third best were placed in
the ``proof test model.'' Until recently, the structure and many components
of this model  were included in an exhibit at the National Air and Space
Museum.   The other model eventually was dismantled. We thank Robert
Ryan of JPL for telling us this.

\bibitem{mass}  
Figures given for the mass of the entire 
Pioneer package range from  under 250 kg to over 315 kg.  
However, we eventually 
found that the total (``wet'') weight at launch was 259 kg (571 lbs),   
including 36 kg of hydrazine (79.4 lbs).  Credit and thanks for 
these numbers are  
due to Randall Rathbun,  Allen Parker, and  Bruce A. Giles  of 
TRW, who checked and rechecked for us including going to their launch 
logs.   Consistent total mass with lower fuel (27 kg) 
numbers were given by  Larry Kellogg of NASA/Ames.   
(We also thank V. J. Slabinski of USNO 
who first asked us about the mass.)

\bibitem{gasuse}  
Information about the gas usage is by this time 
difficult to find or
lost.  During the Extended Mission the collaboration was most concerned with
power to the craft.  The folklore is that 
most of Pioneer 11's propellant was used up going to Saturn and
used very little for Pioneer 10.    
In particular, a Pioneer 10 nominal input mass of 251.883 kg and
a Pioneer 11 mass of 239.73 kg were used by the JPL program and the
Aerospace program used 251.883 for both.   
The 251 number approximates the mass lost during spin down, and the 239
number models the greater fuel usage.  These numbers were not 
changed in the programs.  For reference, we will use 241 kg, the mass with
half the fuel used, as our number with which to calibrate systematics.  

\bibitem{vanallen} We take this number from Ref. \cite{piodoc}, where the
design, boom-deployed moment of inertia is given as 
433.9 slug (ft)$^2$ (= 588.3 kg m$^2$).  This should be a little low since
we know a small amount of mass was added later in the development.  
A much later order-of-magnitude number 
770 kg m$^2$ was  
obtained with a too large mass \cite{mass,gasuse}.  See 
J.~A. Van Allen,  {\it Episodic Rate of Change in
Spin Rate of Pioneer 10,} Pioneer Project Memoranda, 
20 March 1991 and 5 April 1991.  Both numbers are dominated by the RTGs and
magnetometer at the ends of long booms.  

\bibitem{conscan}  Conscan stands for conical scan.  The receiving antenna
is moved in circles of angular size corresponding to one half  of the
beam-width of the incoming signal.  This  procedure, possibly iterated,
allows the correct  pointing direction of the antenna to be found.   When
coupled with a maneuver,  it can also be used to find the correct pointing
direction for the spacecraft antenna.  The precession maneuvers can be
open-loop, for orientation towards or away from Earth-pointing, or
closed-loop, for homing on the uplink radio-frequency transmission from
the Earth.   

\bibitem{rearfront} When a Pioneer antenna points toward the Earth, this 
defines the ``rear''  direction on the spacecraft.  The equipment
compartment placed on the other side of of the antenna defines the
``front'' direction on the spacecraft.  (See Figure \ref{fig:trusters}.)

\bibitem{tele} 
{\it SNAP 19 Pioneer F \& G Final Report},
Teledyne report IESD 2873-172, June, 1973, tech. report No. 
DOE/ET/13512-T1; DE85017964,  gov. doc. \#  E 1.9, and 
S. T. Christenbury, private communications.  

\bibitem{Rconf} F. A. Russo, in: {\it Proceedings 
of the 3rd RTG Working Group Meeting} (Atomic Energy Commission,
Washington, DC, 1972), ed. by  P. A. O'Rieordan, papers \# 15 and 16.

\bibitem{lasher} L. Lasher, Pioneer Project Manager, recently 
reminded us (March 2000)
that not long after launch, the electrical power had decreased to about
155 W, and degraded from there.  [Plots of the available power with time
are available.]

\bibitem{theorypower}
This is a ``theoretical value,'' which does not account for inverter
losses, line losses, and such. It is interesting to note
that at mission acceptance, the total ``theoretical'' power was 175 Watts.

\bibitem{sband} We take the S-band to be defined by the frequencies
1.55-5.20 GHz. We take the X-band to be defined by the frequencies
5.20-10.90 GHz. It turns out there is no consistent international
definition of these bands.  The definitions vary from field to field, with
geography, and over time.  The above definitions are those used by radio
engineers and  are consistent with the DSN usage.   
(Some detailed band definitions can be found at 
{\tt http://www.eecs.wsu.edu/$\sim$hudson/
Teaching/ee432/spectrum.htm}.)
% {\tt http://advanix.net/{\~}{\hskip
% -0pt}neuhaus/fccindex/letter.html})  
[We especially thank Ralph McConahy of DSN
Goldstone on this point.]

\bibitem{dBm}  dBm is used by radio engineers as a measure of received
power. It stands for decibels in milliwatts.  

\bibitem{johnson} 
For a description of the Galileo mission see T. V. Johnson, C. 
M. Yeats, and R. Young, {Space Sci. Rev.} {\bf 60}, 3--21 (1992).  For a 
description of the trajectory design see L. A. D'Amario,  L. E. Bright, 
and A. A. Wolf, {Space Sci. Rev.} {\bf 60}, 22--78 (1992).

\bibitem{LGA} 
The LGA was originally supposed to ``trickle'' down low-rate engineering 
data.  It was also to be utilized in case a fault resulted in the 
spacecraft ``safing''   and shifting to a 
Sun-pointed attitude, resulting in loss of signal from the HGA. 
[``Safing''  refers to a spacecraft entering the so called 
``safe mode.''  This happens  in case  of an emergency  
when  systems are shut down.]

\bibitem{anderson75} J.~D. Anderson, P.~B. Esposito, W. Martin, C.~L.
Thornton, and D.~O. Muhleman, 
%``Experimental Tests of General
%Relativity  Using Time-Delay Data From Mariner 6 and Mariner 7,''
{Astrophys. J.} {\bf 200}, 221  (1975).

\bibitem{Kinman92} P. W. Kinman, 
%   ``Doppler Tracking of Planetary Spacecraft,'' 
IEEE Transactions on Microwave Theory and Techniques {\bf 40}, 1199 (1992).  

\bibitem{genU}  For  descriptions of the Ulysses mission see E. J. Smith 
and R. G. Marsden, Sci. Am. {\bf 278}, No. 1, 74 (1998); 
B. M. Bonnet, Alexander von Humboldt Magazin, No. 72, 27 (1998).

%#% {\bf This is the end of the references for Section 
%#%      \ref{pioneer}.******************} 
 



\bibitem{dsn} A technical description, with a history and photographs, 
of the Deep Space Network can be found at 
{\tt http://deepspace.jpl.nasa.gov/dsn/}.  
The document describing the radio science system is at 
{\tt http://deepspace.jpl.nasa.gov/dsndocs/810-5/ 810-5.html}.

\bibitem{dsn82} 
N.~A. Renzetti, J.~F. Jordan, A.~L. Berman, J.~A. Wackley, T.~P.
Yunck,  {\it The Deep Space Network -- An Instrument for Radio
Navigation of Deep Space Probes}, Jet Propulsion Laboratory  Technical
Report 82-102 (1982).

\bibitem{barnes} J.~A. Barnes, A.~R. Chi, L.~S. Cutler, D.~J. Healey,
D.~B. Leeson, T.~E. McGunigal, J.~A. Mullen, Jr., W.~L. Smith, 
R.~L. Sydnor, R.~F.~C. Vessot, and G.~M.~R. Winkler, 
%``Characterization of Frequency Stability,''  
{IEEE Transactions on Instrumentation and Measurement} {\bf 20},
105 (1971). 

\bibitem{vessot74} R.~F.~C. Vessot, in: {\it Experimental
Gravitation}, ed. B. Bertotti,  
( Academic Press, New York and London, 1974), p.111.

\bibitem{SFJ98}
O.~J. Sovers,  J.~L. Fanselow, and C.~S. Jacobs,
%``Astrometry and geodesy with radio interferometry: experiments,
% models, results,'' 
{Rev. Mod. Phys.}, {\bf 70}, 1393 (1998).

\bibitem{way} One-way data refers to a transmission and reception, only.  
Two-way data is a transmission and reception, followed by a retransmission
and reception at the original transmission site.  This would be, for
example, a transmission from a radio antenna on Earth to a spacecraft and 
then a retransmission back from the spacecraft to the same antenna. 
Three-way refers to the same as two-way, except the final receiving antenna
is different from the original transmitting antenna.  

\bibitem{datatapes}  Much, but not all, of the data we used has been
archived. Since the Extended Pioneer Mission is complete, the resources
have not been available to properly convert the entire data set to
easily accessible format. 

\bibitem{drift}  The JPL and DSN convention for Doppler frequency shift is
$(\Delta \nu)_{\tt DSN} = \nu_0 - \nu$, where $\nu$ is the measured
frequency and 
$\nu_0$ is the reference frequency. It is positive for a spacecraft
receding from the tracking station (red shift), and negative for a
spacecraft approaching the station (blue shift), just the opposite of the
usual convention, $(\Delta \nu)_{\tt usual} = \nu - \nu_0$.  
In consequence, the velocity shift, 
$\Delta v = v - v_0$, has the same sign as 
$(\Delta \nu)_{\tt DSN}$ but the opposite sign to 
$(\Delta \nu)_{\tt usual}$.  Unless otherwise stated, we will use the 
DSN frequency shift convention in this paper.  
We thank Matthew Edwards for asking us about this. 

\bibitem{falsedop} As we will come to in Section \ref{sec:timemodel},  
this property allowed us to test and reject several 
phenomenological models of the anomalous acceleration 
that fit Doppler data well but failed to fit the range data. 

\bibitem{cane} D. L. Cain, JPL Technical Report (1966).  

\bibitem{Moyer71}  T. D.  Moyer,  {\it Mathematical Formulation of the
Double Precision Orbit Determination Program (DPODP)}, 
Jet Propulsion Laboratory  Technical Report 32-1527 (1971). 

\bibitem{Moyer00} 
T. D.  Moyer, {\it Formulation for Observed and
Computed Values of Deep Space Network (DSN) Data Types
for Navigation}, JPL Publication 00-7 (October 2000).

\bibitem{Gelb} A. Gelb, ed.
{\it Applied Optimal Estimation} 
(M.I.T. Press, Cambridge, MA, 1974).

\bibitem{MuhlemanAnderson81} D.~O. Muhleman and J.~D. Anderson,
%``Solar Wind Electron Densities From Viking 
%Dual-Frequency Radio Measurements,'' 
{Astrophys. J.} {\bf 247}, 1093 (1981).

%#% {\bf This is the end of the references to Section \ref{Exp_tech}.
%#% ********************}





\bibitem{massprog} 
Once in deep space, all major forces on the
spacecraft are gravitational. The Principle of Equivalence holds that
the inertial mass ($m_I$) and the gravitational mass ($m_G$) are
equal.   This means the mass of the craft should cancel out in the
dynamical  gravitational equations.  As a result, 
the people who designed early deep-space programs were not as worried 
as we are about having the correct mass.   When  non-gravitational
forces were modeled, an incorrect mass could be accounted for by
modifying other constants.  For example, in the solar radiation
pressure force the effective sizes of the  antenna and the albedo could
take care of mass inaccuracies.  

\bibitem{anderson74} J.~D. Anderson, in: {\it Experimental
Gravitation}, ed. B. Bertotti  
(New York and London, Academic Press, 1974),  p.163.

\bibitem{dsn86} 
J.~D. Anderson, G.~S. Levy, and   N.~A. Renzetti, ``Application of the
Deep Space Network (DSN) to the testing of general relativity,'' in 
{\it Relativity in Celestial Mechanics and Astrometry},  
eds. J. Kovalevsky and V.~A. Brumberg. (Kluwer Academic, 
Dordrecht, Boston, 1986), p. 329.

\bibitem{Newhall83}
X~X Newhall, E.~M. Standish, and J.~G. Williams,
% ``DE 102: A numerically integrated ephemeris of the Moon and planets
% spanning forty-four centuries,'' 
Astron. Astrophys. {\bf 125}, 150
(1983).

\bibitem{Standish92}
E. M. Standish, Jr., X~X Newhall, J. G. Williams, and D. K. Yeomans, 
     ``Orbital ephemeris of the Sun, Moon, and Planets,'' in: 
Ref. \cite{exp_cat}, p. 279. 
%    {\it Explanatory Supplement to the Astronomical Almanac}, ed. 
%    P. K. Seidelmann (University Science Books, Mill Valley, CA), p.279.  
Also see E. M. Standish, Jr. and R. W. Hellings, 
Icarus {\bf 80}, 326 (1989). 

\bibitem{Standish95a}
E. M. Standish, Jr., X~X Newhall, J. G. Williams, and W. M. Folkner, 
{\it JPL Planetary and Lunar Ephemeris, DE403/LE403}, 
Jet Propulsion Laboratory  Internal IOM No. 314.10-127 (1995). 

\bibitem{Will93} C. M. Will, {\it Theory and Experiment in Gravitational
Physics}, (Rev. Ed.) (Cambridge University Press, Cambridge, 1993). 

\bibitem{WillNordtvedt72}
C. M.  Will and K. Nordtvedt, Jr, {Astrophys. J.} {\bf 177}, 757 (1972).

\bibitem{estabrook69}
F.~E. Estabrook,  {Astrophys. J.} {\bf 158}, 81 (1969).

\bibitem{Moyer81}  T. D. Moyer,  
%{\it Transformation From Proper Time on Earth to Coordinate 
%Time in Solar System barycentric Space-Time Frame of Reference,} 
Parts. 1 and 2, Celest. Mech. {\bf 23}, 33, 57 (1981).

\bibitem{exp_cat} P. K. Seidelmann, ed., 
{\it Explanatory Supplement to the Astronomical
Almanac} (University Science Books, Mill
Valley, CA, 1992).

\bibitem{Ma98}
C. Ma,  E. F. Arias, T. M. Eubanks, A. L. Fey, A.-M. Gontier,
C. S. Jacobs, O. J. Sovers, B. A. Archinal, and P. Charlot,
% ``The International Celestial Reference Frame based on VLBI 
% Observations of Extragalactic Radio Sources'',
Astron. J. {\bf 116},  516
%-546 
(1998).

\bibitem{milani} A. Milani, A. M. Nobili, and P. Farinella, {\it
Non-Gravitational Perturbations and Satellite Geodesy}, 
(Adam Hilger, Bristol, 1987).  See, especially,  p. 125.

\bibitem{longuski} J.~M. Longuski, R.~E. Todd, and W.~W. K\"onig,
%{\it Survey of Nongravitational Forces and Space Environmental 
%Torques: Applied to the Galileo},  
J. Guidance, Control, and Dynamics, {\bf 15}, 545 (1992).

\bibitem{MuhlemanEspositoAnderson77}  D.O. Muhleman, P.B. Esposito, and 
J. D. Anderson, Astrophys. J. {\bf 211}, 943 (1977).

\bibitem{slavanote} The propagation speed for the Doppler signal is the
phase velocity, which is greater than $c$.  Hence, the negative sign in
Eq. (\ref{eq:sol_plasma}) applies. The ranging signal propagates at the
group velocity, which is less than $c$.  Hence, 
there the positive sign applies. 

\bibitem{someone} B.-G. Anderssen and S. G. Turyshev, JPL 
Internal IOM 1998-0625, and references therein.

\bibitem{bird} M.~K. Bird, H. Volland, M. P\"{a}tzold, P. Edenhofer , S.~W.
Asmar and J.~P. Brenkle, {Astrophys. J.} {\bf 426}, 373 (1994).  

\bibitem{scunits} The units conversion factor 
for $A,B,C$ from m to cm$^{-3}$  
is $2 N_c(S)/R_\odot = 0.01877$, where $N_c(S) =1.240 \times 10^4 \nu^2$ 
is the S band (in MHz) critical plasma density, and $R_\odot$ is the radius
of the Sun.

\bibitem{Ekelund} These values of parameters $A, B, C$ were kindly provided
to us by   John E. Ekelund of JPL. They represent the best solution for the
solar corona parameters obtained during his simulations of the 
solar conjunction experiments that will be performed with the Cassini
mission spacecraft in 2001 and 2002. 

\bibitem{F10-7}  This model is explained and described at \\
{\tt http://science.msfc.nasa.gov/ssl/pad/solar/ predict.htm}

\bibitem{Muhleman}
These come from the adjustment in the system of data weights (inverse of
the variance on each  measurement) for Mariner 6/7 range
measurements.  Private communication by Inter-office Memorandum  from 
D.~O. Muhleman of Caltech to P.~B. Esposito of JPL, dated 7 July 1971.

\bibitem{null81} G.~W. Null, E.~L. Lau, E. D. Biller, and J.~D. 
Anderson, 
%``Saturn gravity results obtained from Pioneer 11
%tracking data and Earth-based Saturn satellite data,'' 
{Astron. J.} {\bf 86}, 456 (1981).

\bibitem{Laing91} P.~A. Laing, ``Implementation of J2000.0 reference frame
in CHASMP,'' The Aerospace Corporation's  Internal Memorandum \#
91(6703)-1. January 28, 1991.

\bibitem{Lieske76} J.~H. Lieske, 
%``Precession Matrix Based on IAU (1976)
%System of  Astronomical Constants'',  
{ Astron. Astrophys.} {\bf 73}, 282
(1979).  Also, see {\it FK5/J2000.0 for DSM Applications,} Applied
Technology Associates,  6 June 1985.

\bibitem{Standish82} E.~M. Standish,
%``Orientation of the JPL Ephemerides,
%DE 200/LE 200, to   the Dynamical Equinox of J2000'', 
{Astron. Astrophys.} {\bf 114}, 297 (1982)

\bibitem{sherman} J. Sherman and W. Morrison. Ann. Math. Stat. {\bf  21},
124 (1949)

%#% {\bf This is the end of the reference for Section
%#% \ref{navigate}.** *******************} 




\bibitem{jpl} J. D. Anderson, Quarterly Report to NASA/Ames
Research Center,  
{\it Celestial Mechanics Experiment, Pioneer 10/11,} 22 July 1992.
Also see the later quarterly report for the period 1 Oct. 1992 to 
31 Dec. 1992, dated 17 Dec. 1992,  Letter of Agreement ARC/PP017.  
This last, specifically, contains the present 
Figure \ref{fig:correlation}.  

\bibitem{sunearth} We only measure Earth-spacecraft Doppler frequency and, 
as we will discuss in Sec. \ref{radioantbeam}, 
the down link antenna yields a conical beam  of width 3.6 degrees at 
half-maximum power.  Therefore,  
between Pioneer 10's past and present (May 2001) 
distances of  20 to 78 AU, 
the Earth-spacecraft line and Sun-spacecraft line are so close
that one can not resolve whether the force direction is towards
the Sun or if the force direction is towards the Earth.  
If we could have used a longer arc fit that started earlier and hence 
closer, we might have able to separate the Sun direction from the 
Earth direction.

\bibitem{bled} A preliminary discussion of these results appeared in
M. M. Nieto, T. Goldman, J. D. Anderson, E. L. Lau, and
J. P\'erez-Mercader, in: {\it  Proc. Third Biennial Conference
on Low-Energy Antiproton Physics, LEAP'94},
ed. by G. Kernel, P. Krizan, and M. Mikuz
(World Scientific, Singapore, 1995), p. 606.  Eprint hep-ph/9412234.

\bibitem{gleaned} Since both the gravitational and radiation
pressure forces become so large close to the Sun, the anomalous
contribution  close to the Sun 
in Figures  \ref{fig:forces} and \ref{fig:correlation} 
is meant to represent only what anomaly
can be gleaned from the data, not a measurement.  

\bibitem{tap} B. D. Tapley, in {\it Recent Advances in Dynamical
Astronomy}, eds. B. D. Tapley and V. Szebehely (Reidel, Boston, 1973), 
p.396.

\bibitem{chasmp} P. A. Laing, {\it Thirty Years of CHASMP}, Aerospace
report (in preparation).

\bibitem{poeas} P. A. Laing, {\it Software Specification Document, Radio
Science Subsystem, Planetary Orbiter  Error Analysis Study Program (POEAS)},
Jet Propulsion  Laboratory Technical Report DUK-5127-OP-D,
19  February 1981.  POEAS was originally developed
to support the Mariner Mars program.

\bibitem{aero}  P. A. Laing  and A. S. Liu.  NASA Interim Technical
Report, Grant NAGW-4968, 4 October 1996.

\bibitem{aT} Galileo is less sensitive to either an  $a_P$- or an  
$a_t= a_P/c$-model effect than the Pioneers. 
Pioneers have a smaller solar pressure and a longer light travel time. 
Sensitivity to a clock acceleration is 
proportional to the light travel time squared.

\bibitem{uly} T. McElrath, private communication.

\bibitem{mcelrath} T. P. McElrath,  S. W. Thurman, and K. E. Criddle,
% {\it Navigation Demonstrations of Precision Ranging with the 
% Ulysses Spacecraft,, 
in {\it Astrodynamics 1993}, edited by A. K. Misra, V. J. Modi,R. 
Holdaway, and P. M. Bainum (Univelt, San Diego CA, 1994), 
Ad. Astodynamical Sci. {\bf 85}, 
Part II, p. 1635, paper No. AAS 93-687. 
% Paper presented at the AAS/AIAA Astrodynamic Specialist Conference, 
% Victoria, B. C., Canada, August 16--19, 1993, AAS Publications Office, 

\bibitem{ulygas} The gas leaks found in the Pioneers 
are about an order of 
magnitude too small to explain $a_P$.  
Even so, we feel that some systematic or combination of 
systematics (such as heat or gas leaks) 
will likely explain the anomaly.  However, 
such an explanation has yet to be demonstrated.  
We will discuss his point further in Sections 
\ref{recent_results} and  \ref{int-systema}. 

%#% {\bf This is the current reference list up until the end of Sec. 
%#%   \ref{results}.**************}



\bibitem{AU}  More information on the ``Heliocentric Trajectories for 
Selected Spacecraft, Planets, and Comets,'' can be found at 
{\tt http://nssdc.gsfc.nasa.gov/space/ helios/heli.html}. 

\bibitem{I/II}  ODP/\emph{Sigma} took the Interval I/II boundary as 
22 July 1990, the date of a maneuver.  CHASMP took this boundary
date as  31 August 1990, when a clear anomaly in the spin data was seen. 
We have checked, and these choices produce less than one percent
differences in the results.

\bibitem{sigma} J. A. Estefan, L. R. Stavert, F. M. Stienon, A. H. Taylor,
T. M. Wang, and P. J. Wolff, {\it Sigma User's Guide.  Navigation
Filtering/Mapping Program}, JPL document 699-FSOUG/NAV-601
(Revised: 14 Dec. 1998).   

%#% {\bf This is the current reference list up until the end of Sec. 
%#%   \ref{recent_results}.**************}




\bibitem{null76} G. W. Null,  
%``Gravity field of Jupiter and its
%satellites from Pioneer 10 and Pioneer 11 tracking data,'' 
{Astron. J.} {\bf 81}, 1153 (1976).

\bibitem{rad} R.~M. Georgevic,  {\it Mathematical model of the solar
radiation force and torques acting on the components of a
spacecraft,} JPL Technical Memorandum 33-494 (1971).

\bibitem{solar_irr} Data is available at 
{\tt http://www.ngdc.noaa.gov/stp/ SOLAR/IRRADIANCE/irrad.html}

\bibitem{Lambda} 
For an ideal flat surface facing the Sun, $\mathcal{K} = 
(\alpha + 2\epsilon) = (1 +2\mu + 2 \nu)$.  $\alpha$ and $\epsilon$ are,
respectively, the absorption and
reflection coefficients of the spacecraft's surface. ODP uses the second
formulation in terms of reflectivity coefficients, ODP's input $\mu$ and
$\nu$ for Pioneer 10, are obtained from design information and 
early fits to the data. (See the following paragraph.)  These numbers 
by themselves yield $\mathcal{K_0}= 1.71$.  When a first (negative)
correction is made for the  antenna's parabolic surface, 
$\mathcal{K}\rightarrow 1.66$.  

\bibitem{effect} 
There are complicating effects that modify the ideal antenna.  The craft
actually has multiple different-shaped surfaces (such as the RTGs), that are 
composed of different materials oriented at different angles to the spin 
axis, and which degrade with time.  
But far from the Sun, and given $M$ and $A$, the sum of all such
corrections can be subsumed,  for our purposes, in an {\it effective} 
$\mathcal{K}$.  It is still expected to be of order 1.7.   

\bibitem{sunparam}   
Eq. (\ref{eq:srp}) is combined with information on the spacecraft
surface geometry and it's local orientation to determine 
the magnitude of its solar radiation acceleration as it faces the Sun.  
As with  other  non-gravitational forces, an incorrect 
mass in modeling the solar radiation pressure force can be
accounted for by modifying other constants such as the effective sizes of
the  antenna and the albedo. 

\bibitem{edsmith} E. J. Smith, L. Davis, Jr., D. E. Jones, D. S. Colburn,
P. J. Coleman, Jr., P. Dyal, and C. P. Sonnett, 
Science {\bf 183}, 306 (1974); {\it ibid}. {\bf 188}, 451 (1975).  

\bibitem{lorentz}  This result was obtained from a limit for  positive charge 
on the spacecraft \cite{null76}.  No measurement dealt with negative charge, 
but such a charge would have to be proportionally larger to have a
significant effect.

\bibitem{malhotra} R. Malhotra, Astron. J. {\bf 110}, 420 (1995);
{\it ibid.} {\bf 111}, 504 (1996).
% THE ORIGIN OF PLUTOS ORBIT : IMPLICATIONS FOR THE SOLAR-SYSTEM BEYOND
% NEPTUNE
% The phase space structure near Neptune resonances in the Kuiper Belt

\bibitem{liupeale}  A. P. Boss and S. J. Peale, Icarus {\bf 27}, 119 (1976).  

\bibitem{liudust}  A. S. Liu, J. D. Anderson, and E. Lau, Proc. 
AGU (Fall Meeting, San Francisco, 16-18 December 1996), paper \# SH22B-05.  

\bibitem{backman} G. E. Backman, A. Dasgupta, and R. E. Stencel, 
{Astrophys. J.} {\bf 450}, L35 (1995).  Also see S. A. Stern, Astron.
Astrophys.  {\bf 310}, 999 (1996).  

\bibitem{teplitzinfra}  V. L. Teplitz, S. A. Stern, 
J. D. Anderson, D. Rosenbaum, R. J. Scalise, and P. Wentzler, 
{Astrophys. J.} {\bf 516}, 425 (1999). 

\bibitem{gio} J.~D. Anderson, G. Giampieri, P.~A. Laing, and E.~L. Lau, 
work in progress.
% ``Pioneer 10 Encounter with a Trans-Neptunian Object at 56 AU?''
% Paper \#26.04, Annual AAS DPS Meeting \#31, Padova, Italy.
% Bull. Am. Astron. Soc. (1999).  

\bibitem{vessot_clocks} R.~F.~C. Vessot, ``Space experiments
with high stability clocks,'' in proceedings of the ``Workshop on the
Scientific Applications of Clocks in Space,'' 
(November 7-8, 1996. Pasadena, CA). Edited by  L. Maleki. 
JPL Publication 97-15 (JPL, Pasadena, CA, 1997), p. 67.

\bibitem{SoversJacobs96} 
O. J.  Sovers and C. S. Jacobs,   
    {\it Observational Model and Parameter Partials for the JPL VLBI
     Parameter  Estimation Software ``MODEST'' - 1996}, Jet Propulsion
     Laboratory  Technical  Report 83-39, Rev. 6 (1996).

%#% {\bf This ends the Refs. of Section \ref{ext-systema}.****************}





\bibitem{katz} J. I. Katz,  Phys. Rev. Lett. {\bf 83}, 1892 (1999).

\bibitem{s}  
There is an intuitive way to understand this.  Set up a coordinate system 
at the closest axial point of an RTG pair.  Have 
the antenna be in the (+z,-x) direction, and the RTG pair in the positive x
direction. Then from the RTGs the antenna is in 1/4 of a sphere 
(positive z and negative x). The `antenna occupies about 1/3 of 180 degrees
in azimuthal angle.  Its form is the base part of the parabola. Thus, 
it resembles a ``flat'' triangle of the same width, producing another 
factor of $\sim (1/2-2/3)$  compared to the angular size of a rectangle.  
It occupies of order (1/4-1/3) of the latitudinal-angle phase space
angle of 90$^o$.  This yields a total reduction factor of 
$\sim(1/96-2/108)$, or about 1 to 2 \%.  

\bibitem{ss} The value of 1.5\%  is obtained 
by doing an explicit calculation of the solid angle subtended  by the antenna 
from the middle of the RTG modules using the Pioneer's physical dimensions.  
V. J. Slabinski of USNO independently obtained a figure of 1.6\%. 

\bibitem{heatreflect}  Our high estimate of 40 W is not compromised by
imprecise geometry.  If the RTGs were completely in the plane
of the top of the dish, then the maximum factor multiplying the 40 W 
directed power 
would be $\kappa_z = 1$.  This would presume all the energy was
reflected and/or absorbed and re-emitted towards the rear of the
craft. (If the RTGs were underneath the antenna, then the total
factor could ideally go as high as "2", from adding the RTG heat 
going out the opposite direction.)  
The real situation is that the average sine of the latitudinal angle up
to the antenna from the RTGs is about 0.3.  This means that the 
heat gong out the opposite direction might cause an 
effective factor $\kappa_z$ to go as high as 1.3.  However, the real
reflection off of the antenna is not straight backwards.  It is
closer to 45$^o$. The absorbed and re-emitted radiation is also at an
angle to the rotation axis, although smaller.  (This does not even
consider reflected/reemitted heat that 
does not go directly backwards but rather bounces off of the central
compartment.)  So, the original estimate of $\kappa=1$ is a
good bound.   

\bibitem{uskatz} J. D. Anderson, P. A. Laing, E. L. Lau, A. S. Liu,  
M. M. Nieto, and S. G. Turyshev,
Phys. Rev. Lett. {\bf 83}, 1893 (1999).

\bibitem{slusher} We acknowledge R. E. Slusher of Bell Labs for raising this 
possibility.  

\bibitem{camera} B. A. Smith, G. A. Briggs, G. E. Danielson, A. F. Cook,
II, M. E. Davies, G. E. Hunt, H. Masursky, L. A. Soderblom, T. C. Owen,
C. Sagan, and V. E. Suomi, Space Sci. Rev. {\bf 21}, 103 (1977).

\bibitem{plate}  C. E. Kohlhase and P. A. Penzo, 
Space Sci. Rev. {\bf 21}, 77 (1977).

\bibitem{hansen}   We are grateful to C. J. Hansen of JPL, who  kindly
provided us with operational information on the Voyager video cameras.  

\bibitem{neptune} B. A. Smith,  L. A. Soderblom, D. Banfield, C. Barnet, 
T. Basilevsky, R. F. Beebe, K. Bollinger, J. M. Boyce, 
A. Brahic, G. A. Briggs, R. H. Brown, C. Chyba, S. A. Collins, T. Colvin, 
A. F. Cook, II, D. Crisp, S. K. Croft, D. Cruikshank, J. N. Cuzzi,
G. E. Danielson, M. E. Davies, E. De Jong, L. Dones, D. Godfrey, J. Goguen,
I. Grenier, V. R. Haemmerle, H. Hammel, C. J. Hansen, C. P. Helfenstein,
C. Howell, G. E. Hunt, A. P. Ingersoll, T. V. Johnson, J. Kargel, R. Kirk,
D. I. Kuehn, S. Limaye, H. Masursky, A. McEwen, D. Morrison, T. Owen,
W. Owen, J. B. Pollack, C. C. Porco, K. Rages, P. Rogers, D. Rudy,
C. Sagan, J. Schwartz, E. M. Shoemaker, M. Showalter, B. Sicardy,
D. Simonelli, J. Spencer, L. A. Sromovsky, C. Stoker, R. G. Strom,
V. E. Suomi, S. P. Synott, R. J. Terrile, P. Thomas, W. R. Thompson,
A. Verbiscer, and J. Veverka,  
{ Science} {\bf 246}, 1432 {1989}.  

\bibitem{murphy} E. M. Murphy,  Phys. Rev. Lett. {\bf 83}, 1890 (1999).

\bibitem{usmurphy} J. D. Anderson, P. A. Laing, E. L. Lau, A. S. Liu,  
M. M. Nieto, and S. G. Turyshev, 
{ Phys. Rev. Lett.} {\bf 83}, 1891 (1999).

\bibitem{scheffer} L. K. Scheffer, (a) eprint gr-qc/0106010, the original
modification; (b) eprint gr-qc/0107092;  (c) eprint gr-qc/0108054.

\bibitem{usscheffer} J. D. Anderson, P. A. Laing, E. L. Lau,  
M. M. Nieto, and S. G. Turyshev,
eprint gr-qc/0107022.

\bibitem{earlydata} These results were not treated 
for systematics,  used different time-evolving estimation procedures,
were done by three separate JPL navigation specialists, 
separated and smoothed by one of us \cite{jpl}, and definitely 
not analyzed with the care of our recent run (1987.0 to 1998.5).
In particular, the 
first two Pioneer 11 points, included in the early memos \cite{jpl}, were 
after Pioneer 11 encountered Jupiter and then was going
back across the central solar system to encounter Saturn. 

\bibitem{puck} T. K. Keenan, R. A. Kent, and R. N. R. Milford, 
{\it Data Sheets for PMC Radioisotopic Fuel}, Los Alamos Report 
LA-4976-MS (1972), available from NTIS. 
We thank Gary Reinhart for finding this data for us. 

\bibitem{hefromrtg}  Diagrams showing the receptacle and the 
bayonet coupling connector were made by the Deutsch Company of
Banning, CA.  (The O-ring was originally planned to be silicon.)   
Diagrams of the receptacle as mounted on the RTGs were made by Teledyne
Isotopes (now Teledyne Brown Engineering).  Once again we gratefully
acknowledge Ted Christenbury for obtaining these documents for us.   

\bibitem{leaks}  In principle, many things could be the origin of some
spin down: structural deformations due to adjustments or aging, thermal
radiation,  leakage of the helium from the RTGs, etc.  But   in the case
of Pioneer spacecraft none of these  provide an explanation for the spin
history exhibited by the Pioneer 10, especially the large unexpected
changes  among the  Intervals I, II, and  III.

%\bibitem{hanson} R. J. Hanson and C. L. Lawson,
%``Extension and Applications of the Householder  Algorithm for 
%Solving Linear Least Squares Problem,'' 
%Mathematics of  Computation,  {\bf 23}(108), 787 (1969)


%#% {\bf This end the Refs. of Section \ref{int-systema}.**************}




\bibitem{herrick} S. Herrick, {\it Astodynamics} (Van Nostrand
Reinhold Co., London, New York, 1971-72). Vols. 1-2,

\bibitem{Folkner} We thank William Folkner of JPL for his
assistance in producing several test files  and invaluable advice.

\bibitem{bc} D. Brouwer and G.~M. Clemence,
{\it Methods of Celestial Mechanics} 
(Academic Press, New York, 1961).

\bibitem{melbourne} W.G. Melbourne, Scientific American {\bf 234}, 
No. 6, 58 (1976).

\bibitem{myles} We thank E. Myles Standish of JPL, who encouraged
us to address in greater detail the nature of the 
annual/diurnal terms seen in the Pioneer Doppler residuals. 
(This work is currently under way.) 
He also kindly provided us with the accuracies from his 
internal JPL solar system ephemeris, which is continually under 
development.  


new, soon to be 
published, solar system ephemeris.  

%#% {\bf This is the current reference list up until the end of Sec. 
%#%   \ref{Int_accuracy}.************}




%#% {\bf This ends the Refs. of Sec. \ref{budget}.************}





\bibitem{crawford}  D. F. Crawford, eprint astro-ph/9904150. 

\bibitem{courtens} N. Didon, J. Perchoux, and E. Courtens, 
Universit\'e de Montpellier preprint (1999).  

\bibitem{dust} D. A. Gurnett, J. A. Ansher, W. S. Kurth, and L. J. 
Granroth, Geophys. Res. Lett. {\bf 24}, 3125 (1997);
M. Landgraf, K. Augustsson, E. Gr\"un, and A. S. Gustafson, 
{ Science} {\bf 286}, 239 (1999).

\bibitem{photon} Pioneer 10 data yielded another fundamental physics 
result, a limit on the rest mass of the photon.  See L. Davis, Jr., A. S. 
Goldhaber, and M. M. Nieto, { Phys. Rev. Lett.} {\bf 35}, 1402 (1975).

\bibitem{jgscold} G. J. Stephenson, Jr., T. Goldman, and B. H. J. McKellar,
{ Int. J. Mod. Phys. A} {\bf 13}, 2765 (1998), hep-ph/9603392.

\bibitem{physrep} M. M. Nieto and T. Goldman, 
{ Phys. Rep.} {\bf 205}, 221 (1991); {\bf 216}, 343 (1992),
and references therein.  

\bibitem{milgrom} J. Bekenstein and M. Milgrom, {Astrophys. J.}
{\bf 286}, 7 (1984); M. Milgrom and J. Bekenstein, in: 
{\it Dark Matter in the Universe}, 
eds. J. Kormendy and G. R. Knapp 
(Kluwer Academic, Dordrecht, Boston, 1987), p. 319; 
M. Milgrom, { La Recherche} {\bf 19}, 182 (1988).

\bibitem{mann}  P. D. Mannheim, {Astrophys. J.} {\bf 419}, 150 (1993).
Also see  K. S. Wood and R. J. Nemiroff, {Astrophys. J.} {\bf  369}, 54
(1991).

\bibitem{saunders} K. G. Begeman, A. H. Broeils, and R. H. Sanders, 
Mon. Not. R. Astron. Soc. {\bf 249}, 523 (1991); T. G. Breimer and R. H. 
Sanders, Astron. Astrophys. {\bf 274}, 96 (1993).

\bibitem{mil}  M. Milgrom, {Ann. Phys. (NY)}  {\bf 229}, 384 (1994).  
Also see astro-ph/0112069.

\bibitem{Capozzielloetal} S. Capozziello, S. De Martino, S. De Siena, 
and F. Illuminati,  Mod. Phys. Lett. A {\bf 16}, 693 (2001). 
Eprint gr-qc/0104052. 
Also see eprint gr-qc/9901042. 

\bibitem{reasenberg} R. D. Reasenberg,  I. I. Shapiro, P. E.
MacNeil, R. B. Goldstein, J. C. Breidenthal, J. P. Brenkle, D. L.
Cain, T. M. Kaufman, T. A. Komarek, and A. I. Zygielbaum, 
{Astrophys. J.} {\bf 234}, L219 (1979).

\bibitem{mg6} J. D. Anderson,  J. K. Campbell, R. F. Jurgens, E.
L. Lau, X X Newhall, M. A. Slade III, and E. M. Standish, Jr.,
in: {\it Proceedings of the Sixth Marcel Grossmann Meeting on
General Relativity},  Part A, ed. H. Sato and T. Nakamura,
(World Scientific, Singapore, 1992), p. 353.

\bibitem{sanmil} R. H. Sanders, private communication to M. Milgrom (1998).  

\bibitem{pe}  The Principle of Equivalence  figure of merit is $a_P/a_N$.  
This is worse than for laboratory experiments (comparing small objects) or
for the Nordtvedt Effect (large objects of planetary size) \cite{Will93}.
It again emphasizes  that the Earth and Mars do not change positions due 
to $a_P$.   

\bibitem{MunyanezaViollier}   F. Munyaneza and R.~D. Viollier, 
eprint astro-ph/9910566.

\bibitem{rosales} J. L. Rosales and J. L. S\'anchez-Gomez,
eprint gr-qc/9810085.  

\bibitem{guru}  V. Guruprasad, eprints astro-ph/9907363,  gr-qc/0005014,
gr-qc/0005090. 

\bibitem{ostvang} D. {\O}stvang, eprint gr-qc/9910054.

\bibitem{belayev} W. B. Belayev, eprint gr-qc/9903016.

\bibitem{modan} G. Modanese, Nucl. Phys. B {\bf 556}, 397 (1999). 
Eprint gr-qc/9903085.

\bibitem{MansouriNasseriKhorrami} R. Mansouri, F. Nasseri and M. Khorrami,
Phys. Lett. A {\bf 259},194 (1999). Eprint gr-qc/9905052.

\bibitem{Sidharth} B. G. Sidharth, Nuovo Cim. {\bf B115}, 151 (2000).   
Eprint astro-ph/9904088.

\bibitem{ivanov} M. A. Ivanov, Gen. Rel. and Grav. {\bf 33}, 479 (2001). 
eprint astro-ph/0005084.  Also see eprint gr-qc/0009043, a contribution 
to the SIGRAV/2000 Congress.  

\bibitem{foot} R. Foot and R. R. Volkas, Phys. Lett. B {\bf 517}, 13
(2001). Eprint gr-qc/0108051.

\bibitem{rey} J. P. Mbelek and M. Lachi\`eze-Rey, eprint gr-qc/9910105.

\bibitem{novati}  S. Calchi Novati, S. Capozziello, and G. Lambiase,
Grav. Cosmol. {\bf 6}, 173 (2000). 
Eprint astro-ph/0005104.

\bibitem{millipulsar}  R. A. Hulse and J. H. Taylor, Astrophys. J. 
{\bf 195}, L51 (1975); J. H. Taylor and J. M. Weisberg, Astrophys. J. 
{\bf 253}, 908 (1982).

\bibitem{consoli_siringo} M. Consoli and F. Siringo, eprint hep-ph/9910372.

\bibitem{consoli} M. Consoli, eprint hep-ph/0002098.

\bibitem{CapozzielloLambiase} S. Capozziello and G. Lambiase, 
Mod. Phys. Lett. A {\bf 14}, 2193 (1999).
Eprint gr-qc/9910026. 

\bibitem{mann2} P. D. Mannheim and D. Kazanas, Astrophys. J. {\bf 342}, 
635 (1989); P. D. Mannheim, Gen. Rel. Grav. {\bf 25}, 697 (1993); 
P. D. Mannheim, Astrophys. J. {\bf 479}, 659 (1997).

\bibitem{moreau}  J. Wood and W. Moreau, eprint gr-qc/0102056.

\bibitem{BertolamiNunes} O. Bertolami and F.~M. Nunes, 
Phys. Lett. B {\bf 452}, 108 (1999).
Eprint hep-ph/9902439.

%#% {\bf This ends the references of Sec. \ref{newphys}.**********}




\bibitem{GAB} L. Iess, G Giampieri, J. D. Anderson, and B. Bertotti, 
{Class. Quant. Grav.} {\bf 16}, 1487 (1999).

\bibitem{sim} R. Danner and S. Unwin, eds.,
{\it SIM Interferometry Mission: Taking the Measure of
the Universe}, NASA document JPL~400-811 (1999).
Also see {\tt http://sim.jpl.nasa.gov/}

\bibitem{tuck}  
The situation may be analogous to what happened in the
1980's to geophysical exploration.  Mine and tower gravity experiments
seemed to indicate anomalous forces with ranges on the order of km
\cite{stacey}.  
But later analyses showed that the experiments had been so precise that 
small inhomogeneities in the field surveys had introduced 
anomalies in the results at this newly precise level \cite{bartlett}.   
But the very important positive outcome was that geophysicists realized 
the point had been reached where more precise studies of systematics 
were necessary.  

\bibitem{stacey}
F. D. Stacey, G. J. Tuck, G. J. Moore, S. C. Holding, B. D. Goodwin,
and R. Zhou, Rev. Mod. Phys. {\bf 59}, 157 (1987); 
D. H. Eckhardt, C. Jekeli, A. R. Lazarewicz, A. J. Romaides, and
R. W. Sands, Phys. Rev. Lett. {\bf 60}, 2567 (1988).

\bibitem{bartlett}
Measurements were more often taken at easily accessible sites, such as
roads, rather than at more inaccessible cites at different heights, 
such as mountain sides or marshes. See 
D. F. Bartlett and W. L. Tew, Phys. Rev. D {\bf 40}, 673 (1989);
{\it. ibid.}, J. Geophys. Res. [Solid Earth Planet] {\bf 95}, 17363 
(1990); C. Jekeli, D. H. Eckhardt, and  A. J. Romaides, 
Phys. Rev. Lett. {\bf 64}, 1204 (1990).  For a review, see 
Section 4 of Ref. \cite{physrep}.

%#% {\bf This ends the references of Sec. \ref{conclusions}.**********}


%#% {\bf This ends the references, period.*************************}

 
\end{thebibliography}
\end{document}